\documentclass[a4paper]{article}
\pdfoutput=1

\usepackage[widelayout,hyperrefblack,copyright]{./research17}

\renewcommand{\mycopyright}{Li, Moser, Wang, Wigger, revised version, 20 January 2020}

\usepackage{threeparttable}
\usepackage{diagbox}
\usepackage{array}
\usepackage{boldline}
\usepackage{caption}
\usepackage{bbm}
\usepackage{algorithmicx}
\usepackage{algpseudocode}

\allowdisplaybreaks
\usetikzlibrary{datavisualization} 

\usetikzlibrary{spy,backgrounds}
\usetikzlibrary{decorations.markings}
\usepackage{appendix}

\newcommand{\C}{\const{C}}

\newcommand{\U}{\mathscr{U}}
\newcommand{\T}{\mathscr{T}}
\newcommand{\mU}{\set{U}}
\newcommand{\mV}{\set{V}}
\newcommand{\mK}{\set{K}}
\newcommand{\mcU}{\cset{U}}
\newcommand{\rvU}{\mathbb{U}}
\newcommand{\rvUstar}{\rvU^{\star}}

\renewcommand{\Reals}{\mathfrak{R}}
\renewcommand{\indicatorfunction}{\mathop{}\!\mathbbm{1}}
\newcommand{\vol}{\operatorname{vol}}
\newcommand{\cl}{\operatorname{cl}}
\newcommand{\interior}{\operatorname{int}}

\newcommand{\V}{\const{V}_{\mat{H}}}
\newcommand{\F}{\set{F}_{\textnormal{CP}}}
\newcommand{\reg}{\set{R}}

\newcommand{\Xbar}{\bar{\vect{X}}}
\newcommand{\xbar}{\bar{\vect{x}}}
\newcommand{\Xs}{\bar{X}}
\newcommand{\xs}{\bar{x}}

\providecommand{\keywords}[1]{\textbf{Index terms ---} #1}

\definecolor{darkgreen}{rgb}{0, 0.8, 0}
\definecolor{darkcyan}{rgb}{0, 0.8, 0.8}
\definecolor{darkmagenta}{rgb}{0.5, 0, 0.6}

\theoremstyle{definition}
\newtheorem{algorithm}[theorem]{Algorithm}

\begin{document}

\title{On the Capacity of MIMO Optical Wireless Channels}

\author{Longguang Li, Stefan M.~Moser, Ligong Wang, Michèle
  Wigger\thanks{L.~Li was with LTCI, Telecom ParisTech, Université
    Paris-Saclay, 75013 Paris, France, and is now with the Department
    of Electrical and Computer Engineering at National University of
    Singapore, Singapore.
    S.~Moser is with the Signal and Information Processing Lab, ETH
    Zurich, Switzerland and with the Institute of Communications
    Engineering at National Chiao Tung University, Hsinchu, Taiwan.
    L.~Wang is with ETIS---Université Paris Seine, Université de (NUS
    Cergy-Pontoise, ENSEA, CNRS, Cergy-Pontoise, France.
    M.~Wigger is with LTCI, Telecom ParisTech, Université
    Paris-Saclay, 75013 Paris, France.
    This work was presented in part at the IEEE Information Theory
    Workshop, Guangzhou, China, November 2018.  }}

\date{20 January 2020}

\maketitle
\begin{abstract}
  This paper studies the capacity of a general multiple-input
  multiple-output (MIMO) free-space optical intensity channel under a
  per-input-antenna peak-power constraint and a total average-power
  constraint over all input antennas. The focus is on the scenario
  with more transmit than receive antennas. In this scenario,
  different input vectors can yield identical distributions at the
  output, when they result in the same image vector under
  multiplication by the channel matrix. We first determine the most
  energy-efficient input vectors that attain each of these image
  vectors. Based on this, we derive an equivalent capacity expression
  in terms of the image vector, and establish new lower and upper
  bounds on the capacity of this channel.  The bounds match when the
  signal-to-noise ratio (SNR) tends to infinity, establishing the
  high-SNR asymptotic capacity. We also characterize the low-SNR slope
  of the capacity of this channel.

  \medskip
  
\end{abstract}


\section{Introduction}
\label{sec:introduction}

This paper considers an optical wireless communication system where
the transmitter modulates the intensity of optical signals coming from
light emitting diodes (LEDs) or laser diodes (LDs), and the receiver
measures incoming optical intensities by means of photodetectors
\cite{gagliardikarp76_1, coxackermanhelkeybetts97_1,
  leerandaelbreyerkoonen09_1}. Such
\emph{intensity-modulation-direct-detection (IM-DD)} systems are
appealing because of their simplicity and their good performance at
relatively low costs.  As a first approximation, the noise in such
systems can be assumed to be Gaussian and independent of the
transmitted signal \cite[Ch.~1, p.~3]{barry94_1},
\cite{khalighiuysal14_1, haasshapiro03_1}. Inputs are nonnegative and
typically subject to both peak- and average-power constraints, where
the peak-power constraint is mainly due to technical limitations of
the used components and where the average-power constraint is imposed
by battery limitations and safety considerations. We should notice
that, unlike in radio-frequency communication, the average-power
constraint applies directly to the transmitted signal and not to its
square, because the power of the transmitted signal is proportional to
the optical intensity and hence relates directly to the transmitted
signal.

IM-DD systems have been extensively studied in recent years
\cite{khalighiuysal14_1, karunatilakazafarkalavallyparthiban15_1,
  uysalnouri14_1, wigger03_1, lapidothmoserwigger09_7, mckellips04_1,
  thangarajkramerbocherer17_1, rassouliclerckx16_1,
  moserwangwigger17_2, moserwangwigger18_3,
  mosermylonakiswangwigger17_1, chaabanrezkialouini17_2,
  chaabanrezkialouini18_1, dytsogoldenbaumshamaipoor17_1,
  chaabanrezkialouini18_2}, with an increasing interest in
multiple-input multiple-output (MIMO) systems where transmitters are
equipped with $\nt>1$ LEDs or LDs and receivers with $\nr>1$
photodetectors.  Practical transmission schemes for such systems with
different modulation methods, such as pulse-position modulation or LED
index modulation based on orthogonal frequency-division multiplexing,
were presented in \cite{wangzhongfulin09_1,
  basarpanayirciuysalhaas16_1,
  yesilkayabasarmiramirkhanipanayirciuysalhaas17_1}.  Code
constructions were described in \cite{haasshapirotarokh02_1,
  bayakischober10_1, songcheng13_1}.

%

A previous work \cite{mosermylonakiswangwigger17_1} presented upper
and lower bounds on the capacity when the channel matrix is of full
column-rank (so necessarily $\nr\ge\nt$) and determined the asymptotic
capacity at high signal-to-noise ratio (SNR) exactly.  For general
MIMO channels with average-power constraints only, the asymptotic
high-SNR capacity was determined in \cite{chaabanrezkialouini17_2,
  chaabanrezkialouini18_1}. The works \cite{chaabanrezkialouini17_2,
  chaabanrezkialouini18_1} also study general MIMO channels with both
peak- and average-power constraints, but they only determine the
high-SNR pre-log (degrees of freedom), and not the exact asymptotic
capacity. The work in \cite{dytsogoldenbaumshamaipoor17_1} considers
general MIMO channels but with peak-power constraints only.

The works most related to ours are \cite{moserwangwigger17_2,
  moserwangwigger18_3, chaabanrezkialouini18_2}. For the MISO case,
\cite{moserwangwigger17_2, moserwangwigger18_3} show that the optimal
signaling strategy is to rely as much as possible on antennas with
larger channel gains. Specifically, if an antenna is used for active
signaling in a channel use, then all antennas with larger channel
gains should transmit at maximum allowed peak power $\amp$, and all
antennas with smaller channel gains should be silenced, i.e., send
$0$. It is shown that this antenna-cooperation strategy is optimal at
all SNRs.

In \cite{chaabanrezkialouini18_2}, the asymptotic capacity in the
low-SNR regime is considered for general MIMO channels under both a
peak- and an average-power constraint. It is shown that the
asymptotically optimal input distribution in the low-SNR regime puts
the antennas into a certain order and assigns positive mass points
only to input vectors in $\{0,\amp\}^{\nt}$ in such a way that, if a
given input antenna is set to full power $\amp$, then also all
preceding antennas in the specified order are set to $\amp$. This
strategy is reminiscent of the optimal signaling strategy for MISO
channels \cite{moserwangwigger17_2, moserwangwigger18_3}. However,
whereas the optimal order in \cite{chaabanrezkialouini18_2} needs to
be determined numerically, in the MISO case the optimal order
naturally follows the channel strengths of the input
antennas. Furthermore, the order in \cite{moserwangwigger17_2,
  moserwangwigger18_3} is optimal at all SNRs, whereas the order in
\cite{chaabanrezkialouini18_2} is shown to be optimal only in the
asymptotic low-SNR limit.

The current paper focuses on MIMO channels with more transmit than
receive antennas, i.e., more LEDs than photodetectors:
\begin{IEEEeqnarray}{c}
  \nt > \nr > 1.
\end{IEEEeqnarray}
Such a system arises for example when the transmitter is based on an
existing illumination system consisting of a large number of LEDs,
whereas the receiver photodetectors are purchased at additional cost.
	
Our main contributions are as follows:

\begin{enumerate}
\item \emph{Minimum-Energy Signaling:} The optimal signaling strategy
  for MISO channels of \cite{moserwangwigger17_2, moserwangwigger18_3}
  is generalized to MIMO channels with $\nt > \nr >1$. For each
  ``image vector'' $\xbar$ --- an $\nr$-dimensional vector that can be
  produced by multiplying an input vector $\vect{x}$ by the channel
  matrix --- Lemma~\ref{lem:lemma1} identifies the input vector
  $\vect{x}_{\min}$ that induces $\xbar$ with minimum total energy.
  The minimum-energy signaling strategy partitions the image space of
  vectors $\xbar$ into at most $\binom{\nt}{\nr}$ parallelepipeds,
  each one spanned by a different subset of $\nr$ linearly independent
  columns of the channel matrix (see
  Figures~\ref{fig:23_1}--\ref{fig:23_dependent}). In each
  parallelepiped, the minimum-energy signaling sets the $\nt-\nr$
  inputs corresponding to the columns that were not chosen either to
  $0$ or to $\amp$ according to a predescribed rule and uses the $\nr$
  inputs corresponding to the chosen columns for signaling within the
  parallelepiped.

\item \emph{Equivalent Capacity Expression:} Using
  Lemma~\ref{lem:lemma1}, Proposition~\ref{prop:prop1} expresses the
  capacity of the MIMO channel in terms of the random image vector
  $\Xbar$. In particular, the power constraints on the input vector
  are translated into a set of constraints on $\Xbar$.

\item \emph{Maximizing the Trace of the Covariance Matrix:} The
  low-SNR slope of the capacity of the MIMO channel is determined by
  the maximum trace of the covariance matrix of $\Xbar$
  \cite{chaabanrezkialouini18_2}.  Lemmas~\ref{lem:binaryinput},
  \ref{lem:path}, and~\ref{lem:optimalinputtracecov} establish several
  properties for the optimal input distribution that maximizes this
  trace. They restate the result in \cite{chaabanrezkialouini18_2}
  that the covariance-trace maximizing input distribution puts
  positive mass points only on $\{0,\amp\}^{\nt}$ in a way that if an
  antenna is set to $\amp$, then all preceding antennas in a specified
  order are also set to $\amp$. The lemmas restrict the search space
  for finding the optimal antenna ordering and show that the optimal
  probability mass function (PMF) puts nonzero probability to the
  origin and to at most $\nr+1$ other input vectors.
  
\item \emph{Lower Bounds:} Lower bounds on the capacity of the channel
  of interest are obtained by applying the Entropy Power Inequality
  (EPI) \cite{coverthomas06_1} and choosing input vectors that
  maximize the differential entropy of $\Xbar$ under the imposed power
  constraints; see Theorems~\ref{thm:them4} and~\ref{thm:lowerbound2}.

\item \emph{Upper Bounds:} Three capacity upper bounds are derived by
  means of the equivalent capacity expression in
  Proposition~\ref{prop:prop1} and the duality-based upper-bounding
  technique for capacity; see Theorems~\ref{thm:them5},
  \ref{thm:them6}, and~\ref{thm:them7}. Another upper bound uses
  simple maximum-entropy arguments and algebraic manipulations; see
  Theorem~\ref{thm:them9}.


\item \emph{Asymptotic Capacity:} Theorem~\ref{thm:them8} presents the
  asymptotic capacity when the SNR tends to infinity, and Theorem
  \ref{thm:them11} gives the slope of capacity when the SNR tends to
  zero.  (This later result was already proven in
  \cite{chaabanrezkialouini18_2}, but as described above, our results
  simplify the computation of the slope.)

\end{enumerate}

The paper is organized as follows. We end the introduction with a few
notational conventions.  Section \ref{sec:channel-model} provides
details of the investigated channel
model. Section~\ref{sec:minim-energy-sign} identifies the
minimum-energy signaling schemes. Section \ref{sec:equiv-capac-expr}
provides an equivalent expression for the capacity of the
channel. Section~\ref{sec:maxim-vari-sign} shows properties of
maximum-variance signaling schemes.
Section~\ref{sec:capacity-results} presents all new lower and upper
bounds on the channel capacity, and also gives the high- and low-SNR
asymptotics. The paper is concluded in
Section~\ref{sec:conclusion}. Most of the proofs are in the
appendices.

\medskip

\textbf{Notation:} We distinguish between random and deterministic
quantities. A random variable is denoted by a capital Roman letter,
e.g., $Z$, while its realization is denoted by the corresponding small
Roman letter, e.g., $z$.  Vectors are boldfaced, e.g., $\vect{X}$
denotes a random vector and $\vect{x}$ its realization.  All the
matrices in this paper are deterministic, which are denoted in capital
letters, and are typeset in a sans-serif font, e.g., $\mat{H}$. Sets
are denoted by capital letters in a calligraphic font, e.g., $\mU$ or
$\U$ (the latter typically denoting a set of sets), except the set of
real numbers that is designated by $\Reals$.  We further use another
special font for random sets, e.g., $\rvU$.  Constants are typeset
either in small Romans, in Greek letters, or in a special font, e.g.,
$\EE$ or $\amp$. Entropy is typeset as $\HH(\cdot)$, differential
entropy as $\hh(\cdot)$, and mutual information as
$\II(\cdot;\cdot)$. The relative entropy (Kullback--Leibler
divergence) between probability vectors $\vect{p}$ and $\vect{q}$ is
denoted by $\const{D}(\vect{p}\|\vect{q})$. We will use the
$\Lone$-norm, which we indicate by $\| \cdot \|_1$, while
$\|\cdot \|_2$ denotes the $\Ltwo$-norm. The logarithmic function
$\log(\cdot)$ denotes the natural logarithm. The Lebesgue measure of a
set $\set{A} \subseteq\Reals^n$ is denoted by $\vol(\set{A})$.

\section{Channel Model}
\label{sec:channel-model}

Consider an $\nr \times \nt$ MIMO channel
\begin{IEEEeqnarray}{c}
  \vect{Y}= \mat{H}\vect{x}+\vect{Z},
\end{IEEEeqnarray}
where $\vect{x}=\trans{(x_1, \ldots, x_{\nt})}$ denotes the
$\nt$-dimensional real-valued channel input vector, where $\vect{Z}$
denotes the $\nr$-dimensional real-valued noise vector, and where
\begin{IEEEeqnarray}{c}
  \mat{H} = [\vect{h}_1, \vect{h}_2,\ldots, \vect{h}_{\nt}]
\end{IEEEeqnarray}   
is the deterministic real-valued $\nr\times\nt$ channel matrix (hence
$\vect{h}_1,\ldots, \vect{h}_{\nt}$ are $\nr$-dimensional column
vectors).

The channel matrix $\mat{H}$ models the crosstalk among different
channel inputs in terms of optical intensity, and thus, in realistic
situations, its components are nonnegative. However, for the purpose
of generality, in this work we allow the entries of $\mat{H}$ to take
negative values, i.e., they can be any real numbers. The additive
noise $\vect{Z}$ describes random fluctuations caused by thermal noise
and by ambient light arriving at the receiver, where we assume that
the receiver has already removed its expectation of the influence of
the ambient light. Since the expected intensity of ambient light is
typically much larger than its fluctuations, and since the thermal
noise can be negative, the additive noise $Z$ can also be negative
(even when $\mat{H}$ only has nonnegative entries).  In this paper, we
assume that the noise vector is independent of the channel input
$\vect{X}$ and that it has independent standard Gaussian entries,
\begin{IEEEeqnarray}{c}
  \vect{Z} \sim \Normal{0}{\mat{I}}.
\end{IEEEeqnarray}

The channel inputs correspond to optical intensities sent by the LEDs,
hence they are nonnegative:
\begin{IEEEeqnarray}{c}
  x_k\in \Reals_0^+, \quad k=1, \ldots, \nt.
  \label{eq:nonzero}
\end{IEEEeqnarray}
We assume the inputs to be subject to a peak-power (peak-intensity)
and an average-power (average-intensity) constraint:
\begin{IEEEeqnarray}{rCl}
  \subnumberinglabel{eq:constraints}
  \bigPrv{X_k>\amp} & = & 0, \quad \forall\,k\in\{1, \ldots, \nt\},
  \label{eq:peak}
  \\
  \bigE{\|\vect{X}\|_1} & \le & \EE,
  \label{eq:average}
\end{IEEEeqnarray}
for some fixed parameters $\amp,\EE>0$.  As mentioned in the
introduction, the average-power constraint is on the expectation of
the channel input and not on its square. Also note that $\amp$
describes the maximum power of each single LED, while $\EE$ describes
the allowed total average power of all LEDs together.  We denote the
ratio between the allowed average power and the allowed peak power by
$\alpha$:
\begin{IEEEeqnarray}{c}
  \alpha \eqdef \frac{\EE}{\amp}.
  \label{eq:2}
\end{IEEEeqnarray}

Throughout this paper, we assume that
\begin{IEEEeqnarray}{c}
  \nt > \nr \textnormal{ and } \rank{\mat{H}} = \nr.
  \label{eq:40}
\end{IEEEeqnarray}
The second assumption does not incur any loss of generality. Indeed,
if
\begin{IEEEeqnarray}{c}
  \label{eq:40not}
  \nt > \nr  > \rank{\mat{H}},
\end{IEEEeqnarray}
then the receiver can perform the singular-value decomposition (SVD)
$\mat{H} = \mat{U}\mat{\Sigma} \trans{\mat{V}}$, compute
$\trans{\mat{U}}\vect{Y}$, and then discard the entries in
$\trans{\mat{U}}\vect{Y}$ that correspond to zero singular
values. Since $\mat{U}$ is invertible, this new channel has the same
capacity as the original channel, but the rank of the new channel
matrix is equal to the (reduced) number of receive antennas. For more
details, see Appendix~\ref{app:proof_chanequiv}.

The first assumption in \eqref{eq:40}, however, is crucial to the
current setting.  The situation with $\nt \le \nr$ and
$\rank{\mat{H}} = \nt$ was studied in
\cite{mosermylonakiswangwigger17_1}, where it is shown that, when
$\nt<\nr$, the channel can be transformed into an equivalent
$\nt\times\nt$ channel.  More generally, one can show that any channel
with
\begin{IEEEeqnarray}{c}
  \rank{\mat{H}} \eqdef r \le \nt \le \nr
\end{IEEEeqnarray}
can be transformed into an equivalent channel of dimension
$r \times \nt$.  The idea is again to use the SVD
$\mat{H} = \mat{U}\mat{\Sigma} \trans{\mat{V}}$ to transform
$\vect{Y}$ into $\trans{\mat{U}}\vect{Y}$, and to discard the last
$\nr-r$ entries in $\trans{\mat{U}}\vect{Y}$. If $r=\nt$, we end up
with a full-rank square $\nt\times\nt$ channel matrix --- a case that
is not studied in this paper, but that it is the most widely studied
MIMO setup in the literature (e.g.,
\cite{mosermylonakiswangwigger17_1, chaabanrezkialouini18_1}).  If
$r < \nt$, we have transformed the channel into an equivalent one that
satisfies our assumptions \eqref{eq:40}. For more details, we again
refer to Appendix~\ref{app:proof_chanequiv}.

As we shall see, when $\nt>\nr$, it is in general suboptimal to
discard $\nt - \nr$ transmit antennas; instead, the optimal signaling
scheme involves collaboration of \emph{all} transmit antennas.

In this paper we are interested in deriving capacity bounds for a
channel satisfying \eqref{eq:40}. The capacity has the standard
formula
\begin{IEEEeqnarray}{c}
  \label{eq:capacity}
  \C_{\mat{H}}(\amp,\alpha\amp) = \max_{P_{\vect{X}} \textnormal{
      satisfying } \eqref{eq:constraints}} \II(\vect{X};\vect{Y}) .
\end{IEEEeqnarray}

The next proposition shows that, when $\alpha > \frac{\nt}{2}$, the
channel essentially reduces to one with only a peak-power
constraint. The other case where $\alpha\le \frac{\nt}{2}$ will be the
main focus of this paper.

\begin{proposition}
  \label{prop:proposition1}
  If $\alpha > \frac{\nt}{2}$, then the average-power constraint
  \eqref{eq:average} is inactive, i.e.,
  \begin{IEEEeqnarray}{c}
    \C_{\mat{H}}(\amp,\alpha \amp)
    = \C_{\mat{H}}\left(\amp,\frac{\nt}{2} \amp\right), \quad
    \alpha>\frac{\nt}{2}.
  \end{IEEEeqnarray} 
  If $\alpha \leq \frac{\nt}{2}$, then there exists a
  capacity-achieving input distribution $P_{\vect{X}}$ in
  \eqref{eq:capacity} that satisfies the average-power constraint
  \eqref{eq:average} with equality.
\end{proposition}
\begin{IEEEproof}
  See Appendix~\ref{app:prof_proposition1}.
\end{IEEEproof}
\medskip

We can alternatively write the MIMO channel as
\begin{IEEEeqnarray}{c}
  \vect{Y} = \xbar +\vect{Z},
\end{IEEEeqnarray}
where we set
\begin{IEEEeqnarray}{c}
  \xbar \eqdef \mat{H}\vect{x}.
\end{IEEEeqnarray}
We introduce the following notation. For a matrix
$\mat{M}=[\vect{m}_1,\ldots,\vect{m}_k]$, where $\{\vect{m}_i\}$ are
column vectors, define the set
\begin{IEEEeqnarray}{c}
  \reg(\mat{M}) \eqdef \left\{ \sum_{i=1}^k \lambda_i
    \vect{m}_i\colon \lambda_1,\ldots,\lambda_k \in[0,\amp]\right\}. 
\end{IEEEeqnarray}
Note that this set is a \emph{zonotope}. Since the $\nt$-dimensional
input vector $\vect{x}$ is constrained to the $\nt$-dimensional
hypercube $[0,\amp]^{\nt}$, the $\nr$-dimensional image vector $\xbar$
takes value in the zonotope $\reg(\mat{H})$.

For each $\xbar\in \reg(\mat{H})$, let
\begin{IEEEeqnarray}{c}
  \label{setxbar}
  \set{S}(\xbar) \eqdef \bigl\{\vect{x} \in [0,\amp]^{\nt}\colon
  \mat{H}\vect{x} = \xbar \bigr\} 
\end{IEEEeqnarray}
be the set of input vectors inducing $\xbar$.  In the following
section we derive the most energy-efficient signaling method to attain
a given $\xbar$. This will allow us to express the capacity in terms
of $\Xbar = \mat{H}\vect{X}$ instead of $\vect{X}$, which will prove
useful.

\section{Minimum-Energy Signaling}
\label{sec:minim-energy-sign}

The goal of this section is to identify for every
$\xbar\in\reg(\mat{H})$ the minimum-energy choice of input vector
$\vect{x}$ that induces $\xbar$. Since the energy of an input vector
$\vect{x}$ is $\|\vect{x}\|_1$, we are interested in finding an
$\vect{x}_{\min}$ that satisfies
\begin{IEEEeqnarray}{c}
  \|\vect{x}_{\min}\|_1
  = \min_{\vect{x} \in \set{S}(\xbar) } \|\vect{x}\|_1.
  \label{eq:minienergy}
\end{IEEEeqnarray}
We start by describing (without proof) the choice of $\vect{x}_{\min}$
in two different $2\times 3$ examples.
\begin{figure}[htbp]
  \centering
  \begin{tikzpicture}
    [scale=1,
    >={Stealth[scale=1.2]},
    dot/.style={circle,draw=red,fill=red,thick,
      inner sep=0pt,minimum size=1.2mm}]
    \begin{axis}[scale only axis,
      xmin=0,
      xmax=6,
      xtick={0,1,...,6},
      xmajorgrids,
      xlabel={$\xs_1$},        
      ymin=0,
      ymax=6,
      ytick={0,1,...,6},
      ymajorgrids,
      ylabel={$\xs_2$},
      grid style={densely dotted,black}] 
      \coordinate (h) at (0,0);
      \coordinate (h1) at (2.5,1);
      \coordinate (h2) at (2,2);
      \coordinate (h3) at (1,2);

      \draw[draw=black,thick] 
        (h) -- ++(h1) -- ++(h2) -- ++(h3);

      \edef\temp{\noexpand\draw[draw=black,thick] 
        (h) 
        foreach \i in {3,...,1} {
          -- ++(h\i) 
        };
      }
      \temp
      
      \coordinate (g12) at ($(h)$);
      \coordinate (g13) at ($(h2)$);
      \coordinate (g23) at ($(h)$);

      \newcount\mycount
      \global\mycount=0

      \foreach \i in {1,...,2}
        \foreach \jj [evaluate={\j=\jj+1}] in {\i,...,2} {
          \edef\temp{\noexpand\coordinate (s) at (g\i\j);}
          \temp
          \edef\temp{\noexpand\coordinate (a) at (h\i);}
          \temp
          \edef\temp{\noexpand\coordinate (b) at (h\j);}
          \temp
          \edef\temp{\noexpand\draw[fill=red!\the\mycount!yellow,opacity=0.5,draw=black]
            (s) -- ($(s)+(a)$) -- ($(s)+(a)+(b)$) -- ($(s)+(b)$) --
            (s);}
          \temp
          \global\advance\mycount by 33.33          
        }

      \foreach \i in {1,...,3} {
        \edef\temp{\noexpand\draw[very thick,->] (h) -- (h\i);}
        \temp
      }

      \node at (2,0.5) {$\vect{h}_1$};
      \node at (1.8,1.5) {$\vect{h}_2$};
      \node at (0.5,1.5) {$\vect{h}_3$};
      \node at (3.8,3.5) {$\amp\vect{h}_2+\set{D}_{\{1,3\}}$};
      \node at (1.7,2.2) {$\set{D}_{\{2,3\}}$};
      \node at (2.6,1.7) {$\set{D}_{\{1,2\}}$};

      \draw[draw=black,->] (3.9,1.5) to [out=180,in=-30] (3.5,2);
      \node at (4.2,1.5) {$\reg(\mat{H})$};      
    \end{axis}
  \end{tikzpicture}
  
  \caption{The zonotope $\reg(\mat{H})$ for the $2 \times 3$ MIMO
    channel matrix $\mat{H}=[2.5, 2, 1;1, 2, 2]$ given in
    \eqref{eq:example23} and its minimum-energy decomposition into
    three parallelograms. The peak power is assumed to be $\amp = 1$.}
  \label{fig:23_1}
\end{figure}
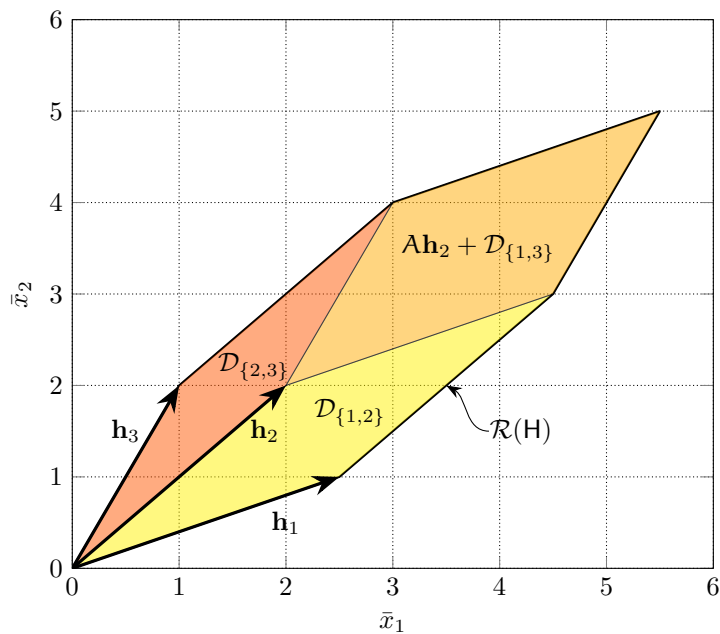

\begin{example}
  \label{example:23}
  Consider the $2\times 3$ MIMO channel matrix
  \begin{IEEEeqnarray}{c}
    \mat{H} =
    \begin{pmatrix}
      2.5 & 2 &  1
      \\
      1 & 2 & 2
    \end{pmatrix}
    \label{eq:example23}
  \end{IEEEeqnarray} 
  composed of the three column vectors $\vect{h}_1=\trans{(2.5,1)}$,
  $\vect{h}_2=\trans{(2,2)}$, and $\vect{h}_3=\trans{(1,2)}$.
  Figure~\ref{fig:23_1} depicts the zonotope $\reg(\mat{H})$ and
  partitions it into three parallelograms based on three different
  forms of $\vect{x}_{\min}$. For any $\xbar$ in the parallelogram
  $\set{D}_{\{1,2\}} \eqdef\reg\bigl(\mat{H}_{\{1,2\}}\bigr)$, where
  $\mat{H}_{\{1,2\}}\eqdef[\vect{h}_1 ,\;\vect{h}_2]$, the
  minimum-energy input $\vect{x}_{\min}$ inducing $\xbar$ has 0 as its
  third component.  Since $\mat{H}_{\{1,2\}}$ has full rank, there is
  only one such input inducing $\xbar$:
  \begin{IEEEeqnarray}{c}
    \label{eq:D1}
    \vect{x}_{\min} =
    \begin{pmatrix}
      \mat{H}_{\{1,2\}}^{-1} \xbar \\
      0
    \end{pmatrix},
    \qquad \textnormal{if } \xbar \in \set{D}_{\{1,2\}}.
  \end{IEEEeqnarray}
  Similarly, for any $\xbar$ in the parallelogram
  $\set{D}_{\{2,3\}}\eqdef\reg\bigl(\mat{H}_{\{2,3\}}\bigr)$, where
  $\mat{H}_{\{2,3\}}\eqdef[\vect{h}_2 , \;\vect{h}_3]$, the
  minimum-energy input $\vect{x}_{\min}$ inducing $\xbar$ has 0 as its
  first component:
  \begin{IEEEeqnarray}{c}
    \label{eq:D2}
    \vect{x}_{\min} =
    \begin{pmatrix}
      0 \\ \mat{H}_{\{2,3\}}^{-1} \xbar
    \end{pmatrix},
    \qquad\textnormal{if } \xbar \in \set{D}_{\{2,3\}}.
  \end{IEEEeqnarray}
  Finally, for any $\xbar$ in the parallelogram
  $\amp \vect{h}_2+ \set{D}_{\{1,3\}}$, where
  $\set{D}_{\{1,3\}}\eqdef\reg\bigl(\mat{H}_{\{1,3\}}\bigr)$ and
  $\mat{H}_{\{1,3\}}\eqdef[\vect{h}_1 , \;\vect{h}_3]$, the
  minimum-energy input $\vect{x}_{\min}$ inducing $\xbar$ has $\amp$
  as its second component:
  \begin{IEEEeqnarray}{c}
    \label{eq:D3}
    \vect{x}_{\min} =
    \begin{pmatrix}
      x_{\min,1} \\ \amp \\ x_{\min,3}
    \end{pmatrix},
    \qquad  \textnormal{if } \xbar \in
    \amp\vect{h}_{2}+\set{D}_{\{1,3\}}, 
  \end{IEEEeqnarray}
  where 
  \begin{IEEEeqnarray}{c+x*}
    \label{eq:inv}
    \begin{pmatrix}
      x_{\min,1} \\ x_{\min,3}
    \end{pmatrix}
    = \mat{H}_{\{1,3\}}^{-1} (\xbar - \amp \vect{h}_2).
  \end{IEEEeqnarray}
  
  These minimum-energy choices of $\vect{x}$ can be understood
  intuitively. Recall that we search for a triple
  $\vect{x} = (x_1,x_2,x_3) \in [0,\amp]^3$ that induces some given
  $\xbar \in \reg(\mat{H})$:
  \begin{IEEEeqnarray}{c}
    x_1\vect{h}_1 + x_2\vect{h}_2 + x_3\vect{h}_3 = \xbar.
    \label{eq:xbarexp}
  \end{IEEEeqnarray}
  Our aim is to find an $\vect{x}$ with minimum $\set{L}_1$ norm,
  i.e., $x_1 + x_2+x_3$ shall be small. Now, we observe that in our
  example we can write
  \begin{IEEEeqnarray}{rCl}
    \vect{h}_2 = 0.5\vect{h}_1 + 0.75\vect{h}_3.
    \label{eq:h2exp}
  \end{IEEEeqnarray}
  Since here the weights add up to more than one, $0.5+0.75>1$, it is
  energy-efficient to reduce $x_1$ and $x_3$ at the cost of increasing
  $x_2$. So, whenever possible, we reduce one of $x_1$ and $x_3$ to
  zero. This covers $\set{D}_{\{1,2\}}$ and $\set{D}_{\{2,3\}}$. The
  third parallelogram $\amp\vect{h}_{2}+\set{D}_{\{1,3\}}$ can only be
  reached if all three components are nonzero. In this case, it is
  best to set the efficient component $x_2$ to its maximal value
  $\amp$ and then use the other two to reach to $\xbar$.
\end{example}

\begin{figure}[htbp]
  \centering
  \begin{tikzpicture}
    [scale=1,
     >={Stealth[scale=1.2]},
     dot/.style={circle,draw=red,fill=red,thick,
       inner sep=0pt,minimum size=1.2mm}]

    \begin{axis}[scale only axis,
      xmin=0,
      xmax=5,
      xtick={0,1,...,5},
      xmajorgrids,
      xlabel={$\xs_1$},        
      ymin=0,
      ymax=4,
      ytick={0,1,...,4},
      ymajorgrids,
      ylabel={$\xs_2$},
      grid style={densely dotted,black}] 
      
      \coordinate (h) at (0,0);
      \coordinate (h1) at (2.5,1);
      \coordinate (h2) at (0.8,0.8);
      \coordinate (h3) at (1,2);
 

      \draw[draw=black,thick] 
        (h) -- ++(h1) -- ++(h2) -- ++(h3);

      \edef\temp{\noexpand\draw[draw=black,thick] 
        (h) 
        foreach \i in {3,...,1} {
          -- ++(h\i) 
        };
      }
      \temp
      
      \coordinate (g12) at ($(h3)$);
      \coordinate (g13) at ($(h)$);
      \coordinate (g23) at ($(h1)$);

      \newcount\mycount
      \global\mycount=0

      \foreach \i in {1,...,2}
        \foreach \jj [evaluate={\j=\jj+1}] in {\i,...,2} {
          \edef\temp{\noexpand\coordinate (s) at (g\i\j);}
          \temp
          \edef\temp{\noexpand\coordinate (a) at (h\i);}
          \temp
          \edef\temp{\noexpand\coordinate (b) at (h\j);}
          \temp
          \edef\temp{\noexpand\draw[fill=red!\the\mycount!yellow,opacity=0.5,draw=black]
            (s) -- ($(s)+(a)$) -- ($(s)+(a)+(b)$) -- ($(s)+(b)$) --
            (s);}
          \temp
          \global\advance\mycount by 33.33          
        }

      \foreach \i in {1,...,3} {
        \edef\temp{\noexpand\draw[very thick,->] (h) -- (h\i);}
        \temp
      }

      \node at (2,0.5) {$\vect{h}_1$};
      \node at (0.9,0.9) {$\vect{h}_2$};
      \node at (0.5,1.5) {$\vect{h}_3$};
      \node at (4.4,2.5) {$\amp\vect{h}_2+\set{D}_{\{1,3\}}$};
      \draw[draw=black,->] (4.3,2.4) to [out=-90,in=-30] (3.4,2.4);
      \node at (2.6,3.6) {$\amp\vect{h}_3+\set{D}_{\{1,2\}}$};
      \draw[draw=black,->] (2.6,3.5) to [out=-90,in=130] (3.1,3.1);
      \node at (1.8,1.5) {$\set{D}_{\{1,3\}}$};
      \draw[draw=black,->] (2.8,0.5) to [out=180,in=-30] (2.4,0.8);
      \node at (3.1,0.5) {$\reg(\mat{H})$};
    \end{axis}
  \end{tikzpicture}
  
  \caption{The zonotope $\reg(\mat{H})$ for the $2 \times 3$ MIMO
    channel matrix $\mat{H}=[2.5, 0.8, 1;1, 0.8, 2]$ given in
    \eqref{eq:example24} and its minimum-energy decomposition into
    three parallelograms. The peak power is $\amp = 1$.}
  \label{fig:23_2}
\end{figure}
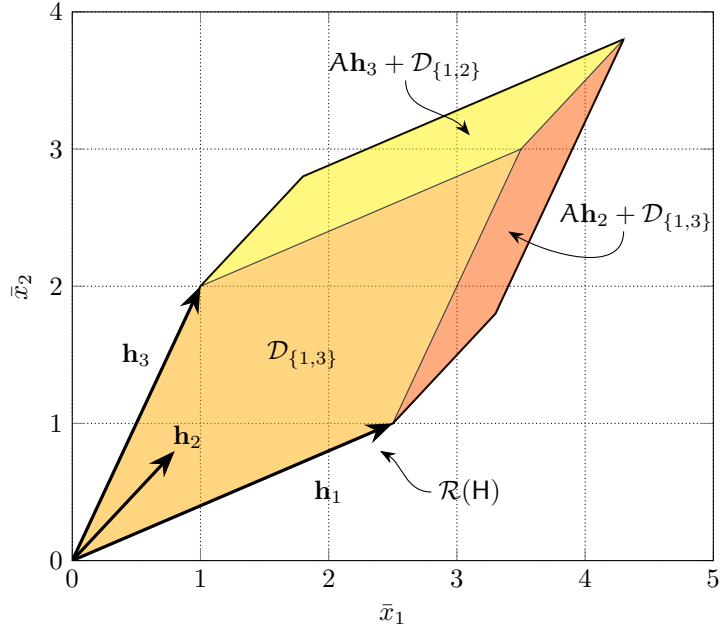

\begin{example}
  \label{example:24}
  We now consider another $2 \times 3$ MIMO channel with channel
  matrix
  \begin{IEEEeqnarray}{c}
    \mat{H} =
    \begin{pmatrix}
      2.5 &  0.8 &  1
      \\
      1 & 0.8 & 2
    \end{pmatrix}.
    \label{eq:example24}
  \end{IEEEeqnarray}  
  Figure~\ref{fig:23_2} depicts the zonotope $\reg(\mat{H})$ and its
  partitioning based on three different forms of $\vect{x}_{\min}$.

  Note that $\vect{h}_1$ and $\vect{h}_3$ are kept the same as in
  Example~\ref{example:23}, but $\vect{h}_2$ is reduced in length. It
  can now be written as
  \begin{IEEEeqnarray}{rCl}
    \vect{h}_2 = 0.2\vect{h}_1 + 0.3\vect{h}_3.
    \label{eq:h2exp2}
  \end{IEEEeqnarray}
  Since here the sum of the weights is less than one, $0.2 + 0.3 < 1$,
  it is energy-efficient to reduce $x_2$ at the cost of increasing
  $x_1$ and $x_3$ in \eqref{eq:xbarexp}. Thus, whenever possible we
  keep $x_2=0$ and only use $x_1$ and $x_3$; this is how we cover
  $\set{D}_{\{1,3\}}$. The other two parallelograms can only be
  reached with all three components being positive. In these cases, it
  is best to set one of $x_1$ and $x_3$ to its maximal value $\amp$
  and use the other two to reach to $\xbar$.
\end{example}

We now generalize Examples~\ref{example:23} and~\ref{example:24} to
formally solve the optimization problem in \eqref{eq:minienergy} for
an arbitrary $\nr\times\nt$ channel matrix $\mat{H}$.  To this end, we
need some further definitions. Denote by $\U$ the set of all choices
of $\nr$ columns of $\mat{H}$ that are linearly independent:
\begin{IEEEeqnarray}{c}
  \U \eqdef \Bigl\{\mU = \{i_1,\ldots, i_{\nr}\} \subseteq
  \{1,\ldots,\nt\} \colon \vect{h}_{i_1}, \ldots,
  \vect{h}_{i_{\nr}} \textnormal{are linearly independent} \Bigr\}.
  \IEEEeqnarraynumspace
\end{IEEEeqnarray}
For every one of these index sets $\mU \in \U$, we denote its
complement by
\begin{IEEEeqnarray}{c}
  \mcU \eqdef \{1, \ldots,\nt\} \setminus \mU;
\end{IEEEeqnarray}
we define the $\nr\times\nr$ matrix $\mat{H}_{\mU}$ containing the
columns of $\mat{H}$ indicated by $\mU$:
\begin{IEEEeqnarray}{c}
  \mat{H}_{\mU} \eqdef [ \vect{h}_i \colon i\in\mU];
\end{IEEEeqnarray}
and we define the $\nr$-dimensional parallelepiped
\begin{IEEEeqnarray}{c}
  \set{D}_{\mU} \eqdef \reg\left(\mat{H}_{\mU}\right).
  \label{eq:5}
\end{IEEEeqnarray}
Notice that
\begin{IEEEeqnarray} {c}
  \vol(\set{D}_{\mU})=\amp^n |\det \mat{H}_{\mU} |,
\end{IEEEeqnarray}
which is positive because the columns of $\mat{H}_\mU$ are linearly
independent.

We shall see (Lemma~\ref{lem:lemma1} ahead) that $\reg(\mat{H})$ can
be partitioned into parallelepipeds that are shifted versions of
$\{\set{D}_{\mU}\}$ in such a way that, within each parallelepiped,
$\vect{x}_{\min}$ has the same form, in a sense similar to
\eqref{eq:D1}--\eqref{eq:inv} in Example~\ref{example:23}.
To specify our partition, we define the $\nr$-dimensional vectors
\begin{IEEEeqnarray}{c}
  \vectg{\gamma}_{\mU,j}
  \eqdef \inv{\mat{H}_{\mU}} \vect{h}_j,
  \quad \mU\in\U, \; j\in\mcU,
  \label{eq:4}
\end{IEEEeqnarray}
and the sum of their components
\begin{IEEEeqnarray}{c}
  a_{\mU,j} \eqdef \trans{\vect{1}_{\nr}} \vectg{\gamma}_{\mU,j}, 
  \quad \mU\in\U, \; j\in\mcU.
  \label{eq:aij}
\end{IEEEeqnarray}
We next choose a set of coefficients
$\{g_{\mU,j}\}_{\mU\in\U, j\in\mcU}$, which are either $0$ or $1$, as
follows.
\begin{itemize}
\item If
  \begin{IEEEeqnarray}{c}
    a_{\mU,j} \neq 1, \quad \forall\, \mU\in\U, \; \forall\,
    j\in\mcU, 
    \label{eq:cond1}
  \end{IEEEeqnarray}
  then let
  \begin{IEEEeqnarray}{c}
    g_{\mU,j} \eqdef
    \begin{cases}
      1 & \textnormal{if } a_{\mU,j} > 1,
      \\
      0 & \textnormal{otherwise}, 
    \end{cases}
    \quad \mU \in \U, \; j\in \mcU. 
    \label{eq:6}
  \end{IEEEeqnarray}
  
\item If \eqref{eq:cond1} is violated, then we have one or several
  ties, i.e., the solution to the minimization problem in
  \eqref{eq:minienergy} is not unique and there exist several
  different but equivalent vectors $\vect{x}_{\min}$.  (To give an
  example in the spirit of Examples~\ref{example:23} and
  \ref{example:24}, consider the channel matrix
  \begin{IEEEeqnarray}{c}
    \mat{H} =
    \begin{pmatrix}
      2.5 & 1.6 &  1
      \\
      1 & 1.6 & 2
    \end{pmatrix},
    \label{eq:example25}
  \end{IEEEeqnarray}  
  for which $\vect{h}_2 = 0.4\vect{h}_1+0.6\vect{h}_3$ with
  $0.4+0.6=1$.)
  
  In order to break such ties, the following algorithm simply picks
  one out of all possible equivalent optimal choices.
  \begin{blockbox}
  \begin{algorithm}
    \label{def:algorithm}
    \mbox{}
    \begin{algorithmic}
      \For{$j \in \{1,\ldots,\nt\}$}
        \For{$\mU \in \U$ such that $\mU \subseteq \{j,\ldots, \nt\}$}
          \If {$j \in \mcU$}
            \begin{IEEEeqnarray}{c}
              g_{\mU,j} \eqdef
              \begin{cases}
                1 & \textnormal{if } a_{\mU,j} \geq 1
                \\
                0 & \textnormal{otherwise} 
              \end{cases}
            \end{IEEEeqnarray}
          \Else
            \For{$k \in \mcU \cap \{j+1, \ldots, \nt\}$}
              \begin{IEEEeqnarray}{c}
                g_{\mU,k} \eqdef
                \begin{cases}
                  1 & \textnormal{if } a_{\mU,k} > 1 \textnormal{ or }
                  \bigl( a_{\mU,k}=1 \textnormal{ and the first
                    component} \\
                  & \qquad\qquad\qquad\;\; \textnormal{of
                    $\vectg{\gamma}_{\mU,j}$ is negative}\bigr) 
                  \\
                  0 & \textnormal{otherwise} 
                \end{cases}
                \IEEEeqnarraynumspace
              \end{IEEEeqnarray}
            \EndFor
          \EndIf
        \EndFor
      \EndFor
    \end{algorithmic}
  \end{algorithm}
  \end{blockbox}
\end{itemize}
Finally, let
\begin{IEEEeqnarray}{c}
  \label{eq:vl}
  \vect{v}_{\mU} \eqdef  \amp \sum_{j \in \mcU}
  g_{\mU,j} \vect{h}_j, \quad \mU \in \U.
\end{IEEEeqnarray}

We are now ready to describe our partition of $\reg(\mat{H})$.

\begin{lemma}
  \label{lem:lemma1}
  For each $\mU \in \U$ let $\set{D}_{\mU}$,
  $\{g_{\mU,j}\}_{j \in \mcU}$, and $\vect{v}_{\mU}$ be as given in
  \eqref{eq:5}, \eqref{eq:6} or Algorithm~\ref{def:algorithm}, and
  \eqref{eq:vl}, respectively.
  \begin{enumerate}
  \item \label{part1} The zonotope $\reg(\mat{H})$ is covered by the
    parallelepipeds $\{ \vect{v}_{\mU} + \set{D}_{\mU}\}_{\mU\in\U}$,
    which overlap only on sets of measure zero:
    \begin{IEEEeqnarray}{c}
      \label{eq:union}
      \bigcup_{\mU\in\U} \bigl( \vect{v}_{\mU} +
      \set{D}_{\mU}\bigr)  = \reg(\mat{H}) 
    \end{IEEEeqnarray}
    and 
    \begin{IEEEeqnarray}{c}
      \label{eq:overlap}
      \vol\Bigl(  \bigl(  \vect{v}_{\mU} +
      \set{D}_{\mU}\bigr) \cap  
      \bigl(\vect{v}_{\mV} + \set{D}_{\mV}\bigr)\Bigr) = 0,
      \qquad \mU, \mV\in\U, \; \mU \neq \mV,
    \end{IEEEeqnarray}
    where we recall that $\vol(\cdot)$ denotes the ($\nr$-dimensional)
    Lebesgue measure.

  \item \label{part2} Fix some $\mU \in \U$ and some
    $\xbar \in \vect{v}_{\mU} + \set{D}_{\mU}$. A vector that
    induces $\xbar$ with minimum energy, i.e., $\vect{x}_{\min}$ in
    \eqref{eq:minienergy}, is given by
    $\vect{x} = \trans{(x_1,\ldots,x_{\nt})}$, where
    \begin{IEEEeqnarray}{c}
      \label{eq:xvalue}
      x_{i}= 
      \begin{cases}
        \amp \cdot g_{\mU,i}  & \textnormal{if } i \in \mcU,
        \\
        \beta_i & \textnormal{if } i \in \mU,
      \end{cases}
    \end{IEEEeqnarray}
    where the vector $\vectg{\beta}=\trans{(\beta_i \colon i\in \mU)}$
    is given by
    \begin{IEEEeqnarray}{c}
      \vectg{\beta} \eqdef \inv{\mat{H}_{\mU}} (\xbar -
      \vect{v}_{\mU}).
      \label{eq:alpha}
    \end{IEEEeqnarray}
  \end{enumerate}
\end{lemma}
\begin{IEEEproof}
  See Appendix~\ref{app:prof_lemma1}.
\end{IEEEproof}
\medskip

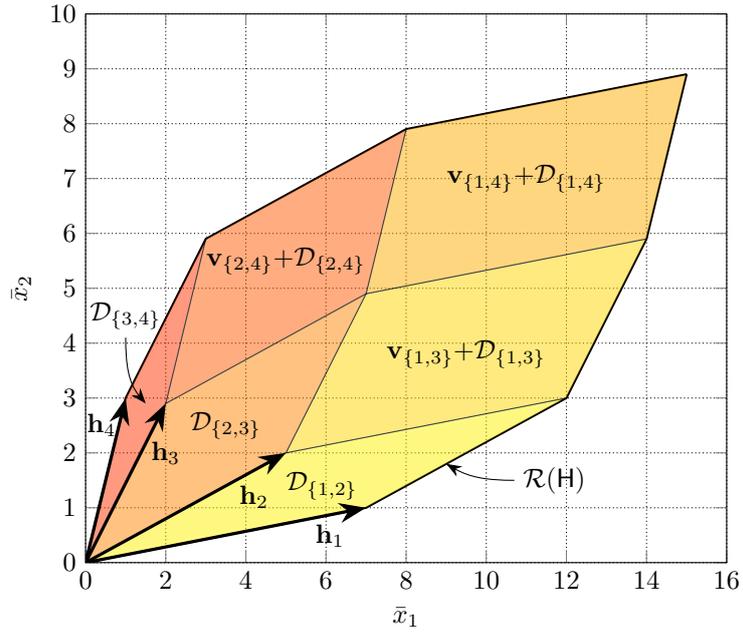
\begin{figure}[htbp]
  \centering
  \begin{tikzpicture}
    [scale=1,
    >={Stealth[scale=1.2]},
    dot/.style={circle,draw=red,fill=red,thick,
      inner sep=0pt,minimum size=1.2mm}]

    \begin{axis}[scale only axis,
      xmin=0,
      xmax=16,
      xtick={0,2,...,16},
      xmajorgrids,
      xlabel={$\xs_1$},        
      ymin=0,
      ymax=10,
      ytick={0,1,...,10},
      ymajorgrids,
      ylabel={$\xs_2$},
      grid style={densely dotted,black}] 
    
      \coordinate (h) at (0,0);
      \coordinate (h1) at (7,1);
      \coordinate (h2) at (5,2);
      \coordinate (h3) at (2,2.9);
      \coordinate (h4) at (1,3);

      \draw[draw=black,thick] 
        (h) -- ++(h1) -- ++(h2) -- ++(h3) -- ++(h4);

      \edef\temp{\noexpand\draw[draw=black,thick] 
        (h) 
        foreach \i in {4,...,1} {
          -- ++(h\i) 
        };
      }
      \temp
      
      \coordinate (g12) at ($(h)$);
      \coordinate (g13) at ($(h2)$);
      \coordinate (g14) at ($(h2)+(h3)$);
      \coordinate (g23) at ($(h)$);
      \coordinate (g24) at ($(h3)$);
      \coordinate (g34) at ($(h)$);

      \newcount\mycount
      \global\mycount=0

      \foreach \i in {1,...,3}
        \foreach \jj [evaluate={\j=\jj+1}] in {\i,...,3} {
          \edef\temp{\noexpand\coordinate (s) at (g\i\j);}
          \temp
          \edef\temp{\noexpand\coordinate (a) at (h\i);}
          \temp
          \edef\temp{\noexpand\coordinate (b) at (h\j);}
          \temp
          \edef\temp{\noexpand\draw[fill=red!\the\mycount!yellow,opacity=0.5,draw=black]
            (s) -- ($(s)+(a)$) -- ($(s)+(a)+(b)$) -- ($(s)+(b)$) --
            (s);}
          \temp
          \global\advance\mycount by 16.67          
        }
        
      \foreach \i in {1,...,4} {
        \edef\temp{\noexpand\draw[very thick,->] (h) -- (h\i);}
        \temp
      }

      \node at (6.1,0.5) {$\vect{h}_1$};
      \node at (4.2,1.2) {$\vect{h}_2$};
      \node at (2.0,2.0) {$\vect{h}_3$};
      \node at (0.4,2.5) {$\vect{h}_4$};
      \node at (5.9,1.4) {$\set{D}_{\{1,2\}}$};
      \node at (3.5,2.5) {$\set{D}_{\{2,3\}}$};
      \node at (1.0,4.5) {$\set{D}_{\{3,4\}}$};
      \node at (9.5,3.8) {$\vect{v}_{\{1,3\}}{+}\set{D}_{\{1,3\}}$};
      \node at (5.0,5.5) {$\vect{v}_{\{2,4\}}{+}\set{D}_{\{2,4\}}$};
      \node at (11,7) {$\vect{v}_{\{1,4\}}{+}\set{D}_{\{1,4\}}$};
      \draw[draw=black,->] (1.0,4.1) to [out=-90,in=130] (1.5,2.9);

      \draw[draw=black,->] (10.7,1.5) to [out=180,in=-30] (9,1.8);
      \node at (11.7,1.5) {$\reg(\mat{H})$};
    \end{axis}
  \end{tikzpicture}
  \caption{Partition of $\reg(\mat{H})$ into the union
    \eqref{eq:union} for the $2 \times 4$ MIMO channel matrix
    $\mat{H}=[7, 5, 2, 1; 1, 2, 2.9,3]$. The peak power is
      $\amp = 1$.}
  \label{fig:24_1}
\end{figure}

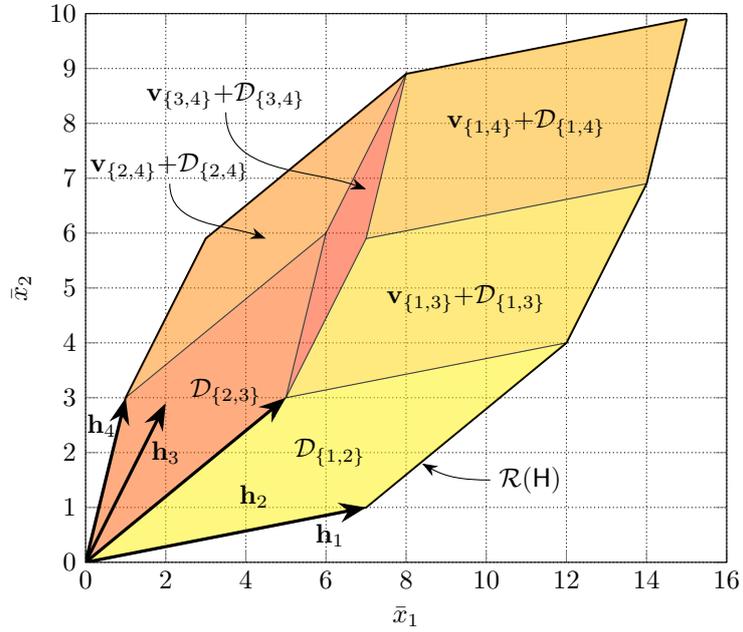
\begin{figure}[htbp]
  \centering
  \begin{tikzpicture}
    [scale=1,
    >={Stealth[scale=1.2]},
    dot/.style={circle,draw=red,fill=red,thick,
      inner sep=0pt,minimum size=1.2mm}]

    \begin{axis}[scale only axis,
      xmin=0,
      xmax=16,
      xtick={0,2,...,16},
      xmajorgrids,
      xlabel={$\xs_1$},        
      ymin=0,
      ymax=10,
      ytick={0,1,...,10},
      ymajorgrids,
      ylabel={$\xs_2$},
      grid style={densely dotted,black}] 
      
      \coordinate (h) at (0,0);
      \coordinate (h1) at (7,1);
      \coordinate (h2) at (5,3);
      \coordinate (h3) at (2,2.9);
      \coordinate (h4) at (1,3);

      \draw[draw=black,thick] 
        (h) -- ++(h1) -- ++(h2) -- ++(h3) -- ++(h4);

      \edef\temp{\noexpand\draw[draw=black,thick] 
        (h) 
        foreach \i in {4,...,1} {
          -- ++(h\i) 
        };
      }
      \temp
      
      \coordinate (g12) at ($(h)$);
      \coordinate (g13) at ($(h2)$);
      \coordinate (g14) at ($(h2)+(h3)$);
      \coordinate (g23) at ($(h4)$);
      \coordinate (g24) at ($(h)$);
      \coordinate (g34) at ($(h2)$);

      \newcount\mycount
      \global\mycount=0

      \foreach \i in {1,...,3} 
        \foreach \jj [evaluate={\j=\jj+1}] in {\i,...,3} {
          \edef\temp{\noexpand\coordinate (s) at (g\i\j);}
          \temp
          \edef\temp{\noexpand\coordinate (a) at (h\i);}
          \temp
          \edef\temp{\noexpand\coordinate (b) at (h\j);}
          \temp
          \edef\temp{\noexpand\draw[fill=red!\the\mycount!yellow,opacity=0.5,draw=black]
            (s) -- ($(s)+(a)$) -- ($(s)+(a)+(b)$) -- ($(s)+(b)$) --
            (s);}
          \temp
          \global\advance\mycount by 16.67          
        }
        
      \foreach \i in {1,...,4} {
        \edef\temp{\noexpand\draw[very thick,->] (h) -- (h\i);}
        \temp
      }

      \node at (6.1,0.5) {$\vect{h}_1$};
      \node at (4.2,1.2) {$\vect{h}_2$};
      \node at (2.0,2.0) {$\vect{h}_3$};
      \node at (0.4,2.5) {$\vect{h}_4$};
      \node at (6.1,2.0) {$\set{D}_{\{1,2\}}$};
      \node at (3.5,3.1) {$\set{D}_{\{2,3\}}$};
      \node at (3.5,8.5) {$\vect{v}_{\{3,4\}}{+}\set{D}_{\{3,4\}}$};
      \node at (9.5,4.8) {$\vect{v}_{\{1,3\}}{+}\set{D}_{\{1,3\}}$};
      \node at (2.1,7.2) {$\vect{v}_{\{2,4\}}{+}\set{D}_{\{2,4\}}$};
      \node at (11,8) {$\vect{v}_{\{1,4\}}{+}\set{D}_{\{1,4\}}$};
      \draw[draw=black,->] (3.5,8.2) to [out=-90,in=160] (7,6.8);
      \draw[draw=black,->] (2.1,6.9) to [out=-90,in=160] (4.5,5.9);

      \draw[draw=black,->] (10.1,1.5) to [out=180,in=-30] (8.4,1.8);
      \node at (11.1,1.5) {$\reg(\mat{H})$};
    \end{axis}
  \end{tikzpicture}
  \caption{Partition of $\reg(\mat{H})$ into the union
    \eqref{eq:union} for the $2 \times 4$ MIMO channel matrix
    $\mat{H}=[7, 5, 2, 1; 1, 3, 2.9, 3]$. The peak power is
      $\amp = 1$.}
  \label{fig:24_2}
\end{figure}

We recall that Figures~\ref{fig:23_1} and \ref{fig:23_2} show the
partition of $\reg(\mat{H})$ into the union \eqref{eq:union} for two
examples of $2\times 3$ channel matrices.
Figures~\ref{fig:24_1}--\ref{fig:fig6} show four more examples. Among
them, Figure~\ref{fig:23_dependent} illustrates a case where $\mat{H}$
contains linearly dependent columns, and Figure~\ref{fig:fig6} shows a
case where $\mat{H}$ has negative entries.

\begin{figure}[htbp]
  \centering
  \begin{tikzpicture}
    [scale=1,
    >={Stealth[scale=1.2]},
    dot/.style={circle,draw=red,fill=red,thick,
      inner sep=0pt,minimum size=1.2mm}]
    \begin{axis}[scale only axis,
      xmin=0,
      xmax=9,
      xtick={0,1,...,9},
      xmajorgrids,
      xlabel={$\xs_1$},        
      ymin=0,
      ymax=6,
      ytick={0,1,...,6},
      ymajorgrids,
      ylabel={$\xs_2$},
      grid style={densely dotted,black}] 
      
      \coordinate (h) at (0,0);
      \coordinate (h1) at (2.5,1.2);
      \coordinate (h2) at (5,2.4);
      \coordinate (h3) at (1,2);
 

      \draw[draw=black,thick] 
        (h) -- ++(h1) -- ++(h2) -- ++(h3);

      \edef\temp{\noexpand\draw[draw=black,thick] 
        (h) 
        foreach \i in {3,...,1} {
          -- ++(h\i) 
        };
      }
      \temp
      
      \coordinate (g12) at ($(h)$);
      \coordinate (g13) at ($(h2)$);
      \coordinate (g23) at ($(h)$);

      \newcount\mycount
      \global\mycount=0

      \foreach \i in {1,...,2}
        \foreach \jj [evaluate={\j=\jj+1}] in {\i,...,2} {
          \edef\temp{\noexpand\coordinate (s) at (g\i\j);}
          \temp
          \edef\temp{\noexpand\coordinate (a) at (h\i);}
          \temp
          \edef\temp{\noexpand\coordinate (b) at (h\j);}
          \temp
          \edef\temp{\noexpand\draw[fill=red!\the\mycount!yellow,opacity=0.5,draw=black]
            (s) -- ($(s)+(a)$) -- ($(s)+(a)+(b)$) -- ($(s)+(b)$) --
            (s);}
          \temp
          \global\advance\mycount by 33.33          
        }
        
      \foreach \i in {1,...,3} {
        \edef\temp{\noexpand\draw[very thick,->] (h) -- (h\i);}
        \temp
      }

      \node at (2,0.65) {$\vect{h}_1$};
      \node at (3.7,1.5) {$\vect{h}_2$};
      \node at (0.4,1.6) {$\vect{h}_3$};
      \node at (7.1,4) {$\amp\vect{h}_2+\set{D}_{\{1,3\}}$};
      \node at (3.2,2.2) {$\set{D}_{\{2,3\}}$};

      \draw[draw=black,->] (4.6,1.5) to [out=180,in=-30] (3.8,1.8);
      \node at (5.1,1.5) {$\reg(\mat{H})$};
    \end{axis}
  \end{tikzpicture}
  
  \caption{The zonotope $\reg(\mat{H})$ for the $2 \times 3$ MIMO
    channel matrix $\mat{H}=[2.5, 5, 1;1.2, 2.4, 2]$ and its
    minimum-energy decomposition into two
    parallelograms. The peak power is $\amp = 1$.}
  \label{fig:23_dependent}
\end{figure}
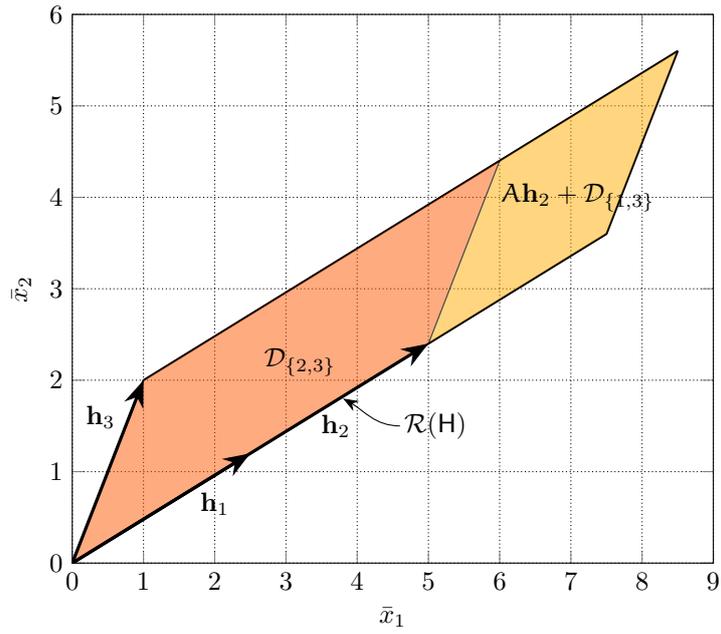

\begin{figure}[htbp]
  \centering
  \begin{tikzpicture}
    [scale=1,
    >={Stealth[scale=1.2]},
    dot/.style={circle,draw=red,fill=red,thick,
      inner sep=0pt,minimum size=1.2mm}]

    \begin{axis}[scale only axis, clip=true,
      xmin=-4,
      xmax=16,
      xtick={-4,-2,...,16},
      xmajorgrids,
      xlabel={$\xs_1$},        
      ymin=-2,
      ymax=7,
      ytick={-2,-1,...,7},
      ymajorgrids,
      ylabel={$\xs_2$},
      grid style={densely dotted,black}] 
    
      \coordinate (h) at (0,0);
      \coordinate (hr1) at (-2,-1.2);
      \coordinate (h11) at (axis direction cs:2,1.2);
      \coordinate (h1) at (axis direction cs: -2,-1.2);
      \coordinate (hr2) at (7,1);
      \coordinate (h2) at (axis direction cs: 7,1);
      \coordinate (hr3) at (5,2);
      \coordinate (h3) at (axis direction cs: 5,2);
      \coordinate (hr4) at (2,2.9);
      \coordinate (h4) at (axis direction cs: 2,2.9);

      \draw[draw=black,thick] 
       (hr1) -- ++ (h2) --++(h3) -- ++(h11)-- ++ (h4);
     
      \edef\temp{\noexpand\draw[draw=black,thick] 
        (hr1) -- ++(h4) -- ++(h11) -- ++(h3) -- ++(h2);
      }
      \temp
      
      \coordinate (g12) at ($(h)$);
      \coordinate (g13) at ($(hr2)$);
      \coordinate (g14) at ($(h)$);
      \coordinate (g23) at ($(h)$);
      \coordinate (g24) at ($(hr3)$);
      \coordinate (g34) at ($(h)$);
      
      \newcount\mycount
      \global\mycount=0

      \foreach \i in {1,...,3}
        \foreach \jj [evaluate={\j=\jj+1}] in {\i,...,3} {
          \edef\temp{\noexpand\coordinate (s) at (g\i\j);}
          \temp
          \edef\temp{\noexpand\coordinate (a) at (h\i);}
          \temp
          \edef\temp{\noexpand\coordinate (b) at (h\j);}
          \temp
          \edef\temp{\noexpand\draw[fill=red!\the\mycount!yellow,opacity=0.5,draw=black]
            (s) -- ($(s)+(a)$) -- ($(s)+(a)+(b)$) -- ($(s)+(b)$) --
            (s);}
          \temp
          \global\advance\mycount by 16.67          
        }
        
      \foreach \i in {1,...,4} {
        \edef\temp{\noexpand\draw[very thick,->] (h) -- (hr\i);}
        \temp
      }
      \node at (0,-0.5) {$\vect{h}_1$};
      \node at (4.2,0.9) {$\vect{h}_2$};
      \node at (2.0,1.2) {$\vect{h}_3$};
      \node at (0.4,1.5) {$\vect{h}_4$};
      \node at (2,-0.2) {$\set{D}_{\{1,2\}}$};
      \node at (6,1.5) {$\set{D}_{\{2,3\}}$};
      \node at (3.6,2.5) {$\set{D}_{\{3,4\}}$};
      \node at (9.5,3.8) {$\vect{v}_{\{2,4\}}{+}\set{D}_{\{2,4\}}$};
      \draw[draw=black,->] (8.6,0.8) to [out=10,in=-30] (8,1.2);
      \node at (11.5,0.8) {$\vect{v}_{\{1,3\}}{+}\set{D}_{\{1,3\}}$};
      \node at (-1,3) {$\set{D}_{\{1,4\}}$};
      \draw[draw=black,->] (-1,2.9) to [out=-90,in=130] (0.4,1.7);

      \draw[draw=black,->] (10.7-7,1.5-2.5) to [out=180,in=-30] (9-7,1.8-2.5);
      \node at (11.7-7,1.5-2.5) {$\reg(\mat{H})$};
    \end{axis}
  \end{tikzpicture}
  
  \caption{Partition of $\reg(\mat{H})$ into the union
    \eqref{eq:union} for the $2 \times 4$ MIMO channel matrix
    $\mat{H}=[-2, 7, 5, 2; -1.2, 1, 2, 2.9]$. The peak power is
    $\amp = 1$.}
  \label{fig:fig6}
\end{figure}
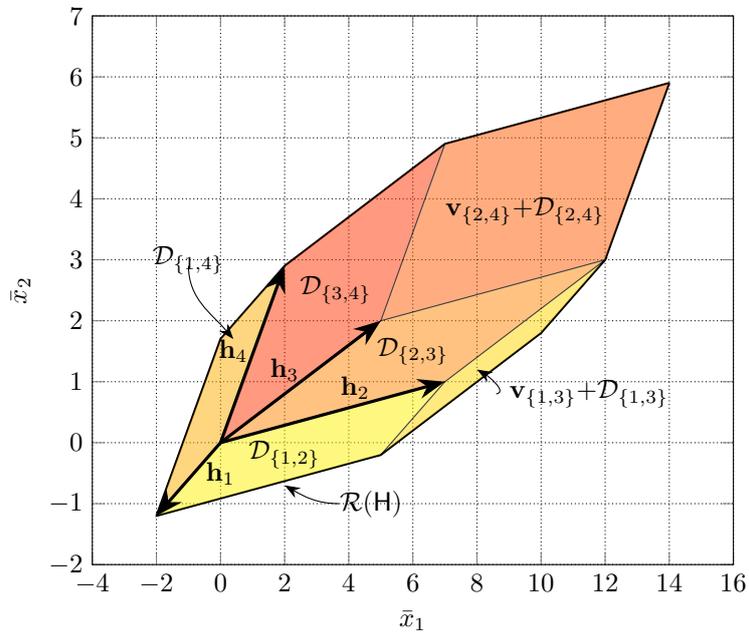

\begin{remark}
  If all $\nr$ column vectors in $\mat{H}$ are linearly independent,
  then the minimum-energy signaling partitions the range of $\xbar$
  into $\binom{\nt}{\nr}$ parallelepipeds. If some column vectors in
  $\mat{H}$ are linearly dependent, the number of parallelepipeds in
  the minimum-energy signaling partitioning will be less than
  $\binom{\nt}{\nr}$. Indeed, the number of parallelepipeds is equal
  to the number of ways of choosing $\nr$ linearly independent column
  vectors from $\mat{H}$.

  Figure~\ref{fig:23_dependent} shows an example of a $2\times 3$ MIMO
  channel with $\vect{h}_1$ and $\vect{h}_2$ being linearly dependent
  ($\vect{h}_2 = 2\vect{h}_1$). Instead of the usual $\binom{3}{2}=3$
  different parallelepipeds, in this example there are only two.

  Note that Lemma~\ref{lem:lemma1} holds true irrespectively of the number
  of parallelepipeds.
\end{remark}

\section{Equivalent Capacity Expression}
\label{sec:equiv-capac-expr}

We are now going to state an alternative expression for the capacity
$\C_{\mat{H}}(\amp,\alpha\amp)$ in terms of $\Xbar$ instead of
$\vect{X}$. To that goal we define for each index set $\mU \in \U$
\begin{IEEEeqnarray}{c}
  \label{eq:al}
  s_{\mU} \eqdef  \sum_{j \in \mcU} g_{\mU,j} , \quad \mU \in \U,
\end{IEEEeqnarray}
which indicates the number of components of the input vector that are
set to $\amp$ in order to induce $\vect{v}_{\mU}$ in \eqref{eq:vl}.
\begin{remark}
  \label{rem:s_values}
  It follows directly from Lemma~\ref{lem:lemma1} that, for every
  $\mU \in \U$,
  \begin{IEEEeqnarray}{c+x*}
    0 \le s_{\mU} \le \nt - \nr.
    \\
    & \nonumber \remarkendhere
  \end{IEEEeqnarray}
\end{remark}
We define a random variable $\rvU$ over $\U$ to indicate which
parallelepiped $\Xbar$ belongs to, i.e.,
\begin{IEEEeqnarray}{c}
  \label{eq:ueqi}
  \big(\rvU = \mU \big)
  \implies \big(\Xbar \in (\vect{v}_{\mU}+\set{D}_{\mU})\big). 
\end{IEEEeqnarray}
The choice of $\rvU$ that satisfies \eqref{eq:ueqi} is not unique
because of the ambiguity at the boundary points of the different
parallelepipeds: when $\xbar$ takes a value on the boundary between
multiple parallelepipeds, $\rvU$ could be randomly chosen among these
parallelepipeds or deterministically set to one of them. However, the
difference between these choices has no influence on our results,
since the set of all boundary points has zero $\nr$-dimensional
Lebesgue measure. For clarity, we shall restrict $\rvU$ to being a
deterministic function of $\Xbar$.

\begin{proposition}
  \label{prop:prop1}
  The capacity $\C_{\mat{H}}(\amp,\alpha\amp)$ as in
  \eqref{eq:capacity} can be written as
  \begin{IEEEeqnarray}{c}
    \label{eq:capacity_alternative}
    \C_{\mat{H}}(\amp,\alpha\amp)= \max_{P_{\Xbar}} \II(\Xbar ;
    \vect{Y}) 
  \end{IEEEeqnarray}
  where the maximization is over all distributions $P_{\Xbar}$ over
  $\reg(\mat{H})$ subject to the power constraint:
  \begin{IEEEeqnarray}{c} 
    \label{eq:new_constraint}
    \BigE[\rvU]{ \amp s_{\rvU} + \bigl\|\inv{\mat{H}_{\rvU}} \bigl(
      \eEcond{\Xbar}{\rvU} -\vect{v}_{\rvU} \bigl) \bigl\|_1 } \leq
    \alpha \amp, 
  \end{IEEEeqnarray} 
  where the random variable $\rvU$ is a function of $\Xbar$ that
  satisfies \eqref{eq:ueqi}.
\end{proposition}
\begin{IEEEproof}
  Notice that $\Xbar$ is a function of $\vect{X}$ and that we have a
  Markov chain $\vect{X} \markov \Xbar \markov \vect{Y}$.  Therefore,
  $\II(\Xbar;\vect{Y}) = \II(\vect{X} ; \vect{Y})$.  Moreover, by
  Lemma~\ref{lem:lemma1}, the range of $\Xbar$ in $\reg(\mat{H})$ can
  be decomposed into the shifted parallelepipeds
  $\{ \vect{v}_{\mU} +\set{D}_{\mU}\}_{\mU \in \U}$. Again by
  Lemma~\ref{lem:lemma1}, for any image point $\xbar$ in
  $\vect{v}_{\mU}+\set{D}_{\mU}$, the minimum energy required to
  induce $\xbar$ is
  \begin{IEEEeqnarray}{c}
    \amp s_{\mU} + \bigl\|\inv{\mat{H}_{\mU}}( \xbar -
    \vect{v}_{\mU})\bigr\|_1.  
  \end{IEEEeqnarray}
  Without loss in optimality, we restrict ourselves to input vectors
  $\vect{x}$ that achieve some $\xbar$ with minimum energy. Then,
  using $p_{\mU}$ to denote $\Prv{\rvU=\mU}$ and by the law of total
  expectation, the average power can be rewritten as
  \begin{IEEEeqnarray}{rCl}
    \bigE{\|\vect{X}\|_1}
    & = & \sum_{\mU \in \U} p_{\mU}
    \bigEcond{\|\vect{X}\|_1}{ \rvU =  \mU}
    \\
    & = & \sum_{\mU \in \U}p_{\mU} \BigEcond{\amp s_{\mU} +
      \bigl\|\inv{\mat{H}_{\mU}}( \Xbar  -
      \vect{v}_{\mU})\bigr\|_1}{\rvU=\mU} 
    \\
    & = & \sum_{\mU \in \U}p_{\mU} \Bigl(\amp s_{\mU} +\BigEcond{
    \bigl\|\inv{\mat{H}_{\mU}}( \Xbar  -
    \vect{v}_{\mU})\bigr\|_1}{\rvU=\mU}\Bigl) 
    \\
    & = & \sum_{\mU \in \U} p_{\mU} \Bigl(\amp s_{\mU} +
    \bigl\|\inv{\mat{H}_{\mU}} \bigl( \eEcond{\Xbar}{\rvU=\mU}
    -  \vect{v}_{\mU}\bigr) \bigr\|_1\Bigl)
    \label{eq:expflip}
    \\
    & = & \BigE[\rvU]{ \amp s_{\rvU} + \bigl\|\inv{\mat{H}_{\rvU}}
      \bigl( \eEcond{\Xbar}{\rvU} -\vect{v}_{\rvU}\bigl)\bigl\|_1 },
  \end{IEEEeqnarray} 
  where \eqref{eq:expflip} follows from the fact that all components
  of $\inv{\mat{H}_{\mU}}( \Xbar - \vect{v}_{\mU})$ are nonnegative.
\end{IEEEproof}
\medskip

\begin{remark}
  The term inside the expectation on the left-hand side (LHS) of
  \eqref{eq:new_constraint} can be seen as a cost function for
  $\Xbar$, where the cost is linear within each of the parallelepipeds
  $\{ \set{D}_{\mU}+\vect{v}_{\mU}\}_{\mU \in \U}$ (but not linear on
  the entire $\reg(\mat{H})$). At very high SNR, the receiver can
  obtain an almost perfect guess of $\rvU$. As a result, our channel
  can be seen as a set of almost parallel channels in the sense of
  \cite[Exercise~7.28]{coverthomas06_1}. Each one of the parallel
  channels is an amplitude-constrained $\nr \times \nr$ MIMO channel,
  with a linear power constraint. This observation will help us obtain
  upper and lower bounds on capacity that are tight in the high-SNR
  limit.  Specifically, for an upper bound, we reveal $\rvU$ to the
  receiver and then apply previous results on full-rank
  $\nr \times \nr$ MIMO channels \cite{mosermylonakiswangwigger17_1}.
  For a lower bound, we choose the inputs in such a way that, on each
  parallelepiped $\set{D}_{\mU}+\vect{v}_{\mU}$, the vector $\Xbar$
  has the high-SNR-optimal distribution for the corresponding
  $\nr \times \nr$ channel.
\end{remark}

\section{Maximum-Variance Signaling}
\label{sec:maxim-vari-sign}

The proofs to the lemmas in this section are given in
Appendix~\ref{app:prof_lemma9}.

As we shall see (Theorem~\ref{thm:them11} ahead and
\cite{chaabanrezkialouini18_2,prelovverdu04_1}), at low SNR the
asymptotic capacity is characterized by the maximum trace of the
covariance matrix of $\Xbar$, which we denote
\begin{IEEEeqnarray}{c}
  \cov{\Xbar} \eqdef \bigE{(\Xbar -\eE{\Xbar})\trans{(\Xbar -
      \eE{\Xbar})}}.
  \label{eq:trace} 
\end{IEEEeqnarray}
In this section we discuss properties of an optimal input distribution
for $\vect{X}$ that maximizes this trace.
Thus,
we are interested in the following maximization problem:
\begin{IEEEeqnarray}{c}
  \max_{P_{\vect{X}} \textnormal{ satisfying } \eqref{eq:constraints}}
  \bigtrace{\cov{\Xbar}}
  \label{eq:optimizationproblem}
\end{IEEEeqnarray}
where the maximization is over all input distributions $P_{\vect{X}}$
satisfying the peak- and average-power constraints given in
\eqref{eq:constraints}.

The following three lemmas show that the optimal input to the
optimization problem in \eqref{eq:optimizationproblem} has certain
structures: Lemma~\ref{lem:binaryinput} shows that it is discrete with
all entries of mass points taking values in $\{0,\amp\}$;
Lemma~\ref{lem:path} shows that the possible values of the optimal
$\vect{X}$ form a ``path'' in $[0,\amp]^{\nt}$ starting from the
origin; and Lemma~\ref{lem:optimalinputtracecov} shows that, under
mild assumptions, this optimal $\vect{X}$ takes at most $\nr+2$
values.

\begin{lemma}
  \label{lem:binaryinput}
  An optimal input to the maximization problem in
  \eqref{eq:optimizationproblem} uses for each component of $\vect{X}$
  only the values $0$ and $\amp$:
  \begin{IEEEeqnarray}{c}
    \bigPrv{X_i \in \{0,\amp\}} = 1, \quad i=1, \ldots, \nt. 
  \end{IEEEeqnarray}
\end{lemma}

\begin{lemma}
  \label{lem:path}
  An optimal input to the optimization problem in
  \eqref{eq:optimizationproblem} is a PMF $P_{\vect{X}}^*$ over a set
  $\{\vect{x}_1^*,\vect{x}_2^*,\ldots\}$ satisfying
  \begin{IEEEeqnarray}{c}
    x_{k,\ell}^* \le x_{k',\ell}^*  \quad \textnormal{for all } k < k',
    \; \ell=1, \ldots, \nt.
    \label{eq:optimalinputc}
  \end{IEEEeqnarray}
  Furthermore, the first point is $\vect{x}_1^*=\vect{0}$, and
  \begin{IEEEeqnarray}{c}
    P_{\vect{X}}^*(\vect{0}) >0.
    \label{eq:optimalinputb}
  \end{IEEEeqnarray}
\end{lemma}
Notice that Lemma~\ref{lem:binaryinput} and the first part of
Lemma~\ref{lem:path} have already been proven in
\cite{chaabanrezkialouini18_2}. A proof is given in the appendix for
completeness.
 
Define $\T$ to be the power set of $\{1,\ldots, \nt\}$ without the
empty set, and define for every $\mV\in\T$ and every
$i\in\{1, \ldots, \nr\}$
\begin{IEEEeqnarray}{c}
  r_{\mV,i} \eqdef  \sum_{k=1}^{\nt} h_{i,k} \I{k \in \mV},
  \quad \forall\, \mV\in\T, \; \forall\, i \in \{1,\ldots, \nr\},
\end{IEEEeqnarray}
with $\I{\cdot}$ denoting the indicator function.  (Here $\mV$
describes a certain choice of input antennas that will be set to
$\amp$, while the remaining antennas will be set to $0$.)  Number all
possible $\mV\in\T$ from $\mV_1$ to $\mV_{\const{T}}$ (where
$\const{T} = 2^{\nt}-1$) and define the matrix
\begin{IEEEeqnarray}{c}
  \label{eq:rank1}
  \mat{R} \eqdef 
  \begin{pmatrix}
    2r_{\mV_1,1} & \cdots & 2r_{\mV_1,\nr} & \abs{\mV_1}
    & \|\vect{r}_{\mV_1}\|_2^2
    \\[1ex] 
    2r_{\mV_2,1} & \cdots & 2r_{\mV_2,\nr} & \abs{\mV_2}
    & \|\vect{r}_{\mV_2}\|_2^2
    \\[1ex] 
    \vdots & \ddots & \vdots & \vdots & \vdots
    \\
    2r_{\mV_{\const{T}},1} & \cdots & 2r_{\mV_{\const{T}},\nr}
    & \abs{\mV_{\const{T}}} & \|\vect{r}_{\mV_{\const{T}}}\|_2^2 
  \end{pmatrix}
\end{IEEEeqnarray}
where
\begin{IEEEeqnarray}{c}
  \vect{r}_{\mV}
  \eqdef \trans{\bigl(r_{\mV,1},r_{\mV,2},\ldots, r_{\mV,\nr}\bigr)},
  \quad \forall\,\mV\in\T.
\end{IEEEeqnarray}

\begin{lemma}
  \label{lem:optimalinputtracecov}
  If every $(\nr+2)\times (\nr+2)$ submatrix $\mat{R}_{\nr+2}$ of
  matrix $\mat{R}$ is of full rank
  \begin{IEEEeqnarray}{c}
    \rank{\mat{R}_{\nr+2}} = \nr+2,
    \label{eq:21}
  \end{IEEEeqnarray}
  then the optimal input to the optimization problem in
  \eqref{eq:optimizationproblem} is a PMF $P_{\vect{X}}^*$ over a set
  $\{\vect{0}, \vect{x}_1^*, \ldots, \vect{x}_{\nr+1}^*\}$ with
  $\nr+2$ points.
  %
\end{lemma}

\begin{table}[htbp]
  \centering
  \caption{Maximum variance distributions for different channel coefficients}
  \label{tab:vmax}
  \vspace{3mm}
  \begin{IEEEeqnarraybox}[\mystrut]{l"l"l"l}
    \hline\hline
    \IEEEeqnarraymulticol{1}{t}{channel gains}
    & \IEEEeqnarraymulticol{1}{c}{\alpha}
    & \IEEEeqnarraymulticol{1}{c}{\max_{P_{\vect{X}}} \bigtrace{\cov{\Xbar}}}
    & \IEEEeqnarraymulticol{1}{c}{P_{{\vect{X}}}\colon
      \max_{P_{\vect{X}}} \bigtrace{\cov{\Xbar}}} 
    \\\hline
    \\[-3mm]
    \mat{H} = 
    \begin{pmatrix} 
      1.3 & 0.6 & 1 & 0.1
      \\ 
      2.1 & 4.5 & 0.7 & 0.5 
    \end{pmatrix} & 1.5 & 16.3687\amp^2
    & P_{\vect{X}}(0,0,0,0) = 0.625,\\[-2mm]
    &\,&\,&P_{\vect{X}}(\amp,\amp,\amp,\amp) = 0.375
    \\[2ex]
    \mat{H} = 
    \begin{pmatrix} 
      1.3 & 0.6 & 1 & 0.1
      \\ 
      2.1 & 4.5 & 0.7 & 0.5 
    \end{pmatrix} & 0.9 & 12.957\amp^2
    & P_{\vect{X}}(0,0,0,0)=0.7,\\[-2mm]
    &\,&\,&P_{\vect{X}}(\amp,\amp,\amp,0)=0.3
    \\[2ex]
    \mat{H} =
    \begin{pmatrix} 
      1.3 & 0.6 & 1 & 0.1
      \\ 
      2.1 & 4.5 & 0.7 & 0.5 
    \end{pmatrix} & 0.6 & 9.9575\amp^2
    & P_{\vect{X}}(0,0,0,0)=0.7438, \\[-2mm]
    &\,&\,&P_{\vect{X}}(\amp,\amp,0,0) =0.1687, \\
    &\,&\,&P_{\vect{X}}(\amp,\amp,\amp,0)=0.0875
    \\[2ex]
    \mat{H} =
    \begin{pmatrix} 
      1.3 & 0.6 & 1 & 0.1
      \\ 
      2.1 & 4.5 & 0.7 & 0.5 
    \end{pmatrix} & 0.3 & 6.0142\amp^2
    & P_{\vect{X}}(0,0,0,0)=0.85, \\[-2mm]
    &\,&\,&P_{\vect{X}}(\amp,\amp,0,0) =0.15
    \\[2ex]
    \mat{H} = 
    \begin{pmatrix} 
      0.9 & 3.2 & 1 & 2.1
      \\ 
      0.5 & 3.5 & 1.7 & 2.5 
      \\
      0.7 & 1.1 & 1.1 & 1.3 
    \end{pmatrix} & 0.9 & 23.8405\amp^2
    & P_{\vect{X}}(0,0,0,0)=0.7755, \\[-4mm]
    &\,&\,&P_{\vect{X}}(\amp,\amp,\amp,\amp)=0.2245
    \\[2ex]
    \mat{H} = 
    \begin{pmatrix} 
      0.9 & 3.2 & 1 & 2.1
      \\ 
      0.5 & 3.5 & 1.7 & 2.5 
      \\
      0.7 & 1.1 & 1.1 & 1.3 
    \end{pmatrix} & 0.75 & 20.8950\amp^2
    & P_{\vect{X}}(0,0,0,0)=0.7772, \\[-4mm]
    &\,&\,&P_{\vect{X}}(\amp,\amp,\amp,0)=0.1413,  \\
    &\,&\,&P_{\vect{X}}(\amp,\amp,\amp,\amp)=0.0815
    \\[2ex]
    \mat{H} = 
    \begin{pmatrix} 
      0.9 & 3.2 & 1 & 2.1
      \\ 
      0.5 & 3.5 & 1.7 & 2.5 
      \\
      0.7 & 1.1 & 1.1 & 1.3 
    \end{pmatrix} & 0.6 & 17.7968\amp^2
    & P_{\vect{X}}(0,0,0,0)=0.8, \\[-4mm]
    &\,&\,&P_{\vect{X}}(\amp,\amp,\amp,0)=0.2
    \\[2mm]\hline\hline
  \end{IEEEeqnarraybox}
\end{table}

\begin{remark}
  Lemmas~\ref{lem:lemma1} and~\ref{lem:binaryinput} together imply
  that the optimal $\Xbar$ in \eqref{eq:optimizationproblem} takes
  value only in the set $\F$ of corner points of the parallelepipeds
  $\{\vect{v}_{\mU} + \set{D}_{\mU}\}$:
  \begin{IEEEeqnarray}{c}
    \F \eqdef \bigcup_{\mU\in\U} \biggl\{ \vect{v}_{\mU} +
    \sum_{i\in\mU} \lambda_i \vect{h}_i \colon \lambda_i\in\{0,\amp\}, \;
    \forall\, i\in\mU \biggr\}.
    \label{eq:cornerpoints}
  \end{IEEEeqnarray}
  Lemmas~\ref{lem:path} and~\ref{lem:optimalinputtracecov} further
  imply that the possible values of this optimal $\Xbar$ form a path
  in $\F$, starting from $\vect{0}$, and containing no more than
  $\nr+2$ points.
\end{remark}

Table~\ref{tab:vmax} shows seven examples of distributions that
maximize the trace of the covariance matrix, which fall into five
different cases in terms of mass-point placement: there are only two
mass points, one at $(0,0,0,0)$ and the other at
$(\amp, \amp, \amp, \amp)$, $(\amp, \amp, \amp,0)$, or
$(\amp,\amp, 0,0)$; or there are three mass points, the first at at
$(0,0,0,0)$, the second at $(\amp,\amp,\amp,0)$, and the third at
either $(\amp,\amp,0,0)$ or $(\amp,\amp,\amp,\amp)$. As expected, the
distributions follow the properties claimed in
Lemmas~\ref{lem:binaryinput},~\ref{lem:path},
and~\ref{lem:optimalinputtracecov}.

\section{Capacity Results}
\label{sec:capacity-results}

Define
\begin{IEEEeqnarray}{c}
  \V \eqdef \sum_{\mU \in \U}
  \abs{\det\mat{H}_{\mU} }.
  \label{eq:totalvolume}
\end{IEEEeqnarray}
Let $\vect{q}$ be a probability vector on $\U$ with entries
\begin{IEEEeqnarray}{c}
  q_{\mU} \eqdef \frac{\abs{\det\mat{H}_{\mU}}}{\V}, \quad
  \mU\in\U. 
  \label{eq:24}
\end{IEEEeqnarray}
Further, define
\begin{IEEEeqnarray}{c}
  \alpha_{\textnormal{th}} \eqdef  \frac{\nr}{2} + \sum_{\mU\in\U} 
  s_{\mU} \> q_{\mU},
  \label{eq:10}
\end{IEEEeqnarray}
where $\{s_{\mU}\}$ are defined in \eqref{eq:al}. Notice that
$\alpha_{\textnormal{th}}$ determines the threshold value for $\alpha$
above which $\Xbar$ can be made uniform over $\reg(\mat{H})$. In fact,
combining the minimum-energy signaling in \eqref{eq:xvalue} with a
uniform distribution for $\Xbar$ over $\reg(\mat{H})$, the expected
input power is
\begin{IEEEeqnarray}{rCl}
  \E{\|\vect{X}\|_1}
  & = & \sum_{\mU \in \U} \Prv{\rvU = \mU} \cdot
  \Econd{\|\vect{X}\|_1}{\rvU = \mU}
  \label{eq:Xv}
  \\ 
  & = & \sum_{\mU \in \U} q_{\mU} \left( \amp s_{\mU} + \frac{\nr
      \amp}{2} \right)
  \label{eq:Xv2}
  \\
  & = & \alpha_{\textnormal{th}} \amp,
  \label{eq:Xv3}
\end{IEEEeqnarray}
where the random variable $\rvU$ indicates the parallelepiped
containing $\Xbar$; see \eqref{eq:ueqi}. Equality \eqref{eq:Xv2} holds
because, when $\Xbar$ is uniform over $\reg(\mat{H})$,
$\Prv{\rvU = \mU}=q_{\mU}$, and because, conditional on $\rvU=\mU$,
using the minimum-energy signaling scheme, the input vector $\vect{X}$
is uniform over $\vect{v}_{\mU}+\set{D}_{\mU}$.
\begin{remark}
  \label{rem:bound}
  Note that
  \begin{IEEEeqnarray}{c}
    \alpha_{\textnormal{th}} \le \frac{\nt}{2},
    \label{eq:15}
  \end{IEEEeqnarray}
  as can be argued as follows. Let $\vect{X}$ be an input that
  achieves a uniform $\Xbar$ with minimum energy. According to
  \eqref{eq:Xv3} it consumes an input power
  $\alpha_{\textnormal{th}} \amp$. Define $\vect{X}'$ as
  \begin{IEEEeqnarray}{c}
    X'_i \eqdef \amp - X_i,  \quad i=1, \ldots, \nt.
  \end{IEEEeqnarray}
  It must consume input power
  $(\nt-\alpha_{\textnormal{th}})\amp$. Note that $\vect{X}'$ also
  induces a uniform $\Xbar$ because the zonotope $\reg(\mat{H})$ is
  point-symmetric. Since $\vect{X}$ consumes minimum energy, we know
  \begin{IEEEeqnarray}{c}
    \E{\|\vect{X}\|_1} \le \E{\|\vect{X}'\|_1},
  \end{IEEEeqnarray}
  i.e.,
  \begin{IEEEeqnarray}{c}
    \alpha_{\textnormal{th}} \amp
    \le (\nt-\alpha_{\textnormal{th}})\amp,
  \end{IEEEeqnarray}
  which implies \eqref{eq:15}.  
\end{remark}

\subsection{Lower Bounds}
\label{sec:lower-bounds}

The proofs to the theorems in this section can be found in
Appendix~\ref{sec:deriv-lower-bounds}.

\begin{theorem}
  \label{thm:them4}
  If $\alpha \geq \alpha_{\textnormal{th}}$, then
  \begin{IEEEeqnarray}{c}
    \C_{\mat{H}}(\amp,\alpha \amp) \geq  \frac{1}{2} \log\left(1+ 
      \frac{\amp^{2\nr} \V^2 }{(2\pi e)^{\nr}}\right).
    \label{eq:lowbnd1} 
  \end{IEEEeqnarray}
\end{theorem}

\begin{theorem}
  \label{thm:lowerbound2}
  If $\alpha < \alpha_{\textnormal{th}}$, then
  \begin{IEEEeqnarray}{c}
    \C_{\mat{H}}(\amp,\alpha \amp) \geq \frac{1}{2}
    \log\left(1+\frac{\amp^{2\nr}\V^2}{(2\pi e)^{\nr}} 
      \ope^{2\nu} \right)
    \label{eq:low_bnd2} 
  \end{IEEEeqnarray}
  with 
  \begin{IEEEeqnarray}{c}
    \nu \eqdef  \sup_{\lambda \in \left(\max \left\{0,\frac{\nr}{2}
          + \alpha - \alpha_{\textnormal{th}}\right\},
        \min\left\{\frac{\nr}{2},\alpha\right\} \right)}
    \left\{\nr\left(1- \log{\frac{\mu}{1-\ope^{-\mu}}}
        - \frac{\mu\ope^{-\mu}}{1-\ope^{-\mu}}\right)
      - \inf_{\vect{p}}  \const{D}(\vect{p}\|\vect{q})\right\},  
    \nonumber\\*
    \label{eq:nu}
  \end{IEEEeqnarray}
  where $\mu$ is the unique solution to the following equation:
  \begin{IEEEeqnarray}{c}
    \frac{1}{\mu} - \frac{\ope^{-\mu}}{1-\ope^{-\mu}}
    = \frac{\lambda}{\nr},
    \label{eq:lambda}
  \end{IEEEeqnarray}
  and where the infimum is over all probability vectors $\vect{p}$ on
  $\U$ such that
  \begin{IEEEeqnarray}{c}
    \sum_{\mU \in \U} p_{\mU} s_{\mU}
    = \alpha-\lambda  
    \label{eq:pconstraint}
  \end{IEEEeqnarray}
  with $\{s_{\mU}\}$ defined in \eqref{eq:al}.
\end{theorem}

The two lower bounds in Theorems~\ref{thm:them4}
and~\ref{thm:lowerbound2} are derived by applying the EPI, and by
maximizing the differential entropy $\hh(\Xbar)$ under the
constraint~\eqref{eq:new_constraint}. When
$\alpha \geq \alpha_{\textnormal{th}}$, choosing $\Xbar$ to be
uniformly distributed on $\reg(\mat{H})$
satisfies~\eqref{eq:new_constraint}, hence we can achieve
$\hh(\Xbar) = \log{\V}$. When $\alpha < \alpha_{\textnormal{th}}$, the
uniform distribution is no longer an admissible distribution for
$\Xbar$. In this case, we first select a PMF over the events
$\{\Xbar \in (\vect{v}_{\mU}+ \set{D}_{\mU})\}_{\mU \in \U}$, and,
given $\Xbar \in \vect{v}_{\mU} + \set{D}_{\mU}$, we choose the inputs
$\{X_{i} \colon i\in\mU\}$ according to a truncated exponential
distribution rotated by the matrix $\mat{H}_{\mU}$. Interestingly, it
is optimal to choose the truncated exponential distributions for all
sets $\mU \in \U$ to have the same parameter $\mu$. This parameter is
determined by the power $\frac{\lambda}{\nr}\amp$ allocated to the
$\nr$ signaling inputs $\{X_i \colon i\in \mU\}$.

\subsection{Upper Bounds}
\label{sec:upper-bounds}

The proofs to the theorems in this section can be found in
Appendix~\ref{sec:deriv-upper-bounds}.

The first upper bound is based on an analysis of the channel with
peak-power constraint only, i.e., the average-power constraint
\eqref{eq:average} is ignored.
\begin{theorem}
  \label{thm:them5}
  For an arbitrary $\alpha$,
  \begin{IEEEeqnarray}{rCl}
    \C_{\mat{H}}(\amp,\alpha \amp)
    & \leq & \sup_{\vect{p}} \Biggl\{\log\V
    -\const{D}(\vect{p}\|\vect{q})  + \sum_{\mU \in \U}
    p_{\mU} \sum_{\ell=1}^{\nr}
    \log\left(\sigma_{\mU,\ell} + \frac{\amp}{\sqrt{2\pi e}}\right)
    \Biggl\}, 
    \label{eq:upp_bnd}  
  \end{IEEEeqnarray}
  where $\sigma_{\mU,\ell}$ denotes the square root of the $\ell$th
  diagonal entry of the matrix
  $\inv{\mat{H}}_{\mU}\invtrans{\mat{H}}_{\mU}$, and where the
  supremum is over all probability vectors $\vect{p}$ on $\U$.
\end{theorem}

The following two upper bounds in Theorems~\ref{thm:them6} and~\ref{thm:them7}
hold only when $\alpha < \alpha_{\textnormal{th}}$.
\begin{theorem}
  \label{thm:them6}
  If $\alpha < \alpha_{\textnormal{th}}$, then
  \begin{IEEEeqnarray}{rCl}
    \C_{\mat{H}}(\amp,\alpha \amp) 
    & \leq & \sup_{\vect{p}} \inf_{\mu > 0} \Biggl\{ \log\V
    -\const{D}(\vect{p}\|\vect{q}) + \sum_{\mU \in \U}p_{\mU}
    \sum_{\ell=1}^{\nr} \log\biggl( \sigma_{\mU,\ell}
    + \frac{\amp}{{\sqrt{2\pi e}}} \frac{1-\ope^{-\mu}}{\mu} \biggr)  
    \nonumber\\
    &&\qquad\qquad +\> \frac{\mu}{\amp\sqrt{2\pi}}
    \sum_{\mU \in \U}p_{\mU}\sum_{\ell=1}^{\nr} \sigma_{\mU,\ell}
    \biggl(1 - \ope^{-\frac{\amp^{2}}{2\sigma_{\mU,\ell}^2}}\biggr)
    + \mu \Biggl( \alpha - \sum_{\mU\in\U} p_{\mU}s_{\mU} \Biggr)      
    \Biggr\},
    \nonumber\\*
    \label{eq:upp2} 
  \end{IEEEeqnarray}
  where the supremum is over all probability vectors $\vect{p}$ on
  $\U$ such that
  \begin{IEEEeqnarray}{c}
    \sum_{\mU \in \U} p_{\mU} s_{\mU} \leq \alpha.  
    \label{eq:pconstraint2}
  \end{IEEEeqnarray}
\end{theorem}

\begin{theorem}
  \label{thm:them7}
  If $\alpha < \alpha_{\textnormal{th}}$, then
  \begin{IEEEeqnarray}{rCl}
    \IEEEeqnarraymulticol{3}{l}{%
      \C_{\mat{H}}(\amp,\alpha \amp)
    }\nonumber\\*\quad%
    & \leq &  \sup_{\vect{p}} \inf_{\delta, \mu > 0} \Biggl\{
    \log\V - \const{D}(\vect{p}\|\vect{q}) + \sum_{\mU \in \U}p_{\mU}
    \sum_{\ell=1}^{\nr} \log \Biggl( \amp
    \cdot \frac{\ope^{\frac{{\mu\delta}}{\amp}} -  
      \ope^{-\mu(1+\frac{\delta}{\amp})}}{\sqrt{2\pi e}\mu \bigl(1 -
      2\bigQf{\frac{\delta}{\sigma_{\mU,\ell}}}\bigr)}  
    \Biggr)  \nonumber\\
    && \qquad\qquad + \sum_{\mU \in \U}p_{\mU}\sum_{\ell=1}^{\nr}
    \Qf{\frac{\delta}{\sigma_{\mU,\ell}}} +  \sum_{\mU \in
      \U}p_{\mU}\sum_{\ell=1}^{\nr} 
    \frac{\delta}{\sqrt{2\pi}\sigma_{\mU,\ell}}
    \ope^{-\frac{\delta^2}{2\sigma_{\mU,\ell}^2}}
    \nonumber\\ 
    && \qquad\qquad + \> \frac{\mu}{\amp\sqrt{2\pi}}
    \sum_{\mU \in \U}p_{\mU} \sum_{\ell=1}^{\nr}
    \sigma_{\mU,\ell}
    \Biggl( \ope^{-\frac{\delta^2}{2\sigma_{\mU,\ell}^2}} -
    \ope^{-\frac{(\amp+\delta)^{2}}{2\sigma_{\mU,\ell}^2}}\Biggr)  
    + \mu \Biggl( \alpha - \sum_{\mU\in\U}p_{\mU} s_{\mU}\Biggr)
    \Biggr\},
    \nonumber\\*
    \label{eq:upp3}
  \end{IEEEeqnarray}
  where $\Qf{\cdot}$ denotes the $\Q$-function associated with the
  standard normal distribution, and the supremum is over all
  probability vectors $\vect{p}$ on $\U$ satisfying
  \eqref{eq:pconstraint2}.
\end{theorem}

The three upper bounds in Theorems~\ref{thm:them5},~\ref{thm:them6}
and~\ref{thm:them7} are derived using the fact that capacity cannot be
larger than over a channel where the receiver observes both $\vect{Y}$
and $\rvU$. The mutual information corresponding to this channel
$\II(\Xbar;\vect{Y},\rvU)$ decomposes as
$\HH(\rvU) + \II(\Xbar;\vect{Y}|\rvU)$, where the term $\HH(\rvU)$
indicates the rate that can be achieved by coding over the choice of
the parallelepiped to which $\Xbar$ belongs, and
$\II(\Xbar;\vect{Y}|\rvU)$ indicates the average rate that can be
achieved by coding over a single parallelepiped. By the results in
Lemma~\ref{lem:lemma1}, we can treat the channel matrix as an
invertible matrix when knowing $\rvU$, which greatly simplifies the
bounding on $\II(\Xbar;\vect{Y}|\rvU)$. The upper bounds are then
obtained by optimizing over the probabilities assigned to the
different parallelepipeds. As we will see later, the upper bounds are
asymptotically tight at high SNR. The reason is that the additional
term
$\II(\Xbar;\vect{Y},\rvU) - \II(\Xbar;\vect{Y}) =
\II(\Xbar;\rvU|\vect{Y})$ vanishes as the SNR grows large. To derive
the asymptotic high-SNR capacity, we also use previous results in
\cite{mosermylonakiswangwigger17_1}, which derived the high-SNR
capacity of this channel when the channel matrix is invertible.

Our next upper bound in Theorem~\ref{thm:them9} is determined by the
maximum trace of the covariance matrix of $\Xbar$ under constraints
\eqref{eq:constraints}.
\begin{theorem}
  \label{thm:them9}
  For an arbitrary $\alpha$,
  \begin{IEEEeqnarray}{c}
    \C_{\mat{H}}(\amp,\alpha\amp)
    \leq \frac{\nr}{2}\log \left(
      1 + \frac{1}{\nr} \max_{P_{\vect{X}}} \bigtrace{\cov{\Xbar}}
    \right), 
    \label{eq:upp4}
  \end{IEEEeqnarray}
  where the maximization is over all input distributions
  $P_{\vect{X}}$ satisfying the power constraints
  \eqref{eq:constraints}.
\end{theorem}
Note that Section~\ref{sec:maxim-vari-sign} provides results that
considerably simplify the maximization in \eqref{eq:upp4}.  In
particular, there exists a maximizing $P_{\vect{X}}$ that is a PMF
over $\vect{0}$ and at most $\nr+1$ other points on $\F$, where $\F$
is defined in \eqref{eq:cornerpoints}.

\subsection{Asymptotic Capacity Expressions}
\label{sec:asympt-capac-expr}

The proofs to the theorems in this section can be found in
Appendix~\ref{sec:deriv-asympt-results}.

\begin{theorem}[High-SNR Asymptotics]
  \label{thm:them8}
  If $\alpha \geq \alpha_{\textnormal{th}}$, then
  \begin{IEEEeqnarray}{c}
    \lim_{\amp \to \infty}  \bigl\{
    \C_{\mat{H}}(\amp,\alpha \amp) - \nr  \log{\amp} \bigr\}
    = \frac{1}{2} \log\left(\frac{\V^2}{(2\pi e)^{\nr}}\right).
    \IEEEeqnarraynumspace
    \label{eq:asycap1}
  \end{IEEEeqnarray}
  If $\alpha < \alpha_{\textnormal{th}}$, then
  \begin{IEEEeqnarray}{c}
    \lim_{\amp \to \infty} \bigl\{ \C_{\mat{H}}(\amp,\alpha \amp) -
    \nr \log{\amp} \bigr\} = \frac{1}{2} \log\left(\frac{\V^2}{(2\pi
        e)^{\nr}}\right) +\nu,
    \label{eq:asycap2}
  \end{IEEEeqnarray}
  where $\nu<0$ is defined in \eqref{eq:nu}--\eqref{eq:pconstraint}.
\end{theorem}

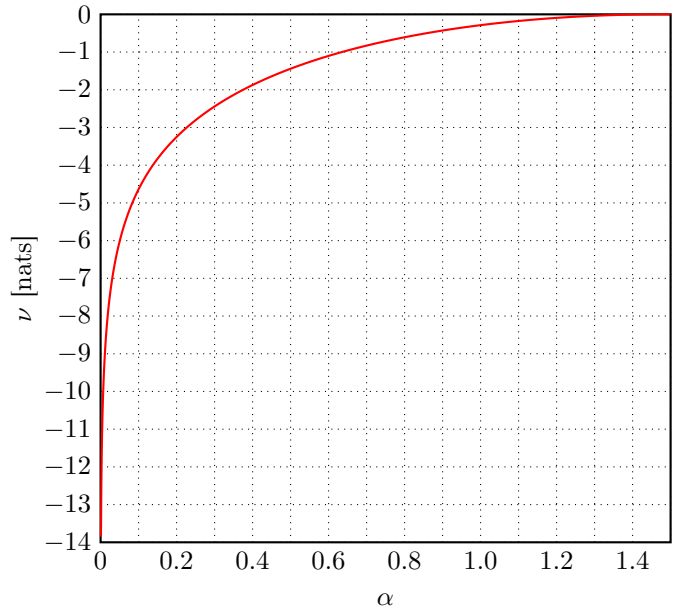
\begin{figure}[htpb]
  \centering
  \begin{tikzpicture}[%
    xscale=0.5,yscale=0.5,
    >={Stealth[scale=1.2]},
    dot/.style={circle,draw=black,fill=black,thick,
      inner sep=0pt,minimum size=1.5mm}]
    
    \draw[thick,-] (0,-14) rectangle (15,0); 
    \node at (7.5,-15.5) {$\alpha$};
    \node [rotate=90] at (-2,-7) {$\nu$ [nats]};
    \draw [dotted] (0,-14) grid (15,0);
    \foreach \i in {0,0.2,0.4,0.6,0.8,1.0,1.2,1.4}{
      \node [below] at (10*\i,-14) {$\i$};
    };
    \foreach \i in {-14,-13,...,0}{
      \node [left] at (0,\i) {$\i$};
    };

    \draw [xscale=10,yscale=1,thick,red] 
      plot file{figs/nu.dat};
  \end{tikzpicture}
  \caption{The parameter $\nu$ in \eqref{eq:nu} as a function of
    $\alpha$, for a $2 \times 3$ MIMO channel with channel matrix
    $\mat{H} = [1, 1.5, 3; 2, 2, 1]$ with corresponding
    $\alpha_{\textnormal{th}}=1.4762$. Recall that $\nu$ is the
    asymptotic capacity gap to the case with no active average-power
    constraint.}
  \label{fig:nu_alpha}
\end{figure}

Recall that $\alpha_{\textnormal{th}}$ is a threshold that determines
whether $\Xbar$ can be uniformly distributed over $\reg(\mat{H})$ or
not. When $\alpha < \alpha_{\textnormal{th}}$, compared with the
asymptotic capacity without active average-power constraint, the
average-power constraint imposes a penalty on the channel
capacity. This penalty is characterized by $\nu$ in
\eqref{eq:asycap2}. As shown in Figure~\ref{fig:nu_alpha}, $\nu$ is a
increasing function of $\alpha$. When
$\alpha < \alpha_{\textnormal{th}}$, $\nu$ is always negative, and
increases to $0$ when $\alpha \geq \alpha_{\textnormal{th}}$.

\begin{theorem}[Low-SNR Asymptotics]
  \label{thm:them11}
  For an arbitrary $\alpha$,
  \begin{IEEEeqnarray}{c}
    \lim_{\amp \downarrow 0}  \frac{\C_{\mat{H}}(\amp,\alpha
      \amp)}{\amp^2} 
    = \frac{1}{2} \max_{P_{\vect{X}}} \bigtrace{\cov{\Xbar}},
    \label{eq:lowSNRasymp}
  \end{IEEEeqnarray}
  where the maximization is over all input distributions
  $P_{\vect{X}}$ satisfying the constraints
  \begin{IEEEeqnarray}{rCl}
    \subnumberinglabel{eq:constraintsnormalized}    
     \bigPrv{X_k\in[0,1]}
    & = & 1, \quad \forall\,k\in\{1, \ldots, \nt\},
    \label{eq:peaknormalized}
    \\
    \bigE{\|\vect{X}\|_1} & \le & \alpha.
    \label{eq:averagenormalized}
  \end{IEEEeqnarray}
\end{theorem}

Again, see the results in Section~\ref{sec:maxim-vari-sign} about
maximizing the trace of the covariance matrix $\cov{\Xbar}$.

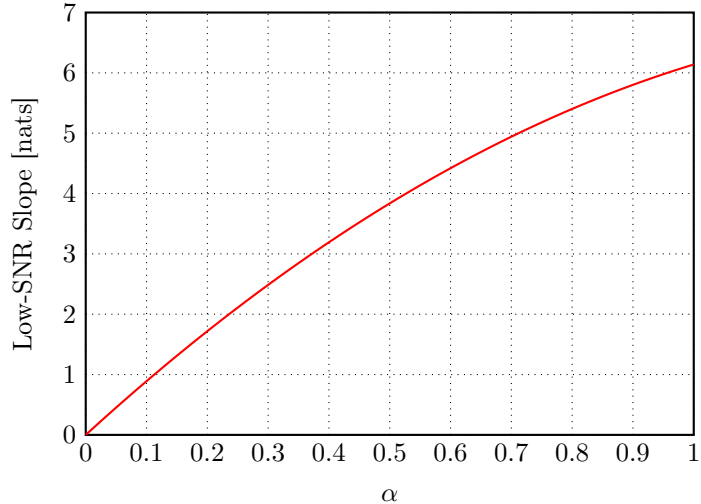
\begin{figure}[!ht]
  \centering
  \begin{tikzpicture}[%
    xscale=0.8,yscale=0.8,
    >={Stealth[scale=1.2]}]
    
    \draw[thick,-] (0,0) rectangle (10,7); 
    \node at (5,-1) {$\alpha$};
    \node [rotate=90] at (-1,3.5) {Low-SNR Slope [nats]};
    \draw [dotted] (0,0) grid (10,7);
    \foreach \i in {0,0.1,0.2,0.3,0.4,0.5,0.6,0.7,0.8,0.9,1}{
      \node [below] at (10*\i,0) {$\i$};
    };
    \foreach \i in {0,1,...,7}{
      \node [left] at (0,\i) {$\i$};
    };

    \draw [xscale=10,yscale=1,thick,red] 
      plot file{figs/slope.dat};
  \end{tikzpicture}
  \caption{Low-SNR slope as a function of $\alpha$, for a $2 \times 3$
    MIMO channel with channel matrix $\mat{H} = [1, 1.5, 3; 2, 2,
    1]$. }
  \label{fig:slope_alpha}
\end{figure}

\begin{example}
  \label{ex:20}
  Figure~\ref{fig:slope_alpha} plots the asymptotic slope, i.e., the
  right-hand side (RHS) of \eqref{eq:lowSNRasymp}, as a function of
  $\alpha$ for a $2\times 3$ MIMO channel.  As we can see, the
  asymptotic slope is strictly increasing for all values of
  $\alpha < \frac{\nt}{2}$.
\end{example}

\subsection{Numerical Results}
\label{sec:numerical-results}

In the following we present some numerical examples of our lower and
upper bounds.

\begin{figure}[htbp]
  \centering
  \begin{tikzpicture}[%
    xscale=0.25,yscale=0.6,
    >={Stealth[scale=1.2]}]

    \draw[thick,-] (-15,0) rectangle (25,14); 
    \node at (5,-1.1) {$\amp$ [dB]};
    \node at (5,14.5) {$2\times 3$, $\alpha=0.9$};
    \node [rotate=90] at (-17.3,7) {Capacity [nats]};
    \draw [dotted] (-15,0) grid [xstep=5,ystep=2] (25,14);
    \foreach \i in {-15,-10,...,25}{
      \node [below] at (\i,0) {$\i$};
    };
    \foreach \i in {0,2,...,14}{
      \node [left] at (-15,\i) {$\i$};
    };

    \draw [xscale=1,yscale=1,thick,red] 
      plot file{figs/b23a09upper19.dat};
    \draw [xscale=1,yscale=1,thick,blue,dashdotted] 
      plot file{figs/b23a09upper16.dat};
    \draw [xscale=1,yscale=1,thick,black,densely dashed] 
      plot file{figs/b23a09upper18.dat};
    \draw [xscale=1,yscale=1,thick,purple,densely dashdotdotted] 
      plot file{figs/b23a09upper17.dat};
    \draw [xscale=1,yscale=1,thick,cyan,dashed] 
      plot file{figs/b23a09lower15.dat};
    \draw [xscale=1,yscale=1,thick,brown,densely dotted] 
      plot file{figs/b23a09lower4P.dat};
    \draw [xscale=1,yscale=1,thick,magenta,densely dotted]
      plot file{figs/b23a09lower3P.dat};
    \draw [xscale=1,yscale=1,thick,green,densely dotted] 
      plot file{figs/b23a09lower2P.dat};

    \filldraw [fill=white] (-14,6.2) rectangle (4,13.4);
    \draw[very thick,red] (-13.5,12.6) -- (-11,12.6)
      node[right,black] {\footnotesize Upper Bound (Th.~\ref{thm:them9})};
    \draw[very thick,blue,dashdotted] (-13.5,11.8) -- (-11,11.8)
      node[right,black] {\footnotesize Upper Bound (Th.~\ref{thm:them5})};
    \draw[very thick,black,densely dashed] (-13.5,11.0) -- (-11,11.0)
      node[right,black] {\footnotesize Upper Bound (Th.~\ref{thm:them7})};
    \draw[very thick,purple,densely dashdotdotted] (-13.5,10.2) -- (-11,10.2)
      node[right,black] {\footnotesize Upper Bound (Th.~\ref{thm:them6})};
    \draw[very thick,cyan,dashed] (-13.5,9.4) -- (-11,9.4)
      node[right,black] {\footnotesize Lower Bound (Th.~\ref{thm:lowerbound2})};
    \draw[very thick,brown,densely dotted] (-13.5,8.6) -- (-11,8.6)
      node[right,black] {\footnotesize Four-Point Lower Bound};
    \draw[very thick,magenta,densely dotted] (-13.5,7.8) -- (-11,7.8)
      node[right,black] {\footnotesize Three-Point Lower Bound};
    \draw[very thick,green,densely dotted] (-13.5,7.0) -- (-11,7.0)
      node[right,black] {\footnotesize Two-Point Lower Bound};
  \end{tikzpicture}
  \caption{Bounds on capacity of $2 \times 3$ MIMO channel with
    channel matrix $\mat{H} = [1, 1.5, 3; 2, 2, 1]$, and
    average-to-peak power ratio $\alpha = 0.9$. Note that the
    threshold of the channel is $\alpha_{\textnormal{th}} = 1.4762.$}
  \label{fig:bounds23_09}
\end{figure}
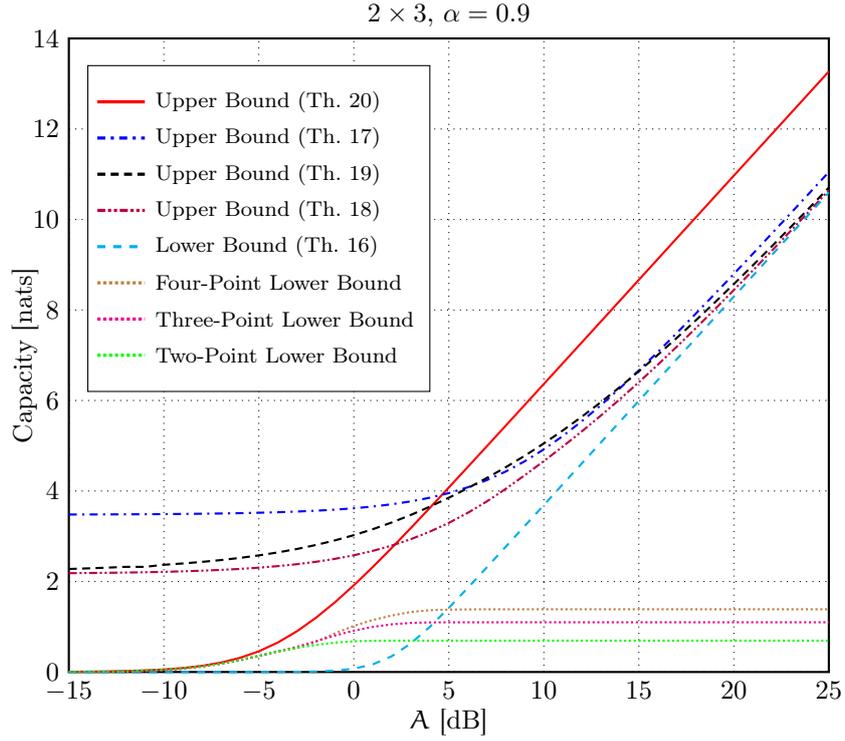

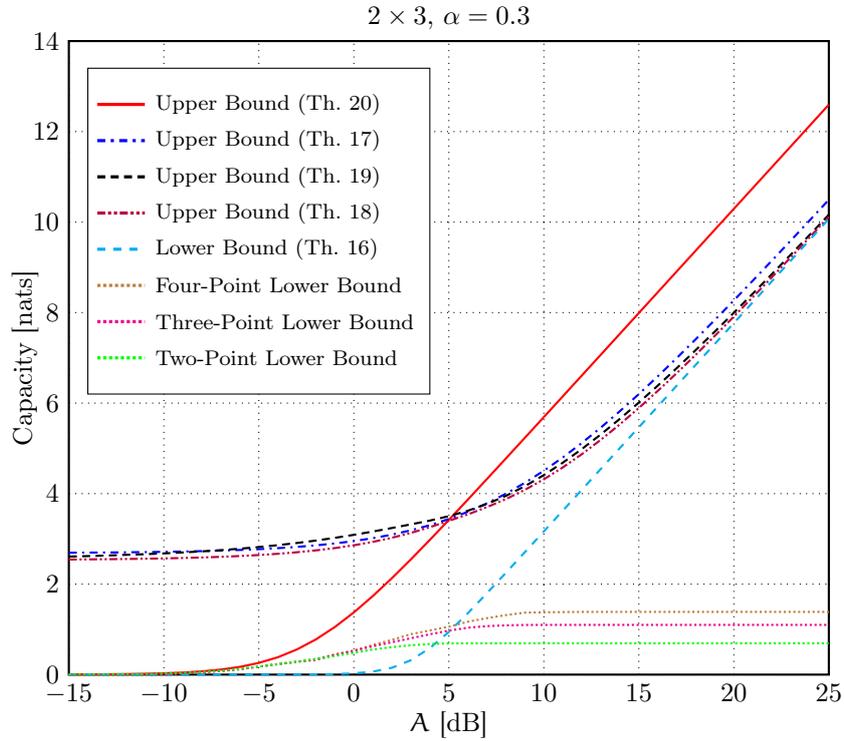
\begin{figure}[htbp]
  \centering
  \begin{tikzpicture}[%
    xscale=0.25,yscale=0.6,
    >={Stealth[scale=1.2]}]

    \draw[thick,-] (-15,0) rectangle (25,14); 
    \node at (5,-1.1) {$\amp$ [dB]};
    \node at (5,14.5) {$2\times 3$, $\alpha=0.3$};
    \node [rotate=90] at (-17.3,7) {Capacity [nats]};
    \draw [dotted] (-15,0) grid [xstep=5,ystep=2] (25,14);
    \foreach \i in {-15,-10,...,25}{
      \node [below] at (\i,0) {$\i$};
    };
    \foreach \i in {0,2,...,14}{
      \node [left] at (-15,\i) {$\i$};
    };

    \draw [xscale=1,yscale=1,thick,red] 
      plot file{figs/b23a03upper19.dat};
    \draw [xscale=1,yscale=1,thick,blue,dashdotted] 
      plot file{figs/b23a03upper16.dat};
    \draw [xscale=1,yscale=1,thick,black,densely dashed] 
      plot file{figs/b23a03upper18.dat};
    \draw [xscale=1,yscale=1,thick,purple,densely dashdotdotted] 
      plot file{figs/b23a03upper17.dat};
    \draw [xscale=1,yscale=1,thick,cyan,dashed] 
      plot file{figs/b23a03lower15.dat};
    \draw [xscale=1,yscale=1,thick,brown,densely dotted] 
      plot file{figs/b23a03lower4P.dat};
    \draw [xscale=1,yscale=1,thick,magenta,densely dotted]
      plot file{figs/b23a03lower3P.dat};
    \draw [xscale=1,yscale=1,thick,green,densely dotted] 
      plot file{figs/b23a03lower2P.dat};

    \filldraw [fill=white] (-14,6.2) rectangle (4,13.4);
    \draw[very thick,red] (-13.5,12.6) -- (-11,12.6)
      node[right,black] {\footnotesize Upper Bound (Th.~\ref{thm:them9})};
    \draw[very thick,blue,dashdotted] (-13.5,11.8) -- (-11,11.8)
      node[right,black] {\footnotesize Upper Bound (Th.~\ref{thm:them5})};
    \draw[very thick,black,densely dashed] (-13.5,11.0) -- (-11,11.0)
      node[right,black] {\footnotesize Upper Bound (Th.~\ref{thm:them7})};
    \draw[very thick,purple,densely dashdotdotted] (-13.5,10.2) -- (-11,10.2)
      node[right,black] {\footnotesize Upper Bound (Th.~\ref{thm:them6})};
    \draw[very thick,cyan,dashed] (-13.5,9.4) -- (-11,9.4)
      node[right,black] {\footnotesize Lower Bound (Th.~\ref{thm:lowerbound2})};
    \draw[very thick,brown,densely dotted] (-13.5,8.6) -- (-11,8.6)
      node[right,black] {\footnotesize Four-Point Lower Bound};
    \draw[very thick,magenta,densely dotted] (-13.5,7.8) -- (-11,7.8)
      node[right,black] {\footnotesize Three-Point Lower Bound};
    \draw[very thick,green,densely dotted] (-13.5,7.0) -- (-11,7.0)
      node[right,black] {\footnotesize Two-Point Lower Bound};
  \end{tikzpicture}
  \caption{Bounds on capacity of the same $2 \times 3$ MIMO channel as
    discussed in Figure~\ref{fig:bounds23_09}, and average-to-peak
    power ratio $\alpha = 0.3$.}
  \label{fig:bounds23_03}
\end{figure}

\begin{example}
  \label{examp:1}
  Figures~\ref{fig:bounds23_09} and~\ref{fig:bounds23_03} depict the
  derived lower and upper bounds for a $2 \times 3$ MIMO channel (same
  channel as in Example~\ref{ex:20}) for $\alpha = 0.9$ and
  $\alpha = 0.3$ (both values are less than
  $\alpha_{\textnormal{th}}=1.4762$), respectively. Both upper bounds
  \eqref{eq:upp2} and \eqref{eq:upp3} match with lower bound
  \eqref{eq:low_bnd2} asymptotically as $\amp$ tends to
  infinity. Moreover, upper bound \eqref{eq:upp_bnd} gives a good
  approximation on capacity when the average-power constraint is weak
  (i.e., when $\alpha$ is close to
  $\alpha_{\textnormal{th}}$). Indeed, \eqref{eq:upp_bnd} is
  asymptotically tight at high SNR when
  $\alpha \geq \alpha_{\textnormal{th}}$. We also plot three numerical
  lower bounds obtained by optimizing $\II(\Xbar;\vect{Y})$ over all
  feasible choices of $\Xbar$ that have positive probability on two,
  three, or four distinct mass points. (One of the mass points is
  always at $\vect{0}$.) In the low-SNR regime, upper bound
  \eqref{eq:upp4} matches well with the two-point numerical lower
  bound. Actually \eqref{eq:upp4} shares the same slope with capacity
  when the SNR tends to zero, which can be seen by comparing
  \eqref{eq:upp4} with Theorem~\ref{thm:them11}.
\end{example}

\begin{figure}[htbp]
  \centering
  \begin{tikzpicture}[%
    xscale=0.25,yscale=0.6,
    >={Stealth[scale=1.2]}]

    \draw[thick,-] (-15,0) rectangle (25,14); 
    \node at (5,-1.1) {$\amp$ [dB]};
    \node at (5,14.5) {$2\times 4$, $\alpha=1.2$};
    \node [rotate=90] at (-17.3,7) {Capacity [nats]};
    \draw [dotted] (-15,0) grid [xstep=5,ystep=2] (25,14);
    \foreach \i in {-15,-10,...,25}{
      \node [below] at (\i,0) {$\i$};
    };
    \foreach \i in {0,2,...,14}{
      \node [left] at (-15,\i) {$\i$};
    };

    \draw [xscale=1,yscale=1,thick,red] 
      plot file{figs/b24a12upper19.dat};
    \draw [xscale=1,yscale=1,thick,blue,dashdotted] 
      plot file{figs/b24a12upper16.dat};
    \draw [xscale=1,yscale=1,thick,black,densely dashed] 
      plot file{figs/b24a12upper18.dat};
    \draw [xscale=1,yscale=1,thick,purple,densely dashdotdotted] 
      plot file{figs/b24a12upper17.dat};
    \draw [xscale=1,yscale=1,thick,cyan,dashed] 
      plot file{figs/b24a12lower15.dat};
    \draw [xscale=1,yscale=1,thick,brown,densely dotted] 
      plot file{figs/b24a12lower4P.dat};
    \draw [xscale=1,yscale=1,thick,magenta,densely dotted]
      plot file{figs/b24a12lower3P.dat};
    \draw [xscale=1,yscale=1,thick,green,densely dotted] 
      plot file{figs/b24a12lower2P.dat};

    \filldraw [fill=white] (-14,6.2) rectangle (4,13.4);
    \draw[very thick,red] (-13.5,12.6) -- (-11,12.6)
      node[right,black] {\footnotesize Upper Bound (Th.~\ref{thm:them9})};
    \draw[very thick,blue,dashdotted] (-13.5,11.8) -- (-11,11.8)
      node[right,black] {\footnotesize Upper Bound (Th.~\ref{thm:them5})};
    \draw[very thick,black,densely dashed] (-13.5,11.0) -- (-11,11.0)
      node[right,black] {\footnotesize Upper Bound (Th.~\ref{thm:them7})};
    \draw[very thick,purple,densely dashdotdotted] (-13.5,10.2) -- (-11,10.2)
      node[right,black] {\footnotesize Upper Bound (Th.~\ref{thm:them6})};
    \draw[very thick,cyan,dashed] (-13.5,9.4) -- (-11,9.4)
      node[right,black] {\footnotesize Lower Bound (Th.~\ref{thm:lowerbound2})};
    \draw[very thick,brown,densely dotted] (-13.5,8.6) -- (-11,8.6)
      node[right,black] {\footnotesize Four-Point Lower Bound};
    \draw[very thick,magenta,densely dotted] (-13.5,7.8) -- (-11,7.8)
      node[right,black] {\footnotesize Three-Point Lower Bound};
    \draw[very thick,green,densely dotted] (-13.5,7.0) -- (-11,7.0)
      node[right,black] {\footnotesize Two-Point Lower Bound};
  \end{tikzpicture}
  \caption{Bounds on capacity of $2 \times 4$ MIMO channel with
    channel matrix $\mat{H} = [1.5, 1, 0.75, 0.5; 0.5, 0.75, 1, 1.5]$,
    and average-to-peak power ratio $\alpha = 1.2$. Note that the
    threshold of the channel is $\alpha_{\textnormal{th}} = 1.947.$}
  \label{fig:bounds24_12}
\end{figure}
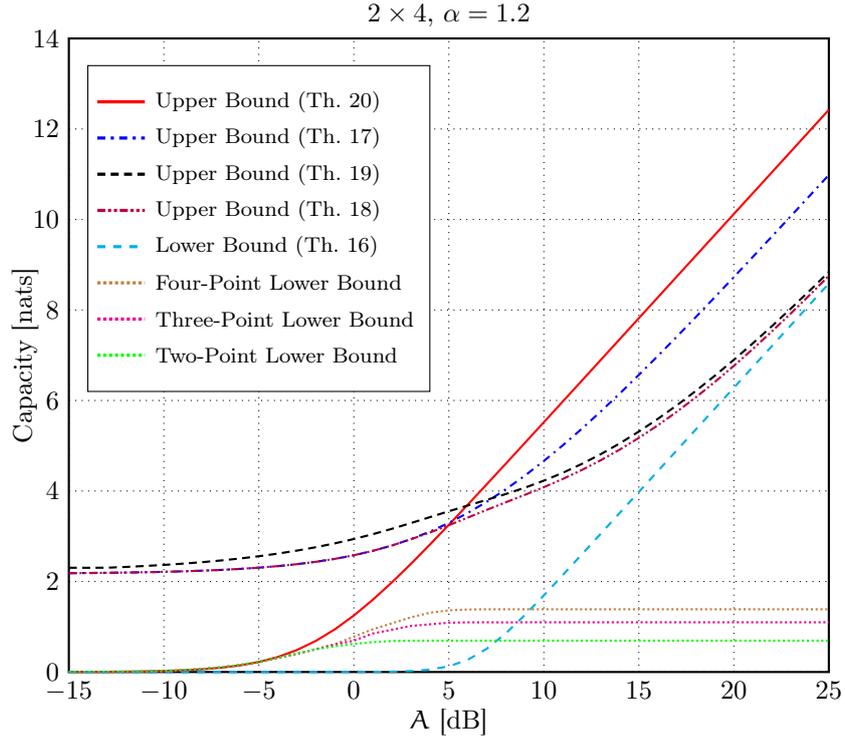

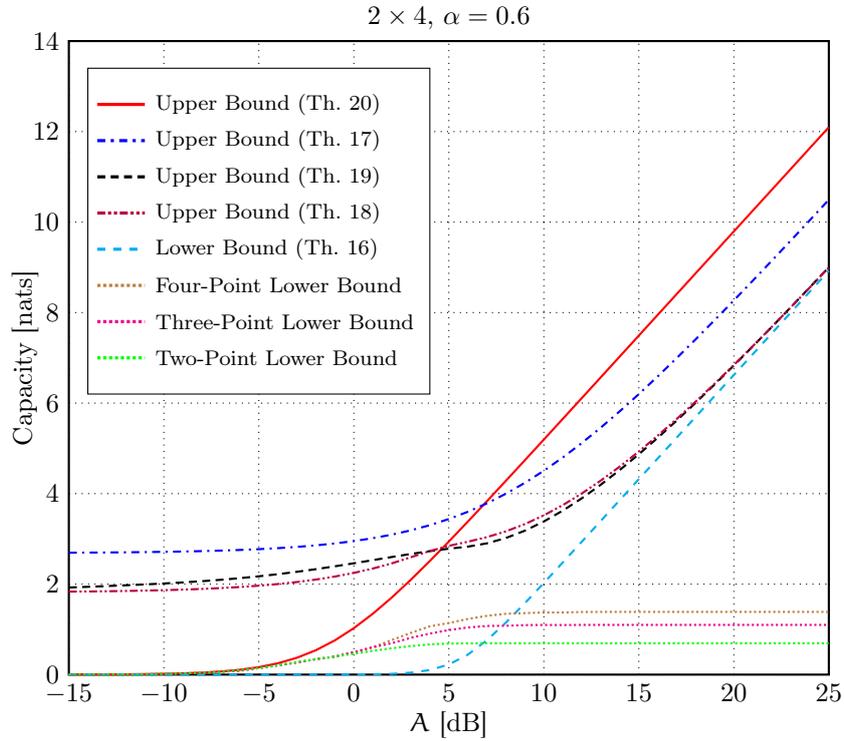
\begin{figure}[htbp]
  \centering
  \begin{tikzpicture}[%
    xscale=0.25,yscale=0.6,
    >={Stealth[scale=1.2]}]

    \draw[thick,-] (-15,0) rectangle (25,14); 
    \node at (5,-1.1) {$\amp$ [dB]};
    \node at (5,14.5) {$2\times 4$, $\alpha=0.6$};
    \node [rotate=90] at (-17.3,7) {Capacity [nats]};
    \draw [dotted] (-15,0) grid [xstep=5,ystep=2] (25,14);
    \foreach \i in {-15,-10,...,25}{
      \node [below] at (\i,0) {$\i$};
    };
    \foreach \i in {0,2,...,14}{
      \node [left] at (-15,\i) {$\i$};
    };

    \draw [xscale=1,yscale=1,thick,red] 
      plot file{figs/b24a06upper19.dat};
    \draw [xscale=1,yscale=1,thick,blue,dashdotted] 
      plot file{figs/b24a06upper16.dat};
    \draw [xscale=1,yscale=1,thick,black,densely dashed] 
      plot file{figs/b24a06upper18.dat};
    \draw [xscale=1,yscale=1,thick,purple,densely dashdotdotted] 
      plot file{figs/b24a06upper17.dat};
    \draw [xscale=1,yscale=1,thick,cyan,dashed] 
      plot file{figs/b24a06lower15.dat};
    \draw [xscale=1,yscale=1,thick,brown,densely dotted] 
      plot file{figs/b24a06lower4P.dat};
    \draw [xscale=1,yscale=1,thick,magenta,densely dotted]
      plot file{figs/b24a06lower3P.dat};
    \draw [xscale=1,yscale=1,thick,green,densely dotted] 
      plot file{figs/b24a06lower2P.dat};

    \filldraw [fill=white] (-14,6.2) rectangle (4,13.4);
    \draw[very thick,red] (-13.5,12.6) -- (-11,12.6)
      node[right,black] {\footnotesize Upper Bound (Th.~\ref{thm:them9})};
    \draw[very thick,blue,dashdotted] (-13.5,11.8) -- (-11,11.8)
      node[right,black] {\footnotesize Upper Bound (Th.~\ref{thm:them5})};
    \draw[very thick,black,densely dashed] (-13.5,11.0) -- (-11,11.0)
      node[right,black] {\footnotesize Upper Bound (Th.~\ref{thm:them7})};
    \draw[very thick,purple,densely dashdotdotted] (-13.5,10.2) -- (-11,10.2)
      node[right,black] {\footnotesize Upper Bound (Th.~\ref{thm:them6})};
    \draw[very thick,cyan,dashed] (-13.5,9.4) -- (-11,9.4)
      node[right,black] {\footnotesize Lower Bound (Th.~\ref{thm:lowerbound2})};
    \draw[very thick,brown,densely dotted] (-13.5,8.6) -- (-11,8.6)
      node[right,black] {\footnotesize Four-Point Lower Bound};
    \draw[very thick,magenta,densely dotted] (-13.5,7.8) -- (-11,7.8)
      node[right,black] {\footnotesize Three-Point Lower Bound};
    \draw[very thick,green,densely dotted] (-13.5,7.0) -- (-11,7.0)
      node[right,black] {\footnotesize Two-Point Lower Bound};
  \end{tikzpicture}
  \caption{Bounds on capacity of the same $2 \times 4$ MIMO channel as
    discussed in Figure~\ref{fig:bounds24_12}, and average-to-peak
    power ratio $\alpha = 0.6$.}
  \label{fig:bounds24_06}
\end{figure}

\begin{example}
  Figures~\ref{fig:bounds24_12} and~\ref{fig:bounds24_06} show similar
  trends in a $2 \times 4$ MIMO channel.  Note that although in the
  $2\times 3$ channel of Figures~\ref{fig:bounds23_09} and
  \ref{fig:bounds23_03} the upper bound \eqref{eq:upp2} is always
  tighter than \eqref{eq:upp3}, this does not hold in general, as can
  be seen in Figure~\ref{fig:bounds24_06}.
\end{example}

\section{Concluding Remarks}
\label{sec:conclusion}

In this paper, we first express capacity as a maximization problem
over distributions for the vector $\Xbar = \mat{H} \vect{X}$. The main
challenge there is to transform the total average-power constraint on
$\vect{X}$ to a constraint on $\Xbar$, as the mapping from $\vect{x}$
to $\xbar$ is many-to-one. This problem is solved by identifying, for
each $\xbar$, the input vector $\vect{x}_{\min}$ that induces this
$\xbar$ with minimum energy. Specifically, we show that the set
$\reg(\mat{H})$ of all possible $\xbar$ can be decomposed into a
number of parallelepipeds such that, for all $\xbar$ within one
parallelepiped, the minimum-energy input vectors $\vect{x}_{\min}$
have a similar form.

At high SNR, the above minimum-energy signaling result allows the
transmitter to decompose the channel into several ``almost parallel''
channels, each of which being an $\nr \times \nr$ MIMO channel in
itself. This is because, at high SNR, the output $\vect{Y}$ allows the
receiver to obtain a good estimate of which of the parallelepipeds
$\Xbar$ lies in. We can then apply previous results on the capacity of
the MIMO channel with full-column rank. The remaining steps in
deriving our results on the high-SNR asymptotic capacity can be
understood, on a high level, as optimizing probabilities and energy
constraints assigned to each of the parallel channels.

In the low-SNR regime, the capacity slope is shown to be proportional
to the trace of the covariance matrix of $\Xbar$ under the given power
constraints. We prove several properties of the input distribution
that maximizes this trace. For example, each entry in $\vect{X}$
should be either zero or the maximum value $\amp$, and the total
number of values of $\vect{X}$ with nonzero probabilities need not
exceed $\nr+2$.

We recall that, although for IM-DD optical channels, the channel
matrix $\mat{H}$ typically only has nonnegative entries, our results
are valid for all real-valued $\mat{H}$.

\section*{Acknowledgment} 

The authors thank Saïd~Ladjal for pointing them to the notion of
zonotopes and related literature. They are also grateful for the
helpful comments of the Associate Editor and the reviewers.

\appendices

\section{Reduction of the Cases $\nt>\nr>\rank{\mat{H}}$ and $\rank{\mat{H}} \le \nt \le \nr$}
\label{app:proof_chanequiv}

We first show that a channel with $\nt>\nr>\rank{\mat{H}}$ can be
reduced to one where the rank of the channel matrix equals the number
of receive antennas, to which our results apply. Denote
\begin{IEEEeqnarray}{c}
  r \eqdef \rank{\mat{H}}.
\end{IEEEeqnarray}
We apply the SVD
\begin{IEEEeqnarray}{c}
  \mat{H} = \mat{U}\mat{\Sigma} \trans{\mat{V}},
  \label{eq:41}
\end{IEEEeqnarray}
where $\mat{U} \in \Reals^{\nr \times \nr}$ and
$\mat{V} \in \Reals^{\nt \times \nt}$ are unitary matrices, and
$\mat{\Sigma} \in \Reals^{\nr \times \nt}$ is a rectangular matrix
whose first $r$ diagonal entries are positive real values, and whose
all other entries are zero.

The receiver can compute the new output
\begin{IEEEeqnarray}{c}
  \tilde{\vect{Y}} \eqdef \trans{\mat{U}} \vect{Y} =
  \mat{\Sigma} \trans{\mat{V}}\vect{X} +\trans{\mat{U}} \vect{Z}, 
\end{IEEEeqnarray}
where the new noise vector
$\tilde{\vect{Z}}\eqdef \trans{\mat{U}} \vect{Z}$ again has
independent zero-mean Gaussian components because $\trans{\mat{U}}$ is
unitary. Moreover, the new channel matrix
$\tilde{\mat{H}}\eqdef \mat{\Sigma} \trans{\mat{V}}$ is of the form
\begin{IEEEeqnarray}{c}
  \tilde{\mat{H}} =
  \begin{pmatrix}
    \hat{\mat{H}} \\
    \mat{0}
  \end{pmatrix}
  \label{eq:46}    
\end{IEEEeqnarray}
with ${\hat{\mat{H}}}$ being an $r \times \nt$ matrix of rank $r$.
Combined with the independence of the new noise components, this
implies that the outputs
$\tilde{Y}_{r+1}, \tilde{Y}_{r+2},\ldots,\tilde{Y}_{\nr}$ are
independent of the first $r$ outputs
$\tilde{Y}_1, \ldots, \tilde{Y}_r$ and the input vector $\vect{X}$,
and hence can be discarded.  The receiver is thus left with outputs
$\tilde{Y}_1, \ldots, \tilde{Y}_r$, in which case the number of
receive antennas and the rank of the new channel matrix
$\hat{\mat{H}}$ both equal $r$. Finally, since we did not change the
transmitter side, the inequality $\nt >r$ remains to hold.

The situation where $\rank{\mat{H}} \le \nt \le \nr$ is handled in the
same way, with the only difference that $\tilde{\mat{H}}$ in
\eqref{eq:46} now is tall instead of wide. If $r=\nt$, $\hat{\mat{H}}$
is a full-rank square $\nt\times\nt$ matrix, resulting in a channel
model that is not considered in this work, but that has been studied
extensively in the literature (e.g.,
\cite{mosermylonakiswangwigger17_1, chaabanrezkialouini18_1}). If
$r < \nt$, we have again obtained a model, to which our results apply.

Be aware that even if $\mat{H}$ only has positive entries,
$\hat{\mat{H}}$ in \eqref{eq:46} may contain negative values.

Note that it is even possible to allow a channel model with dependent
noise: Let $\mat{K}$ be some positive definite matrix and assume that
$\vect{Z}\sim\Normal{\vect{0}}{\mat{K}}$. Since $\mat{K}$ is positive
definite, it can be written as $\mat{K}=\trans{\mat{S}}\mat{S}$ for
some invertible $\nr\times\nr$ matrix $\mat{S}$. Thus,
\begin{IEEEeqnarray}{rCl}
  \II(\vect{X}; \mat{H}\vect{X} + \vect{Z})
  & = & \II(\vect{X}; \invtrans{\mat{S}}\mat{H}\vect{X} +
  \invtrans{\mat{S}}\vect{Z}) 
  \\ 
  & = & \II(\vect{X}; \mat{H}'\vect{X} + \vect{Z}'),
\end{IEEEeqnarray}
where we define $\mat{H}'\eqdef \invtrans{\mat{S}}\mat{H}$ and
$\vect{Z}'\eqdef \invtrans{\mat{S}}\vect{Z}$. Note that
$\rank{\mat{H}'} = \rank{\mat{H}}$ and that
$\vect{Z}' \sim \Normal{\vect{0}}{\mat{I}_{\nr}}$. Therefore, we may
apply the same approach as in \eqref{eq:41}--\eqref{eq:46} to reduce
these cases to an equivalent model that is either square full-rank or
matches the assumptions~\eqref{eq:40} considered in this work.

\section{Proof of Proposition~\ref{prop:proposition1}}
\label{app:prof_proposition1}


Fix a capacity-achieving input $\vect{X}^{\star}$ and let
\begin{IEEEeqnarray}{c}
  \alpha^* \eqdef \frac{\bigE{\|{\vect{X}^{\star}}\|_1}}{\amp}. 
\end{IEEEeqnarray} 
Define $\vect{a} \eqdef \trans{(\amp,\amp,\ldots,\amp)}$ and
\begin{IEEEeqnarray}{c}
  \vect{X}^{\prime} \eqdef \vect{a} - \vect{X}^{\star}.
\end{IEEEeqnarray}
We have 
\begin{IEEEeqnarray}{c}
  \bigE{\|\vect{X}'\|_1} =  \amp (\nt - \alpha^*),
\end{IEEEeqnarray} 
and
\begin{IEEEeqnarray}{rCl}
  \II(\vect{X}^{\star};\vect{Y})
  & = & \II(\vect{X}^{\star};\mat{H}\vect{a}-\vect{Y})
  \\
  & = & \II(\vect{X}^{\star};
  \mat{H}\vect{a}-\mat{H}\vect{X}^{\star}-\vect{Z})  
  \\
  & = & \II\bigl(\vect{a}-\vect{X}^{\star};
  \mat{H}(\vect{a}-\vect{X}^{\star})-\vect{Z}\bigr)  
  \\
  & = & \II\bigl(\vect{a}-\vect{X}^{\star};
  \mat{H}(\vect{a}-\vect{X}^{\star})+\vect{Z}\bigr) 
  \label{eq:minusplus}
  \\
  & = & \II(\vect{X}^{\prime};\mat{H}\vect{X}^{\prime}+\vect{Z})
  \\
  & = & \II(\vect{X}^{\prime};\vect{Y}^{\prime})
  \label{eq:equalentropy}
\end{IEEEeqnarray}
where $\vect{Y}^{\prime} \eqdef \mat{H}\vect{X}^{\prime}+\vect{Z}$,
and where \eqref{eq:minusplus} follows because $\vect{Z}$ is symmetric
around $\vect{0}$ and independent of $\vect{X}^{\star}$.

Define another random vector $\tilde{\vect{X}}$ as follows:
\begin{IEEEeqnarray}{c}
  \label{eq:Xtilde}
  \tilde{\vect{X}} \eqdef
  \begin{cases}
    \vect{X}^{\star} & \textnormal{with probability } p,\\
    \vect{X}^{\prime} & \textnormal{with probability } 1-p.
  \end{cases}
\end{IEEEeqnarray}
Notice that, since $\II(\vect{X};\vect{Y})$ is concave in $P_\vect{X}$
for a fixed channel law, we have
\begin{IEEEeqnarray}{c}
  \II(\tilde{\vect{X}};\tilde{\vect{Y}})
  \ge p\II(\vect{X}^{\star};\vect{Y}) +
  (1-p)\II(\vect{X}^{\prime};\vect{Y}^{\prime}).  
\end{IEEEeqnarray} 
Therefore, by \eqref{eq:equalentropy},
\begin{IEEEeqnarray}{c}
  \label{eq:larger_entropy}
  \II(\tilde{\vect{X}};\tilde{\vect{Y}}) \geq
  \II(\vect{X}^{\star};\vect{Y}) 
\end{IEEEeqnarray}
for all $p\in[0,1]$. Combined with the assumption that
$\vect{X}^{\star}$ achieves capacity, \eqref{eq:larger_entropy}
implies that $\tilde{\vect{X}}$ must also achieve capacity.

We are now ready to prove the two claims in the proposition. We first
prove that for $\alpha >\frac{\nt}{2}$ the average-power constraint is
inactive. To this end, we choose $p=\frac{1}{2}$, which yields
\begin{IEEEeqnarray}{c}
  \bigE{\|\tilde{\vect{X}}\|_1} = \frac{\nt}{2}\amp.
\end{IEEEeqnarray} 
Since $\tilde{\vect{X}}$ achieves capacity (see above), we conclude
that capacity is unchanged if one strengthens the average-power
constraint from $\alpha\amp$ to $\frac{\nt}{2}\amp$.

We now prove that, if $\alpha \leq \frac{\nt}{2}$, then there exists a
capacity-achieving input distribution for which the average-power
constraint is met with equality. Assume that $\alpha^* < \alpha$
(otherwise $\vect{X}^{\star}$ is itself such an input), then choose
\begin{IEEEeqnarray}{c}
  p = \frac{\nt-\alpha^*-\alpha}{\nt-2\alpha^*}.
\end{IEEEeqnarray}
With this choice,
\begin{IEEEeqnarray}{rCl}
  \bigE{\|\tilde{\vect{X}}\|_1}
  & = &   p\bigE{\|\vect{X}^{\star}\|_1}
  + (1-p)\bigE{\|\vect{X}^{\prime}\|_1}
  \\ 
  & = & \bigl( p \alpha^* + (1-p) (\nt  - \alpha^*)\bigr)\amp
  \\
  & = & \alpha \amp.
\end{IEEEeqnarray} 
Hence $\tilde{\vect{X}}$ (which achieves capacity) meets the
average-power constraint with equality.

\section{Proof of Lemma~\ref{lem:lemma1}}
\label{app:prof_lemma1}

First consider the case where Condition \eqref{eq:cond1} is
satisfied. Define for each $\mU\in\U$ the set
\begin{IEEEeqnarray}{c}
  \set{B}_{\mU} \eqdef \bigl\{\vect{x}= (x_1,\ldots, x_{\nt})
  \colon x_{i}\in(0,\amp),  \, \forall\, i\in\mU, \textnormal{ and }
  x_{j} = \amp \cdot g_{\mU,j}, \, \forall\, j \in \mcU \bigr\}, 
  \IEEEeqnarraynumspace
\end{IEEEeqnarray}
where we notice that the interval $(0,\amp)$ is open.  We first
observe that the optimization problem
\begin{IEEEeqnarray}{c}
  \min_{\vect{x}' \in \set{S}(\xbar)} \|\vect{x}'\|_1
  \label{eq:39}
\end{IEEEeqnarray}
has a solution for every $\xbar\in\reg(\mat{H})$. This is because the
minimization is over a compact set and the objective function is
convex and continuous. Furthermore, we have the following lemma.

\begin{lemma}
  \label{lem:KKT}
  Under Condition~\eqref{eq:cond1}, for all $\xbar\in\reg(\mat{H})$
  except a subset of Lebesgue measure zero, the solutions to
  \eqref{eq:39} lie in the union $\bigcup_{\mU \in \U} \set{B}_{\mU}$.
\end{lemma}
\begin{IEEEproof}
  Assume $\vect{x}^{\star}=(x_1^{\star},\ldots, x_{\nt}^{\star})$ is a
  solution to \eqref{eq:39}. It must satisfy the Karush--Kuhn--Tucker
  (KKT) conditions:
  \begin{IEEEeqnarray}{rCl}
    \subnumberinglabel{eq:KTT}
    \vect{1}_{\nt} -\vectg{\mu}^{\star}+ \vectg{\nu}^{\star} +
    \trans{\mat{H}} \vectg{\lambda}^{\star}
    & = & \vect{0}_{\nr},
    \label{eq:K1}
    \\ 
    \mat{H}\vect{x}^{\star} - \xbar
    & = & \vect{0}_{\nr},
    \label{eq:K4}
    \\ 
    - \vect{x}^{\star}  & \leq &\vect{0}_{\nt},
    \label{eq:K5}
    \\ 
    \vect{x}^{\star}  - \amp \cdot \vect{1}_{\nt}
    & \leq & \vect{0}_{\nt},
    \label{eq:K6}
    \\
    \mu_i^{\star} {x}_i^{\star}
    & = & 0,  \quad i\in\{1,\ldots, \nt\},
    \label{eq:K2}
    \\
    \nu^{\star}_i({x}_i^{\star}-\amp )
    & = & 0, \quad i\in\{1,\ldots, \nt\},
    \label{eq:K3}
  \end{IEEEeqnarray} 
  for some Lagrange multipliers
  $\vectg{\mu}^{\star} = (\mu_1^{\star},\ldots, \mu_{\nt}^{\star})$,
  $\vectg{\nu}^{\star}=(\nu_1^{\star},\ldots, \nu_{\nt}^{\star})$, and
  $\vectg{\lambda}^{\star}= (\lambda_1^{\star},\ldots,
  \lambda_{\nr}^{\star})$ satisfying
  \begin{IEEEeqnarray}{rCl}
    \vectg{\mu}^{\star} & \geq & \vect{0}_{\nt},
    \label{eq:K7}
    \\ 
    \vectg{\nu}^{\star} & \geq & \vect{0}_{\nt}.
    \label{eq:K8}
  \end{IEEEeqnarray}
  All vector inequalities above are componentwise.  Let $\mV$ be the
  set of indices corresponding to the components of $\vect{x}^{\star}$
  in $(0,\amp)$:
  \begin{IEEEeqnarray}{c}
    \mV \eqdef \bigl\{ i\in\{1,\ldots,\nt\}\colon x_i^{\star}
    \in(0,\amp)\bigr\}, 
  \end{IEEEeqnarray} 
  and define $\mat{H}_{\mV} \eqdef [\vect{h}_i\colon i\in\mV]$. Notice
  that the set of all image vectors that correspond to
  $\rank{\mat{H}_\mV}<\nr$ has Lebesgue measure zero. Thus, for the
  purpose of this proof, we can ignore those image vectors and assume
  \begin{IEEEeqnarray}{c}
    \label{eq:RankHU}
    \rank{\mat{H}_{\mV}} = \nr.
  \end{IEEEeqnarray}
  Clearly, there must exist some $\mU\in\U$ such that
  $\mU\subseteq \mV$. We next show that $\mV=\mU$ and therefore
  $\mV\in \U$. We show this by contradiction. Assume there exists an
  index $i\in\mV\setminus \mU$. By \eqref{eq:K2} and \eqref{eq:K3},
  \begin{IEEEeqnarray}{c}
    \mu_j^{\star}= \nu_j^{\star} = 0, \quad \forall j \in\mV, 
  \end{IEEEeqnarray}
  and thus by \eqref{eq:K1},
  \begin{IEEEeqnarray}{c}
    \label{eq:implication_KKT}
    \trans{[\mat{H}_{\mU},\vect{h}_i]} \vectg{\lambda}^{\star}
    = - \vect{1}_{\nr+1}. 
  \end{IEEEeqnarray}
  Since $\mU\in\U$, we know that $\mat{H}_{\mU}$ is invertible, hence
  \begin{IEEEeqnarray}{c}
    \label{eq:lambdastar}
    \vectg{\lambda}^{\star} = - \invtrans{\mat{H}_\mU}
    \vect{1}_{\nr}. 
  \end{IEEEeqnarray}
  Plugging this back into \eqref{eq:implication_KKT} yields
  \begin{IEEEeqnarray}{c}
    \trans{\vect{h}_i} \invtrans{\mat{H}_\mU} \vect{1}_{\nr} =
    1,
  \end{IEEEeqnarray}
  which contradicts Condition~\eqref{eq:cond1}. We hence conclude that
  $\mV=\mU\in\U$.

  It remains to show that $x_{j}^{\star} = \amp\cdot g_{\mU,j}$ for
  all $j\in \mcU$. Fix $j \in \mcU$. It follows from \eqref{eq:K2} and
  \eqref{eq:K3} that either $\mu_j=0$ and $x_j^{\star} = \amp$, or
  $\nu_j=0$ and $x_j^{\star} = 0$. To determine between these two
  cases, consider the $j$-th line of \eqref{eq:K1}:
  \begin{IEEEeqnarray}{c}
    \label{eq:124}
    1 - \mu_j^{\star} + \nu_j^{\star} + \vect{h}_j
    \vectg{\lambda}^{\star} = 0. 
  \end{IEEEeqnarray}
  Note that $\vectg{\lambda}^{\star}$ is given by
  \eqref{eq:lambdastar}, so
  \begin{IEEEeqnarray}{c}
    \trans{\vect{h}_j} \vectg{\lambda}^{\star}
    = - \trans{\vect{h}_j} \invtrans{\mat{H}_\mU} \vect{1}_{\nr} =
    -a_{\mU,j}, 
  \end{IEEEeqnarray}
  hence \eqref{eq:124} becomes
  \begin{IEEEeqnarray}{c}
    - \mu_j^{\star} + \nu_j^{\star}  = a_{\mU,j} - 1.
  \end{IEEEeqnarray}
  Since both $\mu_j^{\star}$ and $\nu_j^{\star}$ are nonnegative and
  only one of them can be positive, we conclude that, when
  $a_{\mU,j} > 1$, we must have $\mu_j^{\star}=0$ and
  $x_j^{\star} = \amp$, and when $a_{\mU,j} < 1$, we must have
  $\nu_j^{\star}=0$ and $x_j^{\star}=0$.
\end{IEEEproof} 
\medskip 

We now proceed to prove Lemma~\ref{lem:lemma1} for the case where
Condition~\eqref{eq:cond1} is satisfied.  Notice that the $\Lone$-norm
is continuous, and the correspondence from $\xbar$ to $\set{S}(\xbar)$
is compact valued and both lower and upper hemicontinuous. Berge's
Maximum Theorem \cite{berge97_1} then implies that the correspondence
from $\xbar$ to the solutions to \eqref{eq:39} is nonempty, compact
valued, and upper hemicontinuous. As a consequence, the solutions to
\eqref{eq:39} for all $\xbar\in\reg({\mat{H}})$ are contained in the
closure $\cl\left(\bigcup_{\mU \in \U} \set{B}_{\mU}\right)$. Because
$\U$ is finite, we further have
$\cl\left(\bigcup_{\mU \in \U} \set{B}_{\mU} \right)= \bigcup_{\mU \in
  \U} \cl(\set{B}_{\mU})$. Thus
\begin{IEEEeqnarray}{c}
  \label{eq:image2}
  \reg(\mat{H}) = \left\{ \mat{H}\vect{x}\colon \vect{x} \in
    \bigcup_{\mU \in \U} \cl(\set{B}_{\mU})\right\}. 
\end{IEEEeqnarray}
On the other hand, for each $\mU\in\U$, we have
\begin{IEEEeqnarray}{c}
  \label{eq:image}
  \vect{v}_{\mU}+\set{D}_{\mU} = \bigl\{ \mat{H}\vect{x} \colon
  \vect{x} \in \cl(\set{B}_{\mU}) \bigr\}. 
\end{IEEEeqnarray}
Combining \eqref{eq:image2} and \eqref{eq:image} we obtain
\eqref{eq:union}.

Furthermore, because
\begin{IEEEeqnarray}{c}
  \vol(\vect{v}_\mU + \set{D}_{\mU})
  = \amp^{\nr} \abs{\det \mat{H}_{\mU}} ,\quad \forall\, \mU\in\U,
\end{IEEEeqnarray}
and \cite{shephard74_1,zamirfeder98_1}
\begin{IEEEeqnarray}{c}
  \vol\bigl(\reg(\mat{H})\bigr)
  = \amp^{\nr}\sum_{\mU \in \U} \abs{\det\mat{H}_{\mU}},
\end{IEEEeqnarray} 
we have
\begin{IEEEeqnarray}{c}
  \label{eq:same_volumes}
  \sum_{\mU \in \U} \vol(\vect{v}_\mU +
  \set{D}_{\mU}) = \vol\bigl(\reg(\mat{H})\bigr). 
\end{IEEEeqnarray} 
Combined with \eqref{eq:union}, this proves \eqref{eq:overlap} and
completes the first part of the lemma.

\medskip

We next prove the second part of the lemma. By \eqref{eq:overlap}, for
any $\mU,\mV\in\U$, $\mU\neq \mV$,
\begin{IEEEeqnarray}{c}
  \interior(\vect{v}_\mU + \set{D}_\mU) \cap
  \interior(\vect{v}_\mV + \set{D}_\mV)=\emptyset, 
\end{IEEEeqnarray}
where $\interior(\cdot)$ denotes the interior of a set.  Since
disjoint open sets are also separated, and since
$\cl\left(\interior(\vect{v}_\mV + \set{D}_\mV)\right) = \vect{v}_\mV
+ \set{D}_\mV$, we further obtain
\begin{IEEEeqnarray}{c}
  \interior(\vect{v}_\mU + \set{D}_\mU) \cap
  (\vect{v}_\mV + \set{D}_\mV)=\emptyset.  
\end{IEEEeqnarray}
Combined with Lemma~\ref{lem:KKT} and \eqref{eq:image}, this implies
that, for any $\xbar \in \interior(\vect{v}_\mU+\set{D}_\mU)$, a
solution to \eqref{eq:39} must lie in $\cl(\set{B}_\mU)$ and not in
$\cl(\set{B}_\mV)$ for any $\mV\neq \mU$. Then it is immediate that
the solution is unique and given by \eqref{eq:xvalue}. It remains only
to extend \eqref{eq:xvalue} to the boundaries of
$\{\vect{v}_{\mU}+\set{D}_{\mU}\}_{\mU\in \U}$. This is accomplished
by recalling that the correspondence from $\xbar$ to the solutions to
\eqref{eq:39} is upper hemicontinuous. (One can easily verify that,
for some
$\xbar \in (\vect{v}_\mU+\set{D}_\mU)\cap (\vect{v}_\mV+\set{D}_\mV)$,
computing \eqref{eq:xvalue} for $\mU$ and $\mV$ yields the same
result.) This concludes the proof of the second part of
Lemma~\ref{lem:lemma1}.

\medskip

Finally, we argue that Lemma~\ref{lem:lemma1} holds also when
\eqref{eq:cond1} is violated. Note that if $a_{\mU,j}=1$ for some
$\mU$ and $j$, then the solution to \eqref{eq:39} is not necessarily
unique anymore. To solve this problem, note that
Algorithm~\ref{def:algorithm} can be interpreted as generating a small
perturbation of the matrix $\mat{H}$. We fix some small values
$\eps_1 > \cdots > \eps_{\nt} > 0$ and check through all $a_{\mU,j}$,
$j\in\{1, \ldots, \nt\}$. When we encounter a first tie $a_{\mU,j}=1$,
we multiply the corresponding vector $\vect{h}_j$ by a factor
$(1+\eps_1)$ and thereby break the tie ($\epsilon_1$ is chosen to be
small enough so that it does not affect any other choices). If a
second tie shows up, we use the next perturbation factor $(1+\eps_2)$
(which is smaller than $(1+\eps_1)$, so we do not inadvertently revert
our first perturbation); and so on. Lemma~\ref{lem:lemma1} is then
proven by letting all of $\epsilon_1,\ldots,\epsilon_{\nt}$ go to zero
and invoking Berge's Maximum Theorem. We omit the details.

\section{Proof of Maximum-Variance Signaling Results}
\label{app:prof_lemma9}

\subsection{Proof of Lemma~\ref{lem:binaryinput}}
\label{sec:proof-lemma-refl}

The $i$th diagonal element of $\cov{\Xbar}$
can be decomposed as follows:
\begin{IEEEeqnarray}{rCl}
  \bigl(\cov{\Xbar}\bigr)_{i,i}
  & = & \E{\bigl(\Xs_i-\eE{\Xs_i}\bigr)^2}
  \\
  & = & \E{\left(\sum_{k=1}^{\nt} h_{i,k}\bigl(X_k-\E{X_k}\bigr)\right)^2}
  \\
  & = & \sum_{k=1}^{\nt} h_{i,k}^2 \E{\bigl(X_k-\E{X_k}\bigr)^2} 
  + \sum_{k=1}^{\nt} \sum_{\substack{\ell=1\\\ell\neq k}}^{\nt}
  h_{i,k} h_{i,\ell}  \bigl( \E{X_{k} X_{\ell}} -
  \E{X_{k}}\E{X_{\ell}}\bigr).  
  \IEEEeqnarraynumspace
\end{IEEEeqnarray}
Thus, the objective function in \eqref{eq:optimizationproblem} is
\begin{IEEEeqnarray}{c}
  \sum_{i=1}^{\nr}\sum_{k=1}^{\nt} h_{i,k}^2
  \E{\bigl(X_k-\E{X_k}\bigr)^2} + \sum_{i=1}^{\nr} \sum_{k=1}^{\nt}
  \sum_{\substack{\ell=1\\\ell\neq k}}^{\nt} h_{i,k} h_{i,\ell} \bigl(
  \E{X_{k} X_{\ell}} - \E{X_{k}}\E{X_{\ell}}\bigr).
  \IEEEeqnarraynumspace
  \label{eq:14}
\end{IEEEeqnarray}
If we fix a joint distribution on $(X_1,\ldots,X_{\nt-1})$ and choose
with probability 1 a conditional mean
$\Econd{X_{\nt}}{X_1,\ldots,X_{\nt-1}}$, then the consumed total
average input power is fixed and every summand on the RHS of
\eqref{eq:14} is determined except for
\begin{IEEEeqnarray}{c}
  \E{\bigl(X_{\nt} - \E{X_{\nt}}\bigr)^2}.
\end{IEEEeqnarray} 
This value is maximized --- for any choice of joint distribution on
$(X_1,\ldots,X_{\nt-1})$ and conditional mean
$\Econd{X_{\nt}}{X_1,\ldots,X_{\nt-1}}$ --- if $X_{\nt}$ takes value
only in the set $\{0,\amp\}$.  We conclude that, to maximize the
expression in \eqref{eq:optimizationproblem} subject to a constraint
on the average input power, it is optimal to restrict $X_{\nt}$ to
taking value only in $\{0,\amp\}$.

Repeating this argument for $X_{\nt-1}$, $X_{\nt-2}$, etc., we
conclude that every $X_k$, $k=1,\ldots,\nt$, should take value only in
$\{0,\amp\}$.

\subsection{Proof of Lemma~\ref{lem:path}}
\label{sec:proof-lemma-refl-1}

Some steps in our proof are inspired by
\cite{chaabanrezkialouini18_2}. We start by rewriting the objective
function in \eqref{eq:optimizationproblem} as:
\begin{IEEEeqnarray}{rCl}
  \bigtrace{\cov{\Xbar}}
  & = & \sum_{i=1}^{\nr} \E{\bigl(\Xs_i-\eE{\Xs_i}\bigr)^2}
  \\
  & = & \sum_{i=1}^{\nr} \E{\left(\sum_{k=1}^{\nt} h_{i,k}
      \bigl(X_k-\eE{X_k}\bigr)\right)^2} 
  \\
  & = & \sum_{i=1}^{\nr} \sum_{k=1}^{\nt} \sum_{k'=1}^{\nt} 
  h_{i,k} \, h_{i,k'} \E{\bigl(X_k-\eE{X_k}\bigr)
    \bigl(X_{k'}-\eE{X_{k'}}\bigr)}  
  \\
  & = & \sum_{k=1}^{\nt} \sum_{k'=1}^{\nt} \underbrace{\sum_{i=1}^{\nr} 
    h_{i,k} \, h_{i,k'}}_{\eqdef \kappa_{k,k'}}{} \cdot \Cov{X_k}{X_{k'}}
  \\
  & = & \sum_{k=1}^{\nt} \sum_{k'=1}^{\nt} \kappa_{k,k'}  \Cov{X_k}{X_{k'}}.
\end{IEEEeqnarray}
Thus, we need to maximize $\Cov{X_k}{X_{k'}}$. Assume that we have
fixed the average power $\EE_k$, $k=1, \ldots, \nt$, assigned to each
input antenna, and further assume that we reorder the antennas such
that
\begin{IEEEeqnarray}{c}
  \label{eq:ordering}
  \EE_1 \ge \cdots \ge \EE_{\nt}.
\end{IEEEeqnarray}
Note that since each antenna only uses a binary input
$X_k\in\{0,\amp\}$, the assignment $\E{X_k}=\EE_k$ determines the
probabilities:
\begin{IEEEeqnarray}{c}
  \Prv{X_k=\amp} = \frac{\EE_k}{\amp}
\end{IEEEeqnarray}
and the variances:
\begin{IEEEeqnarray}{c}
  \Cov{X_k}{X_k} = \Var{X_k} = \E{X_k^2} - \EE_k^2 = \EE_k\amp -
  \EE_k^2. 
\end{IEEEeqnarray}
For the covariances with $k < k'$ we obtain
\begin{IEEEeqnarray}{rCl}
  \Cov{X_k}{X_{k'}}
  & = & \E{X_k X_{k'}} - \EE_k\EE_{k'}
  \\
  & = & \amp^2 \Prv{X_k=X_{k'}=\amp} - \EE_k\EE_{k'}
  \\
  & = & \amp^2 \Prv{X_{k'}=\amp}
  \underbrace{\Prvcond{X_k=\amp}{X_{k'}=\amp}}_{\le 1}{} -
  \EE_k\EE_{k'}
  \\
  & \le & \amp \EE_{k'} - \EE_k\EE_{k'}
  \\
  & = & \bigl(\amp-\EE_k\bigr)\EE_{k'}.
\end{IEEEeqnarray}
The upper bound holds with equality if
\begin{IEEEeqnarray}{c}
  \label{eq:condprob}
  \Prvcond{X_k=\amp}{X_{k'}=\amp} = 1.
\end{IEEEeqnarray}
This choice is allowed, because for $k < k'$ the ordering
\eqref{eq:ordering} is compatible with Condition~\eqref{eq:condprob}.
This proves that the mass points can be ordered in such a way that
\eqref{eq:optimalinputc} holds.

We next prove by contradiction that the first mass point must be
$\vect{0}$. By Lemma~\ref{lem:binaryinput}, if
$\vect{x}_1^*\neq \vect{0}$, then $\vect{x}_1^*$ must contain at least
one entry that equals $\amp$. By \eqref{eq:optimalinputc}, that entry
must be $\amp$ for all mass points used by the optimal input. Clearly,
changing its value from $\amp$ to $0$ for all mass points will not
affect the trace of \eqref{eq:trace}, but will reduce the total input
power. Hence we conclude that an input with
$\vect{x}_1^*\neq \vect{0}$ (or with zero probability on $\vect{0}$)
must be suboptimal.

\subsection{Proof of Lemma~\ref{lem:optimalinputtracecov}}
\label{sec:proof-lemma-refl-3}

We investigate the KKT conditions of the optimization problem
\eqref{eq:optimizationproblem}. Using the definition of $\T$ and
$r_{\mV,i}$ we rewrite the objective function of
\eqref{eq:optimizationproblem} as
\begin{IEEEeqnarray}{rCl}
  \bigtrace{\cov{\Xbar}}
  & = & \sum_{i=1}^{\nr} \Bigl(\E{\Xs_i^2} - \bigl(\eE{\Xs_i}\bigr)^2
  \Bigr)
  \\
  & = & \amp^2 \sum_{i=1}^{\nr}  \left(
    \sum_{\mV\in\T} p_{\mV} \> r_{\mV,i}^2 - \left(\sum_{\mV\in\T}
      p_{\mV} \> r_{\mV,i}\right)^2 \right).
\end{IEEEeqnarray}
Taking into account the constraints \eqref{eq:constraints},
the Lagrangian is obtained as:
\begin{IEEEeqnarray}{rCl}
  \const{L}(\vect{p},\mu_0,\mu_1, \vectg{\mu})
  & = & \amp^2 \sum_{i=1}^{\nr}  \left(
    \sum_{\mV\in\T} p_{\mV} \> r_{\mV,i}^2 - \left(\sum_{\mV\in\T}
      p_{\mV} \> r_{\mV,i}\right)^2 \right)
  - \mu_0 \left( \sum_{\mV\in\T} p_{\mV}-1 \right)
  \nonumber\\
  && -\> \mu_1 \left( \sum_{\mV\in\T} p_{\mV}\abs{\mV} -\alpha
  \right)
  - \sum_{\mV\in\T} \mu_{\mV} (0 - p_{\mV}).
\end{IEEEeqnarray}
The KKT conditions for the optimal 
$\{p_{\mK}^*\}_{\mK\in\U}$ are as follows:
\begin{IEEEeqnarray}{rCll}
  \subnumberinglabel{eq:KKT}
  \amp^2\sum_{i=1}^{\nr}  \left(
    r_{\mK,i}^2 - 2 r_{\mK,i} \sum_{\mV\in\T}
    p_{\mV}^* \> r_{\mV,i} \right)
  - \mu_0 - \mu_1 \abs{\mK} + \mu_{\mK} 
  & = & 0,  \quad &  \mK \in\T,
  \label{eq:KKT1}
  \IEEEeqnarraynumspace
  \\ 
  \mu_0 \left( \sum_{\mV\in\T} p_{\mV}^*-1 \right) & = & 0,
  \label{eq:KKT2}
  \\
  \mu_1 \left( \sum_{\mV\in\T} p_{\mV}^*\abs{\mV} -\alpha
  \right) & = & 0,
  \label{eq:KKT3}
  \\
  \mu_{\mK} p_{\mK}^* & = & 0, &  \mK \in\T,
  \label{eq:KKT4}
  \\
  \mu_0 & \ge & 0,
  \label{eq:KKT5}
  \\
  \mu_1 & \ge & 0,
  \label{eq:KKT6}
  \\
  \mu_{\mK} & \geq & 0, &  \mK \in \T,
  \label{eq:KKT7}
  \\
  \sum_{\mV\in\T} p_{\mV}^* & \le & 1,
  \label{eq:KKT8}
  \\
  \sum_{\mV\in\T} p_{\mV}^* \abs{\mV} & \le & \alpha,
  \label{eq:KKT9}
  \\
  p_{\mK}^* & \ge & 0, & \mK\in\T.
  \label{eq:KKT10}
\end{IEEEeqnarray}
We define the vector $\vect{m}=\trans{(m_1,\ldots, m_{\nr})}$
with components
\begin{IEEEeqnarray}{c}
  m_i \eqdef \sum_{\mV \in \T} p_{\mV}^* r_{\mV,i},
  \quad i=1, \ldots, \nr,
  \label{eq:mdef}
\end{IEEEeqnarray}
and rewrite \eqref{eq:KKT1} as 
\begin{IEEEeqnarray}{c}
  \label{eq:KKT1equal}
  \amp^2\|\vect{r}_{\mK}\|_2^2 - 2\amp^2 \trans{\vect{r}_{\mK}} \vect{m}
  - \mu_0 - \mu_1 \abs{\mK} +  \mu_{\mK} = 0, \quad   \mK \in\T. 
\end{IEEEeqnarray}
Since by Lemma~\ref{lem:path} $P_{\vect{X}}^*(\vect{0}) > 0$, it must
hold that \eqref{eq:KKT8} holds with strict inequality and it thus
follows from \eqref{eq:KKT2} that $\mu_0=0$.

Next, assume by contradiction that there exist $\nr+2$ choices
$\mK_1, \ldots, \mK_{\nr+2} \in\T$ with positive probability
$p_{\mK_{\ell}}^*>0$.  Then, by \eqref{eq:KKT4}, $\mu_{\mK_{\ell}}=0$
for all $\ell \in \{1,\ldots,\nr + 2\}$. From \eqref{eq:KKT1equal} we
thus have
\begin{IEEEeqnarray}{c}
  2 \trans{\vect{r}_{\mK_{\ell}}} \vect{m} + \tilde{\mu}_1
  \abs{\mK_{\ell}} = \|\vect{r}_{\mK_{\ell}}\|_2^2, \quad \ell \in
  \{1, \ldots, \nr+2\},
\end{IEEEeqnarray}
with $\tilde{\mu}_1 \eqdef \mu_1/\amp^2$, which can be written in
matrix form:
\begin{IEEEeqnarray}{rCl}
  \begin{pmatrix}
    2r_{\mK_1,1} & \cdots &2r_{\mK_1,\nr} & \abs{\mK_1}
    \\[1ex]
    2r_{\mK_2,1} & \cdots &2r_{\mK_2,\nr} & \abs{\mK_2}
    \\[1ex]
    \vdots & \ddots & \vdots & \vdots
    \\
    2r_{\mK_{\nr+2},1} & \cdots & 2r_{\mK_{\nr+2},\nr} & \abs{\mK_{\nr+2}}  
  \end{pmatrix}
  \begin{pmatrix}
    m_1 \\ m_2 \\ \vdots \\ m_{\nr} \\ \tilde{\mu}_1
  \end{pmatrix}
  =
  \begin{pmatrix}
    \|\vect{r}_{\mK_1}\|_2^2
    \\
    \|\vect{r}_{\mK_2}\|_2^2
    \\
    \vdots
    \\
    \|\vect{r}_{\mK_{\nr+2}}\|_2^2
  \end{pmatrix}. 
  \IEEEeqnarraynumspace
\end{IEEEeqnarray}
This is an over-determined system of linear equations in $\nr+1$
variables $m_1, \ldots, m_{\nr}, \tilde{\mu}_1$, which has a solution
if, and only if,
\begin{IEEEeqnarray}{rCl}
  \IEEEeqnarraymulticol{3}{l}{%
    \rank{ 
      \begin{matrix}
        2r_{\mK_1,1} & \cdots &2r_{\mK_1,\nr} & \abs{\mK_1}
        \\
        2r_{\mK_2,1} & \cdots &2r_{\mK_2,\nr} & \abs{\mK_2}
        \\
        \vdots & \ddots & \vdots & \vdots
        \\
        2r_{\mK_{\nr+2},1} & \cdots & 2r_{\mK_{\nr+2},\nr} & \abs{\mK_{\nr+2}}  
      \end{matrix}
    } } \nonumber\\*\quad%
  & = & \rank{
    \begin{matrix}
      2r_{\mK_1,1} & \cdots &2r_{\mK_1,\nr} & \abs{\mK_1} & \|\vect{r}_{\mK_1}\|_2^2
      \\
      2r_{\mK_2,1} & \cdots &2r_{\mK_2,\nr} & \abs{\mK_2} & \|\vect{r}_{\mK_2}\|_2^2
      \\
      \vdots & \ddots & \vdots & \vdots & \vdots
      \\
      2r_{\mK_{\nr+2},1} & \cdots & 2r_{\mK_{\nr+2},\nr} &
      \abs{\mK_{\nr+2}} & \|\vect{r}_{\mK_{\nr+2}}\|_2^2 
    \end{matrix}
  }.
  \label{eq:ranks}
\end{IEEEeqnarray}
However, since the matrix on the LHS has only $\nr+1$ columns, its
rank can be at most $\nr+1$. The matrix on the RHS, on the other hand,
has by assumption (see \eqref{eq:21}) rank $\nr+2$. This is a
contradiction. We have proven that there exist at most $\nr+1$ values
$p_{\mK}$ with positive values. Together with $\vect{0}$, there are at
most $\nr+2$ mass points in total.

\section{Derivation of the Lower Bounds}
\label{sec:deriv-lower-bounds}

For any choice of the random vector $\Xbar$ over $\reg(\mat{H})$,
the following holds:
\begin{IEEEeqnarray}{rCl}
  \C_{\mat{H}}(\amp, \alpha \amp)
  & \geq & \II( \Xbar ; \Xbar + \vect{Z})
  \\
  & = & \hh(\Xbar+\vect{Z}) - \hh(\vect{Z})
  \\
  & \geq & \frac{1}{2}\log\left(\ope^{2\hh(\Xbar)} +
    \ope^{2\hh(\vect{Z})}\right) -\hh(\vect{Z})
  \label{eq:1}
  \\
  & = & \frac{1}{2}\log\left(1+ \frac{\ope^{2\hh(\Xbar)}}{(2\pi e )^{\nr}}
  \right),
  \label{eq:inequal6}
\end{IEEEeqnarray}
where \eqref{eq:1} follows from the EPI \cite{coverthomas06_1}.

\subsection{Proof of Theorem~\ref{thm:them4}}

We choose $\Xbar$ to be uniformly distributed over $\reg(\mat{H})$. To
verify that this uniform distribution satisfies the average-power
constraint \eqref{eq:new_constraint}, we define
\begin{IEEEeqnarray}{c}
  p_{\mU} \eqdef \Prv{\rvU=\mU}
\end{IEEEeqnarray}
and derive
\begin{IEEEeqnarray}{rCl}
  \IEEEeqnarraymulticol{3}{l}{%
    \E[\rvU]{ \amp s_{\rvU} +  \left\|\inv{\mat{H}_{\rvU}} \bigl(
        \Econd{\Xbar}{\rvU} -\vect{v}_{\rvU} \bigr)\right\|_1}
  }\nonumber\\*\quad%
  & = & \amp\sum_{\mU\in\U} p_{\mU} \> s_{\mU}
  + \sum_{\mU\in\U} p_{\mU} \left\|\inv{\mat{H}_{\mU}} \bigl( 
    \eEcond{\Xbar}{\rvU=\mU} - \vect{v}_{\mU} \bigr)\right\|_1
  \\     
  & = & \amp \sum_{\mU\in\U} q_{\mU} \> s_{\mU}
  + \sum_{\mU\in\U} q_{\mU} \cdot \frac{\nr\amp}{2}  
  \label{eq:uniform}
  \\
  & = & \alpha_{\textnormal{th}} \amp
  \label{eq:9}
  \\
  & \leq & \alpha \amp.
  \label{eq:11}
\end{IEEEeqnarray}
Here, \eqref{eq:uniform} follows because when $\Xbar$ is uniformly
distributed in $\reg(\mat{H})$, we have
\begin{IEEEeqnarray}{c}
  \inv{\mat{H}_{\mU}} \bigl( \eEcond{\Xbar}{\rvU=\mU} - \vect{v}_{\mU}
  \bigr) = \frac{\amp}{2} \cdot \vect{1}_{\nr}
\end{IEEEeqnarray}
and
\begin{IEEEeqnarray}{c}
  p_{\mU} = q_{\mU}, \quad \mU\in\U.
\end{IEEEeqnarray}
Further, \eqref{eq:9} holds because of \eqref{eq:10}, and the last
inequality \eqref{eq:11} holds by the assumption in the theorem.

The uniform distribution of $\Xbar$ results in
\begin{IEEEeqnarray}{c}
  \hh(\Xbar) = \log (\amp^{\nr} \cdot \V),
\end{IEEEeqnarray}
which, by \eqref{eq:inequal6}, leads to \eqref{eq:lowbnd1}.

\subsection{Proof of Theorem~\ref{thm:lowerbound2}}

We choose
\begin{IEEEeqnarray}{c}
  \lambda \in \left(  \max \left\{0,\frac{\nr}{2} +\alpha -
      \alpha_{\textnormal{th}}\right\},
    \min\left\{\frac{\nr}{2},\alpha \right\}\right),
  \label{eq:12}
\end{IEEEeqnarray}
a probability vector $\vect{p}$ satisfying \eqref{eq:pconstraint}, and
$\mu$ as the unique solution to \eqref{eq:lambda}.

Note that such choices are always possible as can be argued as
follows. From \eqref{eq:12} one directly sees that
$0 < \lambda < \frac{\nr}{2}$. Thus,
$0 < \frac{\lambda}{\nr} < \frac{1}{2}$, which corresponds exactly to
the range where \eqref{eq:lambda} has a unique solution.  From
\eqref{eq:12} it also follows that
$\frac{\nr}{2} +\alpha - \alpha_{\textnormal{th}} < \lambda < \alpha$
and thus
\begin{IEEEeqnarray}{c}
  0 < \alpha - \lambda < \alpha_{\textnormal{th}} - \frac{\nr}{2} \le
  \frac{\nt}{2} - \frac{\nr}{2}, 
\end{IEEEeqnarray}
where the last inequality follows from \eqref{eq:15}. So the RHS
of \eqref{eq:pconstraint} takes value within the interval
$\left(0,\frac{\nt-\nr}{2}\right)$. By Remark~\ref{rem:s_values}, the
LHS of \eqref{eq:pconstraint} can take value in the interval
$[0,\nt-\nr]$, which covers the range of the RHS. The existence of
$\vect{p}$ satisfying \eqref{eq:pconstraint} now follows from the
continuity of the LHS of \eqref{eq:pconstraint} in $\vect{p}$.

For each $\mU$ we now pick the probability density function
$f_{\Xbar|\rvU=\mU}$ to be the $\nr$-dimensional product
truncated exponential distribution rotated by the matrix
$\mat{H}_{\mU}$:
\begin{IEEEeqnarray}{c}
  f_{\Xbar|\rvU=\mU}(\xbar)
  =  \frac{1}{\amp^{\nr}
    \abs{\det\mat{H}_{\mU}}} \cdot
  \left(\frac{\mu}{1-\ope^{-\mu}}\right)^{\nr} 
  \ope^{-\frac{\mu}{\amp}\left\|\inv{\mat{H}_{\mU}}(\xbar-
      \vect{v}_{\mU})\right\|_1}. 
  \IEEEeqnarraynumspace
  \label{eq:expdist}
\end{IEEEeqnarray}
Note that this corresponds to the entropy-maximizing distribution
under a total average-power constraint. The average-power constraint
\eqref{eq:new_constraint} is satisfied because
\begin{IEEEeqnarray}{rCl}
  \IEEEeqnarraymulticol{3}{l}{%
    \E[\rvU]{\amp s_{\rvU} + \left\|\inv{\mat{H}_{\rvU}} \bigl(
        \Econd{\Xbar}{\rvU} -\vect{v}_{\rvU} \bigr)\right\|_1}
  }\nonumber\\*\quad%
  & = & \sum_{\mU\in\U} p_{\mU} \Bigl( \amp s_{\mU}
  + \bigl\|\inv{\mat{H}_{\mU}}\bigl(\eEcond{\Xbar}{\rvU=\mU} -
  \vect{v}_{\mU}\bigr)\bigr\|_1 \Bigr) 
  \\     
  & = & \sum_{\mU\in\U} p_{\mU} \left( \amp s_{\mU}
    + \nr \amp
    \left(\frac{1}{\mu}-\frac{\ope^{-\mu}}{1-\ope^{-\mu}}\right)
  \right) 
  \label{eq:16}
  \\
  & = & \sum_{\mU\in\U} p_{\mU} \bigl(\amp s_{\mU} + \amp\lambda
  \bigr)
  \label{eq:17}
  \\
  & = & \amp  \sum_{\mU\in\U} p_{\mU} s_{\mU} + \amp\lambda
  \label{eq:18}
  \\
  & = & \amp(\alpha-\lambda) +  \amp\lambda
  \label{eq:13}
  \\
  & = & \alpha\amp.
  \label{eq:19}
\end{IEEEeqnarray}
Here, \eqref{eq:16} follows from the expected value of the truncated
exponential distribution; \eqref{eq:17} is due to \eqref{eq:lambda};
and \eqref{eq:13} follows from \eqref{eq:pconstraint}.

Furthermore, 
\begin{IEEEeqnarray}{rCl}
  \hh(\Xbar)
  & = & \II(\Xbar ; \rvU) + \hh(\Xbar|\rvU)
  \\
  & = & \HH(\rvU) + \hh(\Xbar|\rvU)
  \label{eq:23}
  \\
  & = & \HH(\vect{p}) + \sum_{\mU\in\U} p_{\mU} \hh(\Xbar|\rvU=\mU)
  \\
  & = & \HH(\vect{p})
  + \sum_{\mU\in\U} p_{\mU} \log \abs{\det\mat{H}_{\mU}}
  + \nr \log{\amp}
  - \nr\log{\frac{\mu}{1-\ope^{-\mu}}}
  \nonumber\\
  && +\> \nr\left(1 -\frac{\mu\ope^{-\mu}}{1-\ope^{-\mu}}\right)  
  \label{eq:expentropy}
  \\
  & = & - \sum_{\mU\in\U} p_{\mU} \log p_{\mU}
  + \sum_{\mU\in\U} p_{\mU} \log\frac{\abs{\det\mat{H}_{\mU}}}{\V}
  + \log\V + \nr \log{\amp}
  \nonumber\\
  && +\>\nr\left(1- \log{\frac{\mu}{1-\ope^{-\mu}}}
    - \frac{\mu \ope^{-\mu}}{1-\ope^{-\mu}}\right)
  \\
  & = &  - \const{D}(\vect{p}\|\vect{q}) + \log\V
  + \nr \log{\amp}
  + \nr\left(1- \log{\frac{\mu}{1-\ope^{-\mu}}}
    - \frac{\mu \ope^{-\mu}}{1-\ope^{-\mu}}\right).
  \label{eq:hxu}
  \IEEEeqnarraynumspace
\end{IEEEeqnarray}
Here, \eqref{eq:23} holds because $\HH(\rvU|\Xbar)=0$;
\eqref{eq:expentropy} follows from the differential entropy of a
truncated exponential distribution; and in \eqref{eq:hxu} we use
the definition of $\vect{q}$ in \eqref{eq:24}.  Then,
\eqref{eq:low_bnd2} follows by plugging \eqref{eq:hxu} into
\eqref{eq:inequal6}.

\section{Derivation of Upper Bounds}
\label{sec:deriv-upper-bounds}

Let $\Xbar^{\star}$ be a maximizer in \eqref{eq:capacity_alternative}
and let $\rvUstar$ be defined by $\Xbar^{\star}$ as in
\eqref{eq:ueqi}. Then,
\begin{IEEEeqnarray}{rCl}
  \C_{\mat{H}}(\amp, \alpha\amp) 
  & = & \II\bigl(\Xbar^{\star}; \Xbar^{\star} +\vect{Z}\bigr)
  \\
  & \leq & \II\bigl(\Xbar^{\star}; \Xbar^{\star} +\vect{Z},\rvUstar\bigr)
  \\
  & \leq & \HH(\rvUstar) + \II\bigl(\Xbar^{\star}; \Xbar^{\star}
  +\vect{Z} \big| \rvUstar\bigr).
  \label{eq:capa_up}
\end{IEEEeqnarray}
For each 
set $\mU\in \U$, we have
\begin{IEEEeqnarray}{rCl}
  \II\bigl(\Xbar^{\star}; \Xbar^{\star} +\vect{Z} \big| \rvUstar=
  \mU\bigr) 
  & = &  \II\bigl(\Xbar^{\star}-\vect{v}_{\mU};
  (\Xbar^{\star}-\vect{v}_{\mU} ) +\vect{Z} \big| \rvUstar= \mU\bigr) 
  \\
  & = & \II\bigl(\inv{\mat{H}}_{\mU}(\Xbar^{\star}-\vect{v}_{\mU});
  \inv{\mat{H}}_{\mU}(\Xbar^{\star}-\vect{v}_{\mU} ) +
  \inv{\mat{H}}_{\mU}\vect{Z} \big| \rvUstar= \mU\bigr)  
  \\
  & = & \II(\vect{X}_{\mU} ; \vect{X}_{\mU} + \vect{Z}_{\mU}|\rvUstar= \mU)
  \label{eq:mutual_inf}
\end{IEEEeqnarray}
where we have defined 
\begin{IEEEeqnarray}{rCl}
  \vect{Z}_{\mU}
  & \eqdef & \inv{\mat{H}_{\mU}}\vect{Z},\label{eq:vectZI}
  \\
  \vect{X}_{\mU}
  & \eqdef & \inv{\mat{H}_{\mU}}(\Xbar^{\star}-\vect{v}_{\mU}).
  \label{eq:XI} 
\end{IEEEeqnarray}
It should be noted that 
\begin{IEEEeqnarray}{c}
  \vect{Z}_{\mU} \sim
  \Normal{0}{\inv{\mat{H}_{\mU}}\invtrans{\mat{H}_{\mU}}}.
\end{IEEEeqnarray}
Moreover, $\vect{X}_{\mU}$ is subject to the following peak- and
average-power constraints:
\begin{IEEEeqnarray}{rCl}
  \subnumberinglabel{eq:sisoconstraints}
  \bigPrv{X_{\mU,\ell} > \amp} & = & 0,
  \quad \forall\,\ell\in\{1, \ldots, \nr\},
  \label{eq:sisopeak}
  \\
  \bigE{\|\vect{X}_{\mU}\|_1} & = & \EE_{\mU},
  \label{eq:sisoaverage}
\end{IEEEeqnarray}
where $\{\EE_{\mU}\colon \mU\in\U\}$ satisfies
\begin{IEEEeqnarray}{c}
  \sum_{\mU \in \U} p_{\mU}(s_{\mU}\amp+\EE_{\mU}) \leq \alpha\amp.
\end{IEEEeqnarray}

To further bound the RHS of \eqref{eq:mutual_inf}, we use the
duality-based upper-bounding technique using a product output
distribution
\begin{IEEEeqnarray}{c}
  R_{\mU}(\vect{y}_{\mU}) = \prod_{\ell=1}^{\nr}
  R_{\mU,\ell}(y_{\mU,\ell} ). 
\end{IEEEeqnarray} 
Denoting by $W_{\mU}(\cdot | \vect{X}_{\mU})$ the transition law of
the $\nr\times \nr$ MIMO channel with input $\vect{X}_{\mU}$ and
output $\vect{Y}_{\mU}\eqdef \vect{X}_{\mU}+\vect{Z}_{\mU}$, and by
$W_{\mU,\ell}(\cdot | X_{\mU,\ell})$ the marginal transition law of
its $\ell$th component, we have:
\begin{IEEEeqnarray}{rCl}
  \IEEEeqnarraymulticol{3}{l}{%
    \II(\vect{X}_{\mU} ;  \vect{X}_{\mU} + \vect{Z}_{\mU} | \rvUstar= \mU) 
  }\nonumber\\*\quad%
  & \leq &  \E[\vect{X}_{\mU}|\rvUstar=\mU]{
    \const{D}\bigl(W_{\mU}(\cdot|\vect{X}_{\mU})\big\| R_{\mU}(\cdot)\bigr)}
  \\ 
  & = &  -\hh\bigl( \vect{X}_{\mU} + \vect{Z}_{\mU} \big|
  \vect{X}_{\mU}, \rvUstar=\mU\bigr)  
  - \E[\vect{X}_{\mU}|\rvUstar=\mU]{\sum_{\ell=1}^{\nr}   
    \bigE[W_{\mU}(\vect{Y}_{\mU} | \vect{X}_{\mU})]{\log
      R_{\mU,\ell}(Y_{\mU,\ell})}}
  \IEEEeqnarraynumspace
  \\ 
  & = & - \frac{\nr}{2} \log 2 \pi e + \log \abs{\det \mat{H}_{\mU}}
  - \sum_{\ell=1}^{\nr}
  \E[X_{\mU,\ell}|\rvUstar=\mU]{\bigE[W_{\mU,\ell}(Y_{\mU,\ell} |  
    X_{\mU,\ell})]{\log R_{\mU,\ell}(Y_{\mU,\ell})}},
  \nonumber\\*
  \label{eq:26}
\end{IEEEeqnarray}
where the last equality holds because
\begin{IEEEeqnarray}{c}
  \hh(\vect{X}_{\mU} +\vect{Z}_{\mU} | \vect{X}_{\mU}, \rvUstar=\mU)
  = \hh(\vect{Z}_{\mU} )
  = \frac{1}{2}\log\bigl((2\pi e)^{\nr}
  \det\inv{\mat{H}}_{\mU}\invtrans{\mat{H}}_{\mU}\bigr).  
  \IEEEeqnarraynumspace 
  \label{eq:gaussentropy}
\end{IEEEeqnarray}
We finally combine \eqref{eq:capa_up} with \eqref{eq:mutual_inf} and
\eqref{eq:26} to obtain
\begin{IEEEeqnarray}{rCl}
  \C_{\mat{H}}(\amp, \alpha\amp) 
  & \leq & \HH(\vect{p}^*) 
  - \sum_{\ell=1}^{\nr} \sum_{\mU\in\U} p^*_{\mU} 
  \E[X_{\mU,\ell}|\rvUstar=\mU]{\bigE[W_{\mU,\ell}(Y_{\mU,\ell} | 
    X_{\mU,\ell})]{\log R_{\mU,\ell}(Y_{\mU,\ell})}}
  \nonumber\\
  &&  + \sum_{\mU\in\U} p^*_{\mU}
  \log\abs{\det\mat{H}_{\mU}} - \frac{\nr}{2} \log 2 \pi e,
  \label{eq:dual} 
\end{IEEEeqnarray}
where $\vect{p}^*$ denotes the probability vector of $\rvUstar$.  The
bounds in Section~\ref{sec:upper-bounds} are then found by picking
appropriate choices for the distribution on the output alphabet
$R_{\mU,\ell}(\cdot)$. We elaborate on this in the following.

\subsection{Proof of Theorem~\ref{thm:them5}}

Inspired by \cite{mckellips04_1} and
\cite{thangarajkramerbocherer17_1}, we choose
\begin{IEEEeqnarray}{c}
  R_{\mU,\ell}(y) =
  \begin{cases} 
    \frac{\beta}{\sqrt{2\pi}\sigma_{\mU,\ell}}
    \cdot \ope^{-\frac{y^2}{2\sigma_{\mU,\ell}^{2}}}
    &  \textnormal{if } y \in (-\infty, 0), 
    \\
    (1-\beta)\cdot \frac{1}{\amp}
    & \textnormal{if } {y} \in [0,\amp],
    \\
    \frac{\beta}{\sqrt{2\pi}\sigma_{\mU,\ell}}
    \cdot \ope^{-\frac{(y-\amp)^2}{2\sigma_{\mU,\ell}^{2}}}
    & \textnormal{if } {y} \in (\amp, \infty), 
  \end{cases}
\end{IEEEeqnarray}
where $\beta \in(0,1)$ will be specified later. Recall that
$\sigma_{\mU,\ell}$ is the square root of the $\ell$th diagonal entry
of the matrix $\inv{\mat{H}}_{\mU}\invtrans{\mat{H}}_{\mU}$, i.e.,
\begin{IEEEeqnarray}{c}
  \label{eq:sigmaI}
  \sigma_{\mU,\ell} = \sqrt{\Var{Z_{\mU,\ell}}}.
\end{IEEEeqnarray}
We notice that 
\begin{IEEEeqnarray}{rCl}
  \IEEEeqnarraymulticol{3}{l}{%
    -\int_{-\infty}^{0} W_{\mU,\ell}(y|x)\log R_{\mU,\ell}(y) \dd y 
  }\nonumber\\*\quad%
  & = &  -\int_{-\infty}^{0}
  \frac{1}{\sqrt{2\pi}\sigma_{\mU,\ell}}
  \ope^{-\frac{(y-x)^2}{2\sigma_{\mU,\ell}^{2}}} 
  \left(\log{\frac{\beta}{\sqrt{2\pi}\sigma_{\mU,\ell}}} -
    \frac{y^2}{2\sigma_{\mU,\ell}^{2}}\right) \dd y  
  \\ 
  & = & -\log\left(\frac{\beta}{\sqrt{2\pi}\sigma_{\mU,\ell}}\right)
  \Qf{\frac{x}{\sigma_{\mU,\ell}}}
  + \frac{1}{2}\Qf{\frac{x}{\sigma_{\mU,\ell}}}
  \nonumber\\
  && +\> \frac{1}{2}\left(\frac{x}{\sigma_{\mU,\ell}}\right)^2
  \Qf{\frac{x}{\sigma_{\mU,\ell}}}
  - \frac{x}{2\sigma_{\mU,\ell}}
  \phi\left(\frac{x}{\sigma_{\mU,\ell}}\right) 
  \IEEEeqnarraynumspace
  \\
  & \leq &  -\left(\log
    \frac{\beta}{\sqrt{2\pi}{\sigma_{\mU,\ell}}}-
    \frac{1}{2}\right)\Qf{{\frac{x}{\sigma_{\mU,\ell}}}}
  \label{eq:25}
  \\
  & = &  - \log \frac{\beta}{\sqrt{2\pi e}{\sigma_{\mU,\ell}}}
  \cdot \Qf{{\frac{x}{\sigma_{\mU,\ell}}}},
  \label{eq:e207}
\end{IEEEeqnarray}
where 
\begin{IEEEeqnarray}{c}
  \phi(x) \eqdef \frac{1}{\sqrt{2\pi}} \ope^{-\frac{x^2}{2}},
\end{IEEEeqnarray}
and where \eqref{eq:25} holds because of \cite[Prop.~A.8]{moser18_50}
\begin{IEEEeqnarray}{c}
  \xi \Q(\xi) \le \phi(\xi), \quad \xi \ge 0.
\end{IEEEeqnarray}
Similarly,
\begin{IEEEeqnarray}{rCl}
  -\int_{\amp}^{\infty} {W_{\mU,\ell}(y|x)\log{{R_{\mU,\ell}(y)} }}
  \dd y 
  & \leq &  - \log \frac{\beta}{\sqrt{2\pi e}\sigma_{\mU,\ell}}
  \cdot \Qf{\frac{\amp-x}{\sigma_{\mU,\ell}}}.
  \label{eq:e208} 
\end{IEEEeqnarray}
Moreover, we have
\begin{IEEEeqnarray}{rCl}
  -\int_{0}^{\amp} {W_{\mU,\ell}(y|x)\log{{R_{\mU,\ell}(y)} }} \dd y
  & = &  -\int_{0}^{\amp} \frac{1}{\sqrt{2\pi}\sigma_{\mU,\ell}}
  \ope^{-\frac{(y-x)^2}{2\sigma_{\mU,\ell}^{2}}}
  \log{\frac{(1-\beta)}{\amp}} \dd y
  \\
  & = & \log \left(\frac{\amp}{1-\beta}\right) \cdot
  \left(1-\Qf{\frac{x}{\sigma_{\mU,\ell}}}
    -\Qf{\frac{\amp-x}{\sigma_{\mU,\ell}}}\right).   
  \IEEEeqnarraynumspace
  \label{eq:20}
\end{IEEEeqnarray}

We choose
\begin{IEEEeqnarray}{c}
  \beta = \frac{\sqrt{2\pi e}\sigma_{\mU,\ell}}{\amp+\sqrt{2\pi
      e}\sigma_{\mU,\ell}}
\end{IEEEeqnarray}
and obtain from \eqref{eq:e207}, \eqref{eq:e208}, and \eqref{eq:20}
\begin{IEEEeqnarray}{c}
  - \E[W_{\mU,\ell}(Y_{\mU,\ell} | X_{\mU,\ell})]{\log
    R_{\mU,\ell}(Y_{\mU,\ell})}  \leq \log\bigl(\amp+\sqrt{2\pi
    e}\sigma_{\mU,\ell}\bigr). 
  \IEEEeqnarraynumspace
  \label{eq:avgupp3}
\end{IEEEeqnarray}
Substituting \eqref{eq:avgupp3} into \eqref{eq:dual} then yields
\begin{IEEEeqnarray}{rCl}
  \C_{\mat{H}}(\amp, \alpha\amp)
  & \leq & \sup_{\vect{p}} \Biggl\{  \HH(\vect{p}) - \frac{\nr}{2}
  \log 2\pi e + \sum_{\mU \in \U}
  p_{\mU} \log \abs{\det \mat{H}_{\mU}}
  \nonumber\\
  && \qquad + \sum_{\mU \in \U} p_{\mU}
  \sum_{\ell=1}^{\nr}\log\Bigl(\amp+\sqrt{2\pi e}
  \sigma_{\mU,\ell}\Bigr)\Biggl\} 
  \\
  & = & \sup_{\vect{p}} \Biggl\{ \HH(\vect{p}) + \sum_{\mU \in \U}
  p_{\mU}\log \frac{\abs{\det\mat{H}_{\mU}}}{\V} + \log\V
  \nonumber\\
  && \qquad + \sum_{\mU \in \U} p_{\mU} \sum_{\ell=1}^{\nr}
  \log\left(\sigma_{\mU,\ell}+\frac{\amp}{\sqrt{2\pi e}}\right) \Biggl\} 
  \IEEEeqnarraynumspace
  \\
  & = & \sup_{\vect{p}} \Biggl\{\log\V - \const{D}(\vect{p}\|\vect{q}) +
  \sum_{\mU \in \U} p_{\mU} \sum_{\ell=1}^{\nr} \log
  \left(\sigma_{\mU,\ell}+\frac{\amp}{\sqrt{2\pi e}}\right) \Biggl\}. 
\end{IEEEeqnarray}

\subsection{Proof of Theorem~\ref{thm:them6}}

We choose 
\begin{IEEEeqnarray}{c}
  R_{\mU,\ell}(y) =
  \begin{cases} 
    \frac{\beta}{\sqrt{2\pi}\sigma_{\mU,\ell}}
    \ope^{-\frac{y^2}{2\sigma_{\mU,\ell}^2}} 
    & \textnormal{if } y \in (-\infty, 0),   
    \\[2mm]
    \frac{1-\beta}{\amp} \cdot \frac{\mu}{1-\ope^{-\mu}}
    \ope^{-\frac{\mu y}{\amp}}  & \textnormal{if } y \in [0,\amp], 
    \\[1mm]
    \frac{\beta}{\sqrt{2\pi}\sigma_{\mU,\ell}}
    \ope^{-\frac{(y-\amp)^2}{2\sigma_{\mU,\ell}^2}}  
    &  \textnormal{if } y \in (\amp, \infty), 
  \end{cases}
\end{IEEEeqnarray}
where $\beta\in(0,1)$ and $\mu>0$ will be specified
later.

We notice that the inequalities in \eqref{eq:e207} and \eqref{eq:e208}
still hold, while 
\begin{IEEEeqnarray}{rCl}
  \IEEEeqnarraymulticol{3}{l}{%
    -\int_{0}^{\amp} {W_{\mU,\ell}(y|x)\log{{R_{\mU,\ell}(y)} }} \dd y
  }\nonumber\\*\quad%
  & = & -\int_{0}^{\amp}
  \frac{1}{\sqrt{2\pi}\sigma_{\mU,\ell}}
  \ope^{-\frac{(y-x)^2}{2\sigma_{\mU,\ell}^2}} 
  \left(\log{\frac{1-\beta}{\amp}}\frac{\mu}{1-\ope^{-\mu}}
    -\frac{\mu}{\amp}y \right) \dd y
  \\
  & = & -\log \left(\frac{1-\beta}{\amp} \frac{\mu}{1-\ope^{-\mu}}\right)
  \left(1-\Qf{\frac{x}{\sigma_{\mU,\ell}}} -
    \Qf{\frac{\amp-x}{\sigma_{\mU,\ell}}} \right)
  \nonumber\\
  && +\> \frac{\mu\sigma_{\mU,\ell}}{\amp}  \left(
    \phi\left(\frac{x}{\sigma_{\mU,\ell}}\right)
    - \phi\left(\frac{\amp-x}{\sigma_{\mU,\ell}}\right) \right)
  + \frac{\mu}{\amp} x \left(1 - \Qf{\frac{x}{\sigma_{\mU,\ell}}}
    - \Qf{\frac{\amp-x}{\sigma_{\mU,\ell}}} \right)
  \IEEEeqnarraynumspace
  \\
  & \leq & -\log\left(\frac{1-\beta}{\amp}
    \frac{\mu}{1-\ope^{-\mu}}\right) 
  \left(1- \Qf{\frac{x}{\sigma_{\mU,\ell}}}
    - \Qf{\frac{\amp-x}{\sigma_{\mU,\ell}}} \right)
  \nonumber \\
  && +\> \frac{\mu\sigma_{\mU,\ell}}{\amp}\left(\phi(0)
    - \phi\left(\frac{\amp}{\sigma_{\mU,\ell}} \right) \right)
  + \frac{\mu}{\amp} x \left(1 -
    2\Qf{\frac{\amp}{2\sigma_{\mU,\ell}}}\right) 
  \label{eq:E206} 
  \\
  & \leq & -\log\left(\frac{1-\beta}{\amp}
    \frac{\mu}{1-\ope^{-\mu}}\right) 
  \left(1-\Qf{\frac{x}{\sigma_{\mU,\ell}}} -
    \Qf{\frac{\amp-x}{\sigma_{\mU,\ell}}} \right)
  \nonumber \\ 
  && +\> \frac{\mu\sigma_{\mU,\ell}}{\amp}\left( \phi(0)
    - \phi\left(\frac{\amp}{\sigma_{\mU,\ell}}\right) \right)
  + \frac{\mu}{\amp} x. 
  \label{eq:E2061} 
\end{IEEEeqnarray}
Here \eqref{eq:E206} follows from the fact that, for $\xi\in[0,\amp]$,
$1-\Qf{\xi}-\Qf{\amp-\xi}$ achieves the maximum value at
$\xi=\frac{\amp}{2}$, and that $\phi(\xi)$ is monotonically
decreasing; and \eqref{eq:E2061} holds because $1-2\Qf{\xi} \le 1$ and
because $x\ge 0$.

Combining \eqref{eq:e207}, \eqref{eq:e208}, and \eqref{eq:E2061}, and
choosing
\begin{IEEEeqnarray}{c}
  \label{eq:beta}
  \beta = \frac{\mu\sqrt{2\pi e}\sigma_{\mU,\ell}}{\amp(1-\ope^{-\mu})
    +\mu\sqrt{2\pi e}\sigma_{\mU,\ell}} 
\end{IEEEeqnarray}
now yield
\begin{IEEEeqnarray}{rCl}
  \IEEEeqnarraymulticol{3}{l}{%
    -\E[W_{\mU,\ell}(Y_{\mU,\ell} | x_{\mU,\ell})]{\log
      R_{\mU,\ell}(Y_{\mU,\ell})}  
  }\nonumber\\*\quad%
  & \leq & \log\left(\sqrt{2\pi
      e}\sigma_{\mU,\ell}+\amp\cdot\frac{1-\ope^{-\mu}}{\mu} \right) 
  + \frac{\mu\sigma_{\mU,\ell}}{\amp \sqrt{2\pi}} 
  \left(1-\ope^{-\frac{\amp^{2} }{2\sigma_{\mU,\ell}^2}}\right) +
  \frac{\mu}{\amp} 
  x_{\mU,\ell}.
  \IEEEeqnarraynumspace 
  \label{eq:avgupp}
\end{IEEEeqnarray}
Substituting \eqref{eq:avgupp} into \eqref{eq:dual}, we have 
\begin{IEEEeqnarray}{rCl}
  \IEEEeqnarraymulticol{3}{l}{%
    \C_{\mat{H}}(\amp, \alpha\amp)
  }\nonumber\\*\;%
  & \leq & \HH(\vect{p}^*) + \sum_{\mU \in \U} p^*_{\mU}\log
  \abs{\det{\mat{H}_{\mU}}}
  - \frac{\nr}{2}\log 2\pi e
  \nonumber\\
  && + \sum_{\mU \in \U} p^*_{\mU} \sum_{\ell=1}^{\nr} \log \left(\sqrt{2\pi
      e}\sigma_{\mU,\ell} + \amp\cdot\frac{1-\ope^{-\mu}}{\mu}\right)
  \nonumber\\
  && +\> \frac{\mu}{\amp \sqrt{2\pi}}
  \sum_{\mU\in\U} p^*_{\mU} \sum_{\ell=1}^{\nr} \sigma_{\mU,\ell} \left(1
    -\ope^{-\frac{\amp^{2}}{2\sigma_{\mU,\ell}^2}}\right)
  \nonumber\\
  && +\> \frac{\mu}{\amp}\sum_{\mU\in\U} p^*_{\mU}
  \sum_{\ell=1}^{\nr} \Econd{X_{\mU,\ell}}{\rvUstar=\mU}
  \label{eq:27}
  \\
  & = & \HH(\vect{p}^*) + \sum_{\mU \in \U} p^*_{\mU}\log
  \frac{\abs{\det{\mat{H}_{\mU}}}}{\V} + \log\V 
  + \sum_{\mU \in \U} p^*_{\mU} \sum_{\ell=1}^{\nr} \log
  \left(\sigma_{\mU,\ell} +
    \frac{\amp}{\sqrt{2\pi e}}\cdot\frac{1-\ope^{-\mu}}{\mu}\right) 
  \nonumber\\
  && +\> \frac{\mu}{\amp \sqrt{2\pi}} \sum_{\mU\in\U} p^*_{\mU}
  \sum_{\ell=1}^{\nr} \sigma_{\mU,\ell} \left(1
    -\ope^{-\frac{\amp^{2}}{2\sigma_{\mU,\ell}^2}}\right)
  + \frac{\mu}{\amp} \sum_{\mU\in\U} p^*_{\mU}
  \left\|\inv{\mat{H}_{\mU}} \bigl(
    \Econd{\vect{X}^{\star}}{\rvUstar=\mU} -\vect{v}_{\mU} \bigr)\right\|_1
  \nonumber\\*
  \label{eq:avgconst2}
  \\
  & \le & \log\V - \const{D}(\vect{p}^*\|\vect{q})
  + \sum_{\mU \in \U} p^*_{\mU} \sum_{\ell=1}^{\nr} \log
  \left(\sigma_{\mU,\ell} +
    \frac{\amp}{\sqrt{2\pi e}} \cdot\frac{1-\ope^{-\mu}}{\mu}\right) 
  \nonumber\\
  && +\> \frac{\mu}{\amp \sqrt{2\pi}} \sum_{\mU\in\U} p^*_{\mU}
  \sum_{\ell=1}^{\nr} \sigma_{\mU,\ell} \left(1
    -\ope^{-\frac{\amp^{2}}{2\sigma_{\mU,\ell}^2}}\right)
  + \mu \left( \alpha - \sum_{\mU\in\U} p^*_{\mU} s_{\mU} \right),
  \label{eq:avgconst}              
\end{IEEEeqnarray}
where \eqref{eq:avgconst2} follows from \eqref{eq:XI}, and
\eqref{eq:avgconst} from
\eqref{eq:new_constraint}. Theorem~\ref{thm:them6} is proven by taking the
supremum over the probability vector $\vect{p}$ and the infimum over
$\mu>0$.

\subsection{Proof of Theorem~\ref{thm:them7}}

We choose
\begin{IEEEeqnarray}{c}
  R_{\mU,\ell}(y)  =
  \begin{cases} 
    \frac{1}{\sqrt{2\pi}\sigma_{\mU,\ell}}
    \ope^{-\frac{y^2}{2\sigma_{\mU,\ell}^2}}
    & \textnormal{if } {y} \in (-\infty,-\delta),  
    \\
    \frac{\mu}{\amp} \cdot
    \frac{1-2\Qf{\frac{\delta}{\sigma_{\mU,\ell}}}}{\ope^{\frac{\mu
          \delta}{\amp}}-   
      \ope^{-\mu(1+\frac{\delta}{\amp})}} 
    \ope^{-\frac{\mu y}{\amp}}
    & \textnormal{if } y \in [-\delta,\amp+\delta], 
    \\
    \frac{1}{\sqrt{2\pi}\sigma_{\mU,\ell}}
    \ope^{-\frac{(y-\amp)^2}{2\sigma_{\mU,\ell}^2}} 
    &\textnormal{if } y \in (\amp+\delta,\infty),
  \end{cases}
\end{IEEEeqnarray}
where $\delta, \mu>0$ are free parameters.  Following the steps in the
proof of \cite[App.~B.B]{lapidothmoserwigger09_7} and bounding
$1-\Qf{\xi_1}-\Qf{\xi_2} \le 1$, we obtain:
\begin{IEEEeqnarray}{rCl}
  \IEEEeqnarraymulticol{3}{l}{%
    -\E[X_{\mU, \ell}|\rvUstar=\mU]{\bigE[W_{\mU,\ell}(Y_{\mU,\ell} |
      X_{\mU,\ell})]{\log R_{\mU,\ell}(Y_{\mU,\ell})}}
  }\nonumber\\*\quad%
  & \leq & \log \left(\amp \cdot
    \frac{\ope^{\frac{{\mu\delta}}{\amp}}-
      \ope^{-\mu(1+\frac{\delta}{\amp})}}{
      \mu\left(1-2\Qf{\frac{\delta}{\sigma_{\mU,\ell}}}\right)} 
  \right)
  + \frac{\delta}{\sqrt{2\pi}\sigma_{\mU,\ell}}
  \ope^{-\frac{\delta^2}{2\sigma_{\mU,\ell}^2}} 
  + \Qf{\frac{\delta}{\sigma_{\mU,\ell}}}  \nonumber\\
  && +\> \frac{\mu\sigma_{\mU,\ell}}{\amp \sqrt{2\pi}}
  \left(\ope^{-\frac{\delta^2}{2\sigma_{\mU,\ell}^2}} -
    \ope^{-\frac{(\amp+\delta)^{2}}{2\sigma_{\mU,\ell}^2}}\right)
  + \frac{\mu}{\amp}\Econd{X_{\mU,\ell}}{\rvUstar=\mU}.  
  \label{eq:avgupp2}
\end{IEEEeqnarray}
Plugging \eqref{eq:avgupp2} into \eqref{eq:dual} and using a
derivation analogous to \eqref{eq:27}--\eqref{eq:avgconst} then
results in the given bound.

\subsection{Proof of Theorem~\ref{thm:them9}}

Using that
\begin{IEEEeqnarray}{c}
  \hh(\vect{Y}) \leq \frac{1}{2}\log\bigl((2\pi e)^{\nr}\det
  \cov{\vect{Y}} \bigl),  
\end{IEEEeqnarray}
where
\begin{IEEEeqnarray}{c}
  \cov{\vect{Y}} = \cov{\Xbar} + \> \mat{I},
\end{IEEEeqnarray}
we have
\begin{IEEEeqnarray}{rCl}
  \C_{\mat{H}}(\amp,\alpha \amp)
  & = & \max_{P_\vect{X}} \bigl\{ \hh(\vect{Y}) - \hh(\vect{Z})
  \bigr\} 
  \\
  & \leq & \max_{P_\vect{X}} \left\{ \frac{1}{2}\log\bigl( (2\pi
    e)^{\nr} \det(\cov{\Xbar}+\>\mat{I})\bigr)
    - \frac{1}{2}\log (2\pi e)^{\nr} \right\}
  \\
  & = & \max_{P_\vect{X}} \frac{1}{2}\log \det(\mat{I} + \cov{\Xbar}) 
  \\
  & \le & \max_{P_\vect{X}} \frac{1}{2}\log \prod_{i=1}^{\nr}
  \bigl(\mat{I} + \cov{\Xbar}\bigr)_{i,i}
  \label{eq:28}
  \\ 
  & = & \max_{P_\vect{X}} \frac{\nr}{2} \sum_{i=1}^{\nr} \frac{1}{\nr}
  \log \Bigl(1 + \bigl(\cov{\Xbar}\bigr)_{i,i}\Bigr) 
  \\ 
  & \le & \max_{P_\vect{X}} \frac{\nr}{2} \log \left(1 +
    \sum_{i=1}^{\nr} \frac{1}{\nr} \bigl(\cov{\Xbar}\bigr)_{i,i}\right) 
  \label{eq:29}
  \\ 
  & = & \max_{P_\vect{X}} \frac{\nr}{2} \log \left(1 +
    \frac{1}{\nr} \bigtrace{\cov{\Xbar}} \right)
  \\ 
  & = &  \frac{\nr}{2} \log \left(1 +  \frac{1}{\nr} \max_{P_\vect{X}}
    \bigtrace{\cov{\Xbar}} \right).
\end{IEEEeqnarray}
Here, \eqref{eq:28} follows from Hadamard's inequality, and
\eqref{eq:29} follows from Jensen's inequality.

\section{Derivation of Asymptotic Results}
\label{sec:deriv-asympt-results}

\subsection{Proof of Theorem~\ref{thm:them8}}

It follows directly from Theorem~\ref{thm:them4} that the RHS of
\eqref{eq:asycap1} is a lower bound to its LHS. To prove the other
direction, using that $\const{D}(\vect{p}\|\vect{q}) \ge 0$, we have
from Theorem~\ref{thm:them5} that
\begin{IEEEeqnarray}{c}
  \C_{\mat{H}}(\amp,\alpha \amp)
  \leq \log\V + \nr\log\left(\sigma_{\textnormal{max}}
    +\frac{\amp}{\sqrt{2\pi e}}\right) 
\end{IEEEeqnarray} 
where
\begin{IEEEeqnarray}{c}
  \sigma_{\textnormal{max}} \eqdef
  \max_{\substack{\mU\in\U\\\ell\in\{1, \ldots, \nr\}}}
  \sigma_{\mU,\ell}. 
  \label{eq:31}
\end{IEEEeqnarray}
This proves that the RHS of \eqref{eq:asycap1} is also an upper bound
to its LHS, and hence completes the proof of \eqref{eq:asycap1}.

Next, we prove \eqref{eq:asycap2}. Again, that its RHS is a lower
bound to its LHS follows immediately from
Theorem~\ref{thm:lowerbound2}. To prove the other direction, we define
for any $\vect{p}$:
\begin{IEEEeqnarray}{c}
  \lambda(\vect{p}) \eqdef \alpha - \sum_{\mU \in\U} p_{\mU} s_{\mU}
  \leq \alpha. 
  \label{eq:35}
\end{IEEEeqnarray}
We then fix $\amp \geq 1$ and choose $\mu$ depending on
$\lambda(\vect{p})$ to be
\begin{IEEEeqnarray}{c}
  \mu = 
  \begin{cases}
    \mu^{*}(\vect{p})
    & \textnormal{if } \amp^{-(1-\zeta)} <
    \frac{\lambda(\vect{p})}{\nr} < \frac{1}{2}, 
    \\ 
    \amp^{1-\zeta}
    & \textnormal{if } \frac{\lambda(\vect{p})}{\nr} \le
    \amp^{-(1-\zeta)}, 
    \\
    \frac{1}{\amp}
    & \textnormal{if } \frac{\lambda(\vect{p})}{\nr} \geq \frac{1}{2}, 
  \end{cases}
  \label{eq:30}
\end{IEEEeqnarray}
where $0 < \zeta < 1$ is a free parameter and $\mu^{*}(\vect{p})$ is
the unique solution to
\begin{IEEEeqnarray}{c}
  \frac{1}{\mu^{*}} - \frac{\ope^{-\mu^{*}}}{1-\ope^{-\mu^{*}}} =
  \frac{\lambda(\vect{p})}{\nr}. 
  \label{eq:33}
\end{IEEEeqnarray}
Note that in the first case of \eqref{eq:30},
\begin{IEEEeqnarray}{c}
  \amp^{-(1-\zeta)}
  < \frac{\lambda(\vect{p})}{\nr}
  = \frac{1}{\mu^*(\vect{p})}
  - \frac{\ope^{-\mu^*(\vect{p})}}{1-\ope^{-\mu^*(\vect{p})}}
  < \frac{1}{\mu^*(\vect{p})},   
\end{IEEEeqnarray}
i.e.,
\begin{IEEEeqnarray}{c}
  \mu^*(\vect{p}) < \amp^{1-\zeta},
\end{IEEEeqnarray}
and thus the choice \eqref{eq:30} makes sure that in all three cases,
irrespective of $\vect{p}$:
\begin{IEEEeqnarray}{c}
  \mu \leq \amp^{1-\zeta}, \quad \textnormal{for } \amp \geq 1.
  \label{eq:34}
\end{IEEEeqnarray}
Then, for $\amp \ge 1$, the upper bound \eqref{eq:upp2} can be
loosened as follows:
\begin{IEEEeqnarray}{c}
  \C_{\mat{H}}(\amp,\alpha \amp) 
  \leq \frac{1}{2} \log\left(\frac{\amp^{2\nr}\V^2}{(2\pi
      e)^{\nr}}\right) + f(\amp)  +\sup_{\vect{p}}
  g(\amp,\vect{p},\mu) 
  \label{eq:upperg}
\end{IEEEeqnarray}
where 
\begin{IEEEeqnarray}{rCl}
  f(\amp)
  & \eqdef & \frac{\nr\sigma_{\textnormal{max}}}{\amp^{\zeta} \sqrt{2\pi}} 
  \left(1-\ope^{-\frac{\amp^2}{2\sigma^2_{\textnormal{min}}}}\right),
  \\
  g(\amp,\vect{p},\mu)
  & \eqdef & \nr\log\left(
    \frac{\sqrt{2\pi e}\sigma_{\textnormal{max}}}{\amp}
    + \frac{1-\ope^{-\mu}}{\mu}\right)
  + \mu\lambda(\vect{p}) - \const{D}(\vect{p}\|\vect{q}) 
\end{IEEEeqnarray}
with $\sigma_{\textnormal{max}}$ defined in \eqref{eq:31} and with
\begin{IEEEeqnarray}{c}
  \sigma_{\textnormal{min}} \eqdef
  \min_{\substack{\mU\in\U\\\ell\in\{1, \ldots, \nr\}}}
  \sigma_{\mU,\ell}. 
\end{IEEEeqnarray}
Note that
\begin{IEEEeqnarray}{c}
  \lim_{\amp \to \infty} f(\amp) = 0.
  \label{eq:36}
\end{IEEEeqnarray}

Next, we upper-bound $g(\amp,\vect{p},\mu)$ individually for each
of the three different cases in \eqref{eq:30} to obtain a bound of
the form
\begin{IEEEeqnarray}{c}
  g(\amp, \vect{p},\mu) \leq 
  \begin{cases}
    g_1(\amp)
    & \textnormal{if } \amp^{-(1-\zeta)} <
    \frac{\lambda(\vect{p})}{\nr} < \frac{1}{2}, 
    \\ 
    g_2(\amp)
    & \textnormal{if } \frac{\lambda(\vect{p})}{\nr} \le
    \amp^{-(1-\zeta)}, 
    \\
    g_3(\amp)
    & \textnormal{if } \frac{\lambda(\vect{p})}{\nr} \geq \frac{1}{2}, 
  \end{cases}
\end{IEEEeqnarray}
for three functions $g_1$, $g_2$, and $g_3$ that only depend on $\amp$
but not on $\vect{p}$ or $\mu$.  Thus, we shall then obtain the bound
\begin{IEEEeqnarray}{c}
  \label{eq:37}
  g(\amp, \vect{p}, \mu) \leq \max \{g_1(\amp), g_2(\amp),
  g_3(\amp)\}, \quad \amp \ge 1. 
\end{IEEEeqnarray}
The functions $g_1$, $g_2$, and $g_3$ are introduced in the following.

For the first case where $\frac{\lambda(\vect{p})}{\nr}\in
\bigl(\amp^{-(1-\zeta)},\frac{1}{2}\bigr)$, we have
\begin{IEEEeqnarray}{rCl}
  g(\amp,\vect{p},\mu)
  & = & \nr\log\left( \frac{\sqrt{2\pi e}\sigma_{\textnormal{max}}}{\amp}
    + \frac{1-\ope^{-\mu^{*}(\vect{p})}}{\mu^{*}(\vect{p})}\right)
  + \mu^{*}(\vect{p})\lambda(\vect{p}) - \const{D}(\vect{p}\|\vect{q})
  \\
  & = & \nr\log\left(1+\frac{\mu^{*}(\vect{p})}{1-\ope^{-\mu^{*}(\vect{p})}}
    \cdot \frac{\sqrt{2\pi e}\sigma_{\textnormal{max}}}{\amp}\right)
  \nonumber\\
  && +\> \nr\left(1 -
    \log\left(\frac{\mu^{*}(\vect{p})}{1-\ope^{-\mu^{*}(\vect{p})}} 
    \right)
    - \frac{\mu^{*}(\vect{p})
      \ope^{-\mu^{*}(\vect{p})}}{1-\ope^{-\mu^{*}(\vect{p})}} 
  \right)
  - \const{D}(\vect{p}\|\vect{q}) 
  \label{eq:32}
  \\
  & \le & \sup_{\vect{p}\colon \frac{\lambda(\vect{p})}{\nr}\in
    \left(\amp^{\zeta-1},\frac{1}{2}\right)} \Biggl\{
  - \const{D}(\vect{p}\|\vect{q})
  + \nr\log\left(1+\frac{\mu^{*}(\vect{p})}{1-\ope^{-\mu^{*}(\vect{p})}}
    \cdot \frac{\sqrt{2\pi e}\sigma_{\textnormal{max}}}{\amp}\right)
  \nonumber\\
  && \qquad\qquad\qquad\qquad
  +\> \nr\left(1 -
    \log\left(\frac{\mu^{*}(\vect{p})}{1-\ope^{-\mu^{*}(\vect{p})}} 
    \right)
    - \frac{\mu^{*}(\vect{p})
      \ope^{-\mu^{*}(\vect{p})}}{1-\ope^{-\mu^{*}(\vect{p})}} 
  \right) \Biggr\}  
  \IEEEeqnarraynumspace
  \\
  & \eqdef & g_1(\amp).
  \label{eq:g1}
\end{IEEEeqnarray}
Here, in \eqref{eq:32} we have used \eqref{eq:33}.

For the second case where
$\frac{\lambda(\vect{p})}{\nr} \le \amp^{-(1-\zeta)}$, we use this
inequality in combination with \eqref{eq:30} to bound
\begin{IEEEeqnarray}{c}
  \mu \lambda(\vect{p}) \le \amp^{1-\zeta} \cdot \nr \amp^{-(1-\zeta)}
  = \nr. 
\end{IEEEeqnarray}
Because $\const{D}(\vect{p}\|\vect{q}) \ge 0$, we thus obtain
\begin{IEEEeqnarray}{rCl}
  g(\amp,\vect{p},\mu)
  & \le & \nr\log\left( \frac{\sqrt{2\pi
        e}\sigma_{\textnormal{max}}}{\amp}+\frac{1-\ope^{-\mu}}{\mu}\right) 
  + \nr 
  \\
  & = & \nr\log\left(\frac{\sqrt{2\pi e}\sigma_{\textnormal{max}}}{\amp}
    + \frac{1-\ope^{-\amp^{1-\zeta}}}{\amp^{1-\zeta}}\right)
  + \nr
  \\
  & \eqdef & g_2(\amp).
  \label{eq:g2}
\end{IEEEeqnarray}

For the third case where
$\frac{\lambda(\vect{p})}{\nr} \ge \frac{1}{2}$, we have
\begin{IEEEeqnarray}{rCl}
  g(\amp,\vect{p},\mu)
  & = & \nr\log\left(
    \frac{\sqrt{2\pi e} \sigma_{\textnormal{max}}}{\amp}
    + \frac{1-\ope^{-\frac{1}{\amp}}}{\frac{1}{\amp}}\right)
  + \frac{\lambda(\vect{p})}{\amp} 
  - \const{D}(\vect{p}\|\vect{q})
  \\
  & \le & \nr\log\left(
    \frac{\sqrt{2\pi e} \sigma_{\textnormal{max}}}{\amp}
    + \frac{1-\ope^{-\frac{1}{\amp}}}{\frac{1}{\amp}}\right)
  + \frac{\alpha}{\amp} 
  - \inf_{\vect{p}\colon \frac{\lambda(\vect{p})}{\nr} >
    \frac{1}{2}} \const{D}(\vect{p}\|\vect{q})  
  \\       
  & \eqdef & g_3(\amp).
  \label{eq:g3}
\end{IEEEeqnarray}
Here, we used \eqref{eq:35} to bound $\lambda(\vect{p}) \le \alpha$.

We have now established \eqref{eq:37} for the three functions defined
in \eqref{eq:g1}, \eqref{eq:g2}, and \eqref{eq:g3}, respectively. We
now analyze the maximum in \eqref{eq:37} when $\amp \to \infty$. Since
$g_2(\amp)$ tends to $-\infty$ as $\amp \to \infty$, and since
$g_1(\amp)$ and $g_3(\amp)$ are both bounded from below for
$\amp \ge 1$, we know that, for large enough $\amp$, $g_2(\amp)$ is
strictly smaller than $\max \{ g_1(\amp), g_3(\amp)\}$.

We next look at $g_3(\amp)$ when $\amp\to\infty$. Note that
\begin{IEEEeqnarray}{c}
  \lim_{\amp\to\infty} \frac{1-\ope^{-\frac{1}{\amp}}}{\frac{1}{\amp}}
  = 1,
\end{IEEEeqnarray}
therefore
\begin{IEEEeqnarray}{rCl}
  \lim_{\amp\to\infty} g_3(\amp)
  & = & - \inf_{\vect{p}\colon \frac{\lambda(\vect{p})}{\nr} >
    \frac{1}{2}} \const{D}(\vect{p}\|\vect{q})
  \\
  & = & - \inf_{\vect{p} \colon \alpha - \sum_{\mU \in\U} p_{\mU}
    s_{\mU} \ge \frac{\nr}{2}} \const{D}(\vect{p}\|\vect{q})
  \\ 
  & = & - \inf_{\vect{p} \colon \alpha - \sum_{\mU \in\U} p_{\mU}
    s_{\mU} = \frac{\nr}{2}} \const{D}(\vect{p}\|\vect{q}), \label{eq:250}
\end{IEEEeqnarray}
where the last equality holds because given
$\alpha < \alpha_{\textnormal{th}}$, an optimal $\vect{p}$ will meet
the constraint with equality.

It remains to investigate the behavior of $g_1(\amp)$ when
$\amp \to \infty$. To this end, we define
\begin{IEEEeqnarray}{rCl}
  \tilde{g}_1(\amp,\vect{p})
  & \eqdef & - \const{D}(\vect{p}\|\vect{q}) + \nr\log\left(
    1+\frac{\mu^{*}(\vect{p})}{1-\ope^{-\mu^{*}(\vect{p})}} \cdot
    \frac{\sqrt{2\pi e}\sigma_{\textnormal{max}}}{\amp}\right) 
  \nonumber\\
  && +\> \nr \left( 1
    - \log\left(\frac{\mu^{*}(\vect{p})}{1-\ope^{-\mu^{*}(\vect{p})}}\right)
    - \frac{\mu^{*}(\vect{p})
      \ope^{-\mu^{*}(\vect{p})}}{1-\ope^{-\mu^{*}(\vect{p})}}\right),
\end{IEEEeqnarray}
and note that, for any fixed $\vect{p}$,
\begin{IEEEeqnarray}{rCl}
  \Delta(\amp,\vect{p})
  & \eqdef & \tilde{g}_1(\amp,\vect{p}) -
  \lim_{\amp \rightarrow \infty} \tilde{g}_1(\amp,\vect{p})  
  = \log\left(
    1+\frac{\mu^{*}(\vect{p})}{1-\ope^{-\mu^{*}(\vect{p})}} \cdot
    \frac{\sqrt{2\pi e}\sigma_{\textnormal{max}}}{\amp}\right). 
  \IEEEeqnarraynumspace
\end{IEEEeqnarray}
Since, when $\amp \to \infty$,
\begin{IEEEeqnarray}{rCl}
  \abs{\Delta(\amp,\vect{p})}
  & \leq & \log\left( 1 + \abs{\frac{1}{1-\ope^{-\amp^{1-\zeta}}}
      \cdot \frac{\sqrt{2\pi e}\sigma_{\textnormal{max}}}{\amp^{\zeta}}}
  \right)  \to \log(1) = 0,
\end{IEEEeqnarray}
we see that $\tilde{g}_1(\amp,\vect{p})$ converges uniformly over
$\vect{p}$ as $\amp \to \infty$, and therefore we are allowed to
interchange limit and supremum:
\begin{IEEEeqnarray}{rCl}
  \IEEEeqnarraymulticol{3}{l}{%
    \lim_{\amp\to\infty} g_1(\amp)
  }\nonumber\\*\quad%
  & = & \lim_{\amp\to\infty} \sup_{\vect{p}\colon
    \frac{\lambda(\vect{p})}{\nr}\in 
    \left(\amp^{\zeta-1},\frac{1}{2}\right)} \tilde{g}_1(\amp,\vect{p})
  \\
  & = & \sup_{\vect{p}\colon \frac{\lambda(\vect{p})}{\nr}\in
    \left(0,\frac{1}{2}\right)} \lim_{\amp\to\infty}
  \tilde{g}_1(\amp,\vect{p}) 
  \\
  & = & \sup_{\vect{p}\colon \frac{\lambda(\vect{p})}{\nr}\in
    \left(0,\frac{1}{2}\right)} \Biggl\{ 
  \nr\left(1-\log{\frac{\mu^{*}(\vect{p}) }{1-\ope^{-\mu^{*}(\vect{p}) }}} -
    \frac{\mu^{*}(\vect{p}) 
      \ope^{-\mu^{*}(\vect{p}) }}{1-\ope^{-\mu^{*}(\vect{p}) }}\right)
  - \const{D}(\vect{p}\|\vect{q}) \Biggl\}
    \label{eq:lam_con3}
  \\
  & = & \sup_{\vect{p}\colon \lambda(\vect{p})\in
    \left(\max\{0,\frac{\nr}{2} +\alpha - 
      \alpha_{\textnormal{th}}\},\min\{\frac{\nr}{2},\alpha\}\right)}
  \Biggl\{ \nr\left(1-\log{\frac{\mu^{*}(\vect{p})
      }{1-\ope^{-\mu^{*}(\vect{p}) }}} - 
    \frac{\mu^{*}(\vect{p}) 
      \ope^{-\mu^{*}(\vect{p}) }}{1-\ope^{-\mu^{*}(\vect{p}) }}\right)
  \nonumber\\
  &&  \hspace{58mm} -\> \const{D}(\vect{p}\|\vect{q}) \Biggl\} 
  \label{eq:lam_con2}
  \\
  & = & \nu.
  \label{eq:lam_con}
\end{IEEEeqnarray}
Here, in \eqref{eq:lam_con2} we are allowed to restrict the
supremum\footnote{Notice that because of the supremum and continuity,
  we can restrict to the open interval instead of the closed
  interval.} to
$\lambda(\vect{p}) \in ( \frac{\nr}{2}+\alpha-\alpha_{\textnormal{th}},
\alpha )$ because of \eqref{eq:35} and because
\begin{IEEEeqnarray}{c}
  \lambda(\vect{q}) \eqdef \alpha - \sum_{\mU \in \U}
  s_{\mU}q_{\mU}
  = \alpha - \alpha_{\textnormal{th}} + \frac{\nr}{2}
\end{IEEEeqnarray}
and for any $\vect{p}$ such that
$ \lambda(\vect{p}) \leq \lambda(\vect{q})$ the objective function in
\eqref{eq:lam_con3} is smaller than for $\vect{p}=\vect{q}$. In fact,
$-\const{D}(\vect{p}\| \vect{q})$ is clearly maximized for
$\vect{p}=\vect{q}$ and
\begin{IEEEeqnarray}{rCl} 
  \mu^*(\vect{p})\mapsto \nr\left(1-\log{\frac{\mu^{*}(\vect{p})
      }{1-\ope^{-\mu^{*}(\vect{p}) }}} - 
    \frac{\mu^{*}(\vect{p}) 
      \ope^{-\mu^{*}(\vect{p}) }}{1-\ope^{-\mu^{*}(\vect{p}) }}\right)  
\end{IEEEeqnarray}
is decreasing in $\mu^*(\vect{p})$, which is a decreasing function of
$\lambda(\vect{p})$; see \eqref{eq:33}.  Finally, \eqref{eq:lam_con}
follows from the definition of $\nu$ in \eqref{eq:nu}.

It is straightforward to see that $\nu$ is larger than the RHS of
\eqref{eq:250}.  Therefore,
\begin{IEEEeqnarray}{c}
  \lim_{\amp\to\infty} \max \{g_1(\amp),g_2(\amp),g_3(\amp)\} = \nu.
  \label{eq:258}
\end{IEEEeqnarray} 
Combining \eqref{eq:upperg} with \eqref{eq:36}, \eqref{eq:37}, and
\eqref{eq:258} proves the theorem.

\subsection{Proof of Theorem~\ref{thm:them11}}

From \cite[Corollary $2$]{prelovverdu04_1}, it is known that the
capacity is lower-bounded as
\begin{IEEEeqnarray}{c}
  \C_{\mat{H}}(\amp,\alpha\amp)
  \geq \frac{1}{2} \max_{P_{\Xbar}}\bigtrace{\cov{\Xbar}}
  + o\biggl( \max_{P_{\Xbar}} \bigtrace{\cov{\Xbar}} \biggr).
\end{IEEEeqnarray} 
For an upper bound, we use that
\begin{IEEEeqnarray}{c}
  \log(1+\xi) \le \xi, \quad \xi > 0,
\end{IEEEeqnarray}
and obtain from Theorem~\ref{thm:them9} that
\begin{IEEEeqnarray}{c}
  \C_{\mat{H}}(\amp,\alpha\amp) \le \frac{1}{2} \max_{P_{\Xbar}}
  \bigtrace{ \cov{\Xbar}}. 
\end{IEEEeqnarray}
The theorem is proven by normalizing $\Xbar$ by $\amp$, which
results in a factor $\amp^2$, and by then letting $\amp$ go to zero.

\bibliographystyle{./myIEEEtran}
\bibliography{./defshort1,./biblio1}

\end{document}